\documentclass[array,eqsecnum,aps,prd,showpacs]{revtex4}
\usepackage[dvips]{graphicx}

\newcommand{\nn}{\nonumber}
\newcommand{\ovl}[1]{\overline{#1}}

\newcommand{\p}{\partial}
\newcommand{\dfrac}{\displaystyle\frac}
\newcommand{\pslash}{p\kern-1ex /}
\newcommand{\lslash}{l\kern-1ex /}
\newcommand{\kslash}{k\kern-1ex /}
\newcommand{\dslash}{\p\kern-1.2ex /}
\newcommand{\Dslash}{{\cal D}\kern-1.5ex /}
\newcommand{\Aslash}{A\kern-1.2ex /}

\newcommand{\tr}{{\rm tr}}

\newcommand{\re}{{\rm Re}}
\newcommand{\im}{{\rm Im}}

\newcommand{\ket}[1]{\left| #1\right\rangle}

\newcommand{\vev}[1]{\left\langle #1 \right\rangle}
\newcommand{\VEV}[3]{\left\langle #1\left| #2 \right| #3\right\rangle}

\begin{document}
\draft

\title{
\vspace{-1.5cm}
\begin{flushright}
{\normalsize UTHEP-445}\\
{\normalsize UTCCP-P-109}\\
\end{flushright}
Calculation of Non-Leptonic Kaon Decay Amplitudes 
from $K\to\pi$ Matrix Elements in Quenched Domain-Wall QCD}

\author{J.~Noaki$^1$}
\altaffiliation{present address:
        RIKEN BNL Research Center, Brookhaven National
        Laboratory, Upton, NY 11973, USA}
\author{S.~Aoki$^{1}$} 
\author{Y.~Aoki$^{1,2}$}
\altaffiliation{present address:
        RIKEN BNL Research Center, Brookhaven National
        Laboratory, Upton, NY 11973, USA}
\author{R.~Burkhalter$^{2}$} 
\author{S.~Ejiri$^{2}$}
\altaffiliation{present address: Department of Physics, University of Wales Swansea, Singleton Park, Swansea SA2 8PP, UK}
\author{M.~Fukugita$^{3}$}
\author{S.~Hashimoto$^{4}$} 
\author{N.~Ishizuka$^{1,2}$} 
\author{Y.~Iwasaki$^{1,2}$} 
\author{T.~Izubuchi$^{5}$}
\altaffiliation{presently on leave at Brookhaven National Laboratory,
Upton, NY 11973, USA} 
\author{K.~Kanaya$^{1,2}$} 
\author{T.~Kaneko$^{4}$} 
\author{Y.~Kuramashi$^{4}$} 
\author{V.~Lesk$^{2}$}
\author{K.~I.~Nagai$^{2}$}
\altaffiliation{present address: Theory Division, CERN, 
CH-1211 Geneva 23, Switzerland}
\author{M.~Okawa$^{4}$}
\author{Y.~Taniguchi$^{1}$} 
\author{A.~Ukawa$^{1,2}$} 
\author{T.~Yoshi\'e$^{1,2}$\\
(CP-PACS Collaboration)}
\affiliation{
  ${}^{1}$Institute of Physics, University of Tsukuba,
  Tsukuba, Ibaraki 305-8571, Japan \\
  ${}^{2}$Center for Computational Physics, University of Tsukuba,
  Tsukuba, Ibaraki 305-8577, Japan \\
  ${}^{3}$Institute for Cosmic Ray Research, University of Tokyo,
  Kashiwa 277-8582, Japan\\
  ${}^{4}$High Energy Accelerator Research Organization (KEK),
  Tsukuba, Ibaraki 305-0801, Japan\\
  ${}^{5}$Institute of Theoretical Physics, Kanazawa University,
  Ishikawa 920-1192, Japan
}
\date{March 11, 2002}

\begin{abstract}
We explore application of the domain wall fermion formalism of lattice QCD 
to calculate the $K\to\pi\pi$ decay amplitudes in terms of the 
$K^+\to\pi^+$ and $K^0\to 0$ hadronic matrix elements through relations 
derived in chiral perturbation theory.  Numerical simulations are carried out 
in quenched QCD using domain-wall fermion action for quarks and an 
RG-improved gauge action for gluons on a $16^3\times 32\times 16$ and 
$24^3\times 32\times 16$ lattice at $\beta=2.6$ corresponding to 
the lattice spacing $1/a\approx 2$~GeV.   Quark loop contractions 
which appear in Penguin diagrams are calculated by the random noise method, 
and the $\Delta I=1/2$ matrix elements which require 
subtractions with the quark loop contractions are obtained with a 
statistical accuracy of about 10\%. We investigate the chiral 
properties required of the $K^+\to\pi^+$ matrix elements. Matching the 
lattice matrix elements to those in the continuum 
at $\mu=1/a$ using the perturbative renormalization factor to one loop 
order, and running to the scale $\mu=m_c=1.3$ GeV with the renormalization 
group for $N_f=3$ flavors, we calculate all the matrix elements needed 
for the decay amplitudes.

With these matrix elements, the $\Delta I=3/2$ decay amplitude $\re A_2$ 
shows a good agreement with experiment after an extrapolation to the 
chiral limit.  The $\Delta I=1/2$ amplitude $\re A_0$, on the other hand, 
is about 50--60\% of the experimental one even after chiral extrapolation. 
In view of the insufficient enhancement of the $\Delta I=1/2$ contribution, 
we employ the experimental values for the real parts of the decay amplitudes 
in our calculation of $\varepsilon'/\varepsilon$.  
The central values of our result indicate that 
the $\Delta I=3/2$ contribution is larger than the $\Delta I=1/2$ 
contribution so that $\varepsilon'/\varepsilon$ is negative and 
has a magnitude of order $10^{-4}$.  
We discuss in detail possible systematic uncertainties, which
seem too large for a definite conclusion on the value of
$\varepsilon'/\varepsilon$.

\end{abstract}



\maketitle

\section{INTRODUCTION}

Understanding non-leptonic weak processes of kaon, 
in particular the $K\to\pi\pi$ decay, 
represents one of the keys to establishing the Standard Model and probing 
the physics beyond it.
This decay exhibits two significant phenomena, namely, the $\Delta I=1/2$
rule, which is a large enhancement of the decay mode with $\Delta I=1/2$
relative to that with $\Delta I=3/2$, and the direct CP violation 
~\cite{ALA-HARA99,FANTI99}, which is naturally built in the Model for 
three or more families of quarks~\cite{KobayashiMaskawa}.
While both of these phenomena are well established by experiments, 
theoretical calculations with sufficient reliability that allow examinations 
of the Standard Model predictions against the experimental results are 
yet to be made. The main reason for this status is the 
difficulty in calculating the hadronic matrix elements of local operators 
which appear in the effective weak Hamiltonian for the decay amplitudes. 
At the energy scales relevant for these operators, analytic treatments
such as the $1/N_c$ expansion are not sufficiently powerful to reliably 
evaluate the effect of the strong interactions in the matrix elements.
In fact, the $\Delta I=1/2$ rule, which is supposed to arise from QCD 
effects, has not been quantitatively explained by analytic methods so far.
With these backgrounds, Monte Carlo simulations of lattice QCD provide 
a hopeful method for the calculation of the decay amplitudes.

A natural framework for theoretical calculations of the decay amplitudes 
is provided by the effective weak Hamiltonian, $H_W$, 
which follows from an operator product expansion (OPE) of weak currents  
~\cite{Buras-lec}:
\begin{eqnarray}
 H_W=\frac{G_F}{\sqrt{2}}V_{\rm us}V^*_{\rm ud}
\sum_{i}^{}W_i(\mu)Q_i(\mu)\label{OPE_HW} .
\end{eqnarray} 
Here the Wilson coefficients $W_i$ contain effects of the energy
scales higher than $\mu$ so that they can be calculated perturbatively.
Non-perturbative QCD effects are contained in the matrix elements of 
the local operators $Q_i$, 
and the calculation of these matrix elements, often called hadronic matrix 
elements (HME), is the task of lattice QCD 
~\cite{cabibbo84,brower84,bernard84,lellouch}. 
Our aim in this paper is to report on our attempt to obtain these matrix 
elements through numerical simulations of lattice QCD
using the domain wall formalism~\cite{Kaplan92,Shamir93,Shamir95}
for quarks.

The amplitudes for $K\to\pi\pi$ decay with $\Delta I=1/2$ and $3/2$ are
written as the matrix elements of $H_W$, 
\begin{eqnarray}
\VEV{(\pi\pi)_I}{H_W}{K^0}\equiv A_Ie^{i\delta_I}\label{amplitudes},
\end{eqnarray}
where the subscript $I=0$ or $2$ denotes the isospin of the final state
corresponding to $\Delta I=1/2$ or $3/2$, and $\delta_I$ is the 
phase shift from final state interactions $\pi\pi\to\pi\pi$ caused by 
QCD effects.  The $\Delta I =1/2$ rule, which is one of the focuses of 
our calculation, is described by the ratio of isospin amplitudes $A_I$:
\begin{equation}
\omega^{-1}\equiv\frac{\re A_0}{\re A_2}\approx 22.2.
\end{equation} 
Another focus is the parameter $\varepsilon'/\varepsilon$ of 
direct CP violation in the Standard Model.  The recent experimental 
results are 
\begin{eqnarray}
\frac{\varepsilon'}{\varepsilon}\equiv
\frac{\omega}{\sqrt{2}|\varepsilon|}\left[
\frac{\im A_2}{\re A_2}-\frac{\im A_0}{\re A_0}\right]
=\left\{\begin{array}{l}
 (20.7\pm 2.8)\cdot 10^{-4} ({\rm KTeV})~[1]        \\
 (15.3\pm 2.6)\cdot 10^{-4} ({\rm NA48})~[2]
\end{array}
\right. .
\label{epep}
\end{eqnarray}

In the numerical simulation of lattice QCD, matrix elements are 
generally extracted from Euclidean correlation functions
of the relevant operators and those which create the initial and final
states in their lowest energy levels. For sufficiently large Euclidean time
distances, excited states damp out and the matrix elements of the lowest
energy states are left.  In fact, the kaon B-parameter $B_K$ has been 
successfully obtained from the three-point correlation function of $K^0$ and 
$\overline{K}^0$ and an insertion of the $\Delta S=2$ weak Hamiltonian 
~\cite{Kuramashi99}.
However, in the calculation of the four-point function,
$\vev{\pi(t_2)\pi(t_1) H_W(t_H) K(t_K)}$,
necessary for the $K\to\pi\pi$ decay, there is a severe limitation as 
pointed out by Maiani and Testa~\cite{MaianiTesta90}. 
They have shown that it is difficult to obtain the matrix elements unless 
the momentum of each of the two pions in the final state is set to zero. 

One of the ways to overcome the difficulty pursued in the past is 
to calculate the matrix elements with the two pions at rest, allowing a 
nonzero energy transfer $\Delta E=2m_\pi-m_K$ at the weak operator.
This generally causes mixings of unphysical lower dimension operators
through renormalization, which has to be removed. (See Ref.~\cite{lellouch} 
and references therein.)
Furthermore, the unphysical amplitudes 
obtained with $\Delta E\ne 0$ need to be extrapolated to physical ones 
by use of some effective theories such as chiral perturbation 
theory. Due to these problems and numerical difficulties of extracting
reasonable signals from four-point functions, this 
approach has not been successful for the $\Delta I=1/2$ amplitude 
despite many efforts over the years~\cite{Historic-reviews}. 
For the $\Delta I=3/2$ amplitude for which the operator mixing is absent, 
on the other hand, a recent study has obtained a result in agreement with 
experiment~\cite{JLQCD-ReA2}.

Several proposals have been presented over the years for extracting the 
physical amplitude from the four-point functions 
~\cite{Ciuchini-etal96,Lellouch-Luscher,Marctinelli-etal2001}. 
Feasibility studies for implementing them in practical simulations are 
yet to come, however.

In this paper we explore a method proposed by Bernard {\it et al.}
\cite{Bernard85} which is alternative to calculating the three-point 
function.
In this method, which we shall call as reduction method, chiral perturbation 
theory ($\chi$PT) is used to relate the matrix elements for 
$K\to\pi\pi$ to those for $K\to\pi$ and $K\to 0$ (vacuum), 
and the latter amplitudes are calculated in lattice QCD. 
Since this calculation involves only 
three- and two-point correlation functions, the Maiani-Testa problem 
mentioned above is avoided.  Statistical fluctuations are also expected to be 
diminished compared with the case of four-point correlation functions. 

Early attempts with this method~\cite{Historic-reviews} encountered 
large statistical fluctuations in the correlation functions so that  
meaningful results were difficult to obtain. 
For the Wilson fermion action or its ${\cal O}(a)$ improved version, there
is an added difficulty that the mixing of operators of wrong chirality 
caused by explicit chiral symmetry breaking of the 
action has to be removed. The mixing problem has been resolved only for the 
$\Delta I=3/2$ operators so far~\cite{Gupta-etal,Donini-etal99,Lell-David}. 
 
The first results on the $\Delta I=1/2$ rule and $\varepsilon'/\varepsilon$ 
calculated with this method were recently reported~\cite{PK} using the  
staggered fermion action which keeps $U(1)$ subgroup of chiral 
symmetry.  In this work, however, a large dependence of the $\Delta I=3/2$ 
amplitude on the meson mass was seen, which made the chiral extrapolation 
difficult. Moreover, large uncertainties due to perturbative renormalization 
factors depending on the value of the matching point were reported. Hence 
clear statements on the viability of the method were difficult to 
make from this work. 

In this article we report on our attempt to apply the domain-wall fermion 
formalism of lattice QCD~\cite{Kaplan92,Shamir93,Shamir95} to the 
calculation of $K\to\pi\pi$ decay amplitudes in the context of the 
reduction method.  A major advantage of this approach over the conventional 
fermion formalisms is that full chiral symmetry can be expected to be 
realized for sufficiently large lattice sizes in the fifth dimension. 
Good chiral property of one of the $K\to\pi$ matrix elements, equivalent 
to the kaon $B$ parameter, was observed in the pioneering application of the 
formalism~\cite{Blum-Soni}.  Detailed investigations into the
realization of the chiral limit have been made in the quenched 
approximation for the plaquette and a renormalization group (RG) improved 
gluon actions~\cite{AIKT00,CPPACSdwqcd,RBCcollab}.  
It was found that the use of RG-improved action leads to much better chiral 
properties compared to the case of the plaquette action for similar lattice 
spacings~\cite{CPPACSdwqcd}. This prompts us to adopt the RG-improved action 
in our simulation. 

Another possible advantage of the domain wall formalism is ${\cal O}(a^2)$ 
scaling violation from the fermion sector as opposed to ${\cal O}(a)$ 
for the Wilson case.  
Indeed our domain wall fermion calculation of $B_K$~\cite{BK-CPPACS} 
exhibits only small scaling violation.  The magnitude of violation 
is much smaller compared to the staggered fermion case~\cite{BK-JLQCD} 
which is also expected to be ${\cal O}(a^2)$.  
An improved scaling behavior may be enhanced with the use of the 
RG-improved gluon action.  

This article is organized as follows. 
In Section \ref{basic-chpt}, we summarize the main points of 
the $\chi$PT reduction method.
For the construction of the formulae which relate the matrix elements 
for $K\to\pi$ and the $K\to\pi\pi$ decay amplitudes, 
the relations between the four quark 
operators $Q_i$  and $\chi$PT operators are considered at tree level 
on the basis of chiral transformation properties. 
The necessity of chiral symmetry on the lattice is emphasized. 
In Section~\ref{actions} we summarize the details of our numerical simulation
procedure. We discuss the form of lattice actions and 
the choice of optimal set of simulation parameters from the point of view of 
chiral properties. 
Some of the technical issues are also explained including 
renormalization of the four-quark operators and RG-running 
of the matrix elements to the relevant energy scale. 
The numerical results are reported in Section~\ref{result-HME} and
\ref{result-Phys}. The former contains results of hadronic matrix elements. 
In particular, we show that the subset of 
$K\to\pi$ matrix elements which are expected to vanish in the chiral 
limit satisfy this requirement. 
We then present the physical matrix elements and combine them 
with the Wilson coefficients, which are already
calculated perturbatively.  This leads us to results for the  
$\Delta I=1/2$ rule and $\varepsilon'/\varepsilon$. 
Our conclusions are given in Section~\ref{conclusions}.

A preliminary report of the present work was presented in 
Ref.~\cite{CPPACSepep}.  We refer to Refs.~\cite{Blum00,Mawhiney00} 
for a similar attempt, and Refs.~\cite{Vranas-ple,lellouch} 
for reviews.
 
\section{CHIRAL PERTURBATION THEORY REDUCTION METHOD}\label{basic-chpt}

\subsection{LOCAL OPERATORS}
We carry out our analyses choosing the energy scale $\mu$ in the OPE for 
the weak Hamiltonian (\ref{OPE_HW}) equal to the charm quark mass 
$m_c=1.3$ GeV.  In this case only $u,d$ and $s$ quarks appear in 
the local four-quark operators.  
Conventionally these operators are written as 
\begin{eqnarray}
 Q_1 &=&[\bar{s}_a\gamma_\mu(1-\gamma_5)u_b]
               [\bar{u}_b\gamma_\mu(1-\gamma_5)d_a] \label{defQ1} ,\\
 Q_2 &=&[\bar{s}_a\gamma_\mu(1-\gamma_5)u_a]
               [\bar{u}_b\gamma_\mu(1-\gamma_5)d_b]\label{defQ2} ,\\
 Q_3 &=&[\bar{s}_a\gamma_\mu(1-\gamma_5)d_a]
               \sum_{q}[\bar{q}_b\gamma_\mu(1-\gamma_5)q_b]
\label{defQ3} ,\\
 Q_4 &=&[\bar{s}_a\gamma_\mu(1-\gamma_5)d_b]
               \sum_q[\bar{q}_b\gamma_\mu(1-\gamma_5)q_a]\label{defQ4} ,\\
 Q_5 &=&[\bar{s}_a\gamma_\mu(1-\gamma_5)d_a]
               \sum_q[\bar{q}_b\gamma_\mu(1+\gamma_5)q_b]\label{defQ5} ,\\
 Q_6 &=&[\bar{s}_a\gamma_\mu(1-\gamma_5)d_b]
               \sum_q[\bar{q}_b\gamma_\mu(1+\gamma_5)q_a]\label{defQ6} ,\\
 Q_7 &=&\frac{3}{2}[\bar{s}_a\gamma_\mu(1-\gamma_5)d_a]
               \sum_q e_q[\bar{q}_b\gamma_\mu(1+\gamma_5)q_b]\label{defQ7} ,\\
 Q_8 &=&\frac{3}{2}[\bar{s}_a\gamma_\mu(1-\gamma_5)d_b]
               \sum_q e_q[\bar{q}_b\gamma_\mu(1+\gamma_5)q_a]\label{defQ8} ,\\
 Q_9 &=&\frac{3}{2}[\bar{s}_a\gamma_\mu(1-\gamma_5)d_a]
               \sum_q e_q[\bar{q}_b\gamma_\mu(1-\gamma_5)q_b]\label{defQ9} ,\\
 Q_{10} &=&\frac{3}{2}[\bar{s}_a\gamma_\mu(1-\gamma_5)d_b]
               \sum_q e_q[\bar{q}_b\gamma_\mu(1-\gamma_5)q_a] ,\label{defQ10}
\end{eqnarray}
where the indices $a,b$ denote color, and
the summation over $q$ appearing in $Q_3$ to $Q_{10}$ runs 
over the three light flavors, $q=u,d,s$, with the charge $e_u=2/3$ and 
$e_d=e_s=-1/3$.

With the use of Fierz rearrangements, one can derive the relations, 
\begin{eqnarray}
Q_4&=&Q_2+Q_3-Q_1 ,\label{Q4rel}\\
Q_9&=&\frac{3}{2}Q_1-\frac{1}{2}Q_3 ,\label{Q9rel}\\
Q_{10}&=&\frac{3}{2}Q_2-\frac{1}{2}Q_4=Q_2-\frac{1}{2}Q_3+\frac{1}{2}Q_1 .
\label{Q10rel}
\end{eqnarray}
Hence $Q_4$, $Q_9$, $Q_{10}$ are not independent operators.  
We emphasize that these relations do not hold in general 
d-dimensions where Fierz rearrangements cannot be used.

In terms of the irreducible representations of chiral $SU(3)_L\otimes SU(3)_R$ 
group, $Q_i$'s are classified as
\begin{eqnarray}
Q_1,Q_2,Q_9,Q_{10}\ \ :
&& (27_L,1_R)\oplus(8_L,1_R)\label{27-1rep} ,\\
Q_3,Q_4,Q_5,Q_6\ \ : 
& & (8_L,1_R) , \label{8-1rep}\\
Q_7,Q_8\ \ \hspace{1cm}: 
& & (8_L,8_R).
\label{8-8rep}
\end{eqnarray}
The operators $Q_i (i=1,\cdots, 10)$ are invariant under 
CPS symmetry, {\it i.e.,} the product of CP transformation and 
$d\leftrightarrow s$ interchange.
A basis of operators which are irreducible under chiral symmetry and 
invariant under CPS is given by 
\begin{eqnarray}
(8_L,1_R)&:& X_1=(\bar{s}d)_L(\bar{u}u)_L-(\bar{s}u)_L(\bar{u}d)_L ,
\label{CPSop-x1}\\
(8_L,1_R)&:& X_2=(\bar{s}d)_L\left[(\bar{u}u)_L+2(\bar{d}d)_L+
2(\bar{s}s)_L\right]+(\bar{s}u)_L(\bar{u}d)_L ,\label{CPSop-x2}\\
(27_L,1_R)&:& X_3=(\bar{s}d)_L\left[2(\bar{u}u)_L-(\bar{d}d)_L-(\bar{s}s)_L
\right]+2(\bar{s}u)_L(\bar{u}d)_L ,\label{CPSop-x3}\\
(8_L,1_R)&:& Y_1=(\bar{s}d)_L\left[(\bar{u}u)_R+(\bar{d}d)_R+(\bar{s}s)_R
\right],\ Y_1{}^c ,\label{CPSop-y1}\\
(8_L,8_R)&:& Y_2=(\bar{s}d)_L\left[2(\bar{u}u)_R-(\bar{d}d)_R-(\bar{s}s)_R
\right],\ Y_2{}^c, \label{CPSop-y2}
\end{eqnarray}
where $(\bar{s}{d})_L=\bar{s}\gamma_\mu(1-\gamma_5)d$ and
$(\bar{s}{d})_R=\bar{s}\gamma_\mu(1+\gamma_5)d$. 
The color and spinor indices are summed 
within each current except for $Y_i{}^c$ for which 
the color summation is taken across the two currents.
While $X_i$'s have the Lorentz structure of $L\otimes L$,
$Y_i$'s have that of $L\otimes R$.
All the independent local operators are written as linear
combinations of these operators :
\begin{eqnarray}
Q_1&=&\frac{1}{2}X_1+\frac{1}{10}X_2 +\frac{1}{5}X_3 \label{Q1} ,\\ 
Q_2&=&-\frac{1}{2}X_1+\frac{1}{10}X_2 +\frac{1}{5}X_3 \label{Q2} ,\\ 
Q_3&=&\frac{1}{2}X_1+\frac{1}{2}X_2 ,\\
Q_5&=&Y_1 ,               \\
Q_6&=&Y_1{}^c ,           \\
Q_7&=&\frac{1}{2}Y_2 ,    \\
Q_8&=&\frac{1}{2}Y_2^c .
\end{eqnarray}
The expressions for the dependent operators $Q_{4,9,10}$ are 
easily derived using (\ref{Q4rel})--(\ref{Q10rel}).

The final states in the $K\to\pi\pi$ decay can have either  
isospin $I=0$ or $2$, {\it i.e.} $\Delta I=1/2$ or $3/2$. Hence 
$Q_i$'s are decomposed as
\begin{eqnarray}
Q_i=Q_i^{(0)}+Q_i^{(2)}. \label{division-Q}
\end{eqnarray}
This decomposition is accomplished by constructing another basis of 
irreducible representations with the intrinsic isospin $I$.
The details are described in Appendix~\ref{app_split}.

\subsection{ CHIRAL PERTURBATION THEORY }

In the low energy region of strong interactions, the octet of pseudo
scalar mesons $\pi^0,\pi^\pm,K^0,\ovl{K}^0,K^\pm,\eta$ play a principal role 
as the Nambu-Goldstone bosons of spontaneously broken chiral symmetry 
$SU(3)_L\otimes SU(3)_R\to SU(3)_V$. 
In chiral perturbation theory ($\chi$PT) as a low energy effective theory of 
QCD, these Nambu-Goldstone boson fields are used to parametrize 
the broken axial symmetry, and we collect them in a $3\times 3$ matrix,  
\begin{eqnarray}
\Sigma &=&(e^{ i \Phi / f })\ ,\\
\Phi&=& \sum_a \lambda^a \phi^a = \left[
        \begin{array}{ccc}
        \frac{1}{\sqrt{2}}\pi^0+\frac{1}{\sqrt{6}}\eta^0 & \pi^+ & K^+\\
        \pi^- &-\frac{1}{\sqrt{2}}\pi^0+\frac{1}{\sqrt{6}}\eta^0 &K^0\\
        K^- & \overline{K^0} & -\frac{2}{\sqrt{6}}\eta^0
        \end{array}
        \right],
\end{eqnarray}
where $\lambda^a$ are Gell-Mann matrices, and $f$ is the decay constant.  
Under $SU(3)_L\otimes SU(3)_R $ chiral transformation, 
$\Sigma \in SU(3)$ transforms as 
\begin{eqnarray}
\Sigma \to g_R\Sigma g_L^\dagger\ ,\ \ 
\Sigma^\dagger \to g_L\Sigma^\dagger g_R^\dagger. \label{chiral-tr}
\end{eqnarray}
The chiral Lagrangean to the lowest order, with the additional mass term, is 
given by 
\begin{eqnarray}
 {\cal L}_\chi=\frac{f^2}{4}\tr(\p_\mu \Sigma^\dagger\p_\mu\Sigma)
  - \frac{f^2}{4} \tr[ M(\Sigma^\dagger+\Sigma) ],\label{chiLag}
\end{eqnarray}
where $M=(2B_0) \cdot {\rm diag}[m_u,m_d,m_s]$ denotes 
the quark mass matrix and $B_0$ is a parameter.
In terms of $\Sigma$, the left- and right-handed currents are given by 
\begin{eqnarray}
(L_\mu)^i_j=\frac{i}{2}f^2(\p_\mu\Sigma^\dagger\cdot\Sigma)^j_i\ ,\ 
(R_\mu)^i_j=\frac{i}{2}f^2(\p_\mu\Sigma\cdot\Sigma^\dagger)^j_i,
\label{def_of_LR}
\end{eqnarray}
respectively. 

The idea of $\chi$PT reduction method by Bernard {\it et al.}~\cite{Bernard85} 
is to relate the hadronic matrix elements for $K\to\pi\pi$ decays to those 
for $K\to\pi$ and $K\to 0$ (vacuum) using $\chi$PT, and calculate the latter 
through numerical simulations of lattice QCD. 
As the first step of $\chi$PT reduction method, we construct operators
in $\chi$PT which correspond to $X_i$'s and $Y_i$'s in QCD {\it i.e.,} those 
which transform under the same irreducible representations of 
$SU(3)_L\otimes SU(3)_R$ and invariant under CPS symmetry. 
In the following, we discuss the case of 
$\{ (27_L,1_R),(8_L,1_R)\}$ and $(8_L,8_R)$ representations separately. 

\subsection{REDUCTION METHOD FOR ${\bf (27_L,1_R)}$ AND ${\bf (8_L,1_R)}$ OPERATORS}
\label{almostQs}

For the irreducible representations $(27_L,1_R)$ and $(8_L,1_R)$,  
which cover $Q_1,\dots,Q_6,Q_9$ and $Q_{10}$, the product of left handed 
currents $(L_\mu)^i_j(L_\mu)^k_l$ is one of the candidates for the operator 
to the lowest order in $\chi$PT.  An explicit form of the operators,
which are also CPS invariant, is given by 
\begin{eqnarray}
( 8_L,1_R) &:& {\bf A} = (L_\mu)^i_3(L_\mu)^2_i ,\label{chiopA}\\
(27_L,1_R) &:& {\bf C} = 3(L_\mu)^2_3(L_\mu)^1_1 + 2(L_\mu)^1_3(L_\mu)^2_1 .
\label{chiopC}
\end{eqnarray}
where ${\bf A}$ corresponds to $X_1$ or $X_2$, while ${\bf C}$ is 
the counterpart of $X_3$. The latter is decomposed into two parts with
$I=0$ and $2$ in the same way as $X_3$ (see Appendix~\ref{app_split}):
\begin{equation}
{\bf C} = \frac{1}{3}{\bf C}^{(0)}+\frac{5}{3}{\bf C}^{(2)},
\end{equation}
where 
\begin{eqnarray}
{\bf C}^{(0)}&=&(L_\mu)^1_1(L_\mu)^2_3+(L_\mu)^2_1(L_\mu)^1_3
+2(L_\mu)^2_2(L_\mu)^2_3-3(L_\mu)^3_3(L_\mu)^2_3 ,\\
{\bf C}^{(2)}&=& (L_\mu)^1_1(L_\mu)^2_3+(L_\mu)^2_1(L_\mu)^1_3
-(L_\mu)^2_2(L_\mu)^2_3 .
\end{eqnarray}

In addition to the operators above, there is another $(8_L,1_R)$ operator 
which is allowed from CPS invariance:
\begin{eqnarray}
(8_L,1_R) \ : \
{\bf B} 
&=& (\Sigma M+M\Sigma^\dagger)^2_3   \cr
&=&   B_0 ( m_s + m_d ) \left( \Sigma + \Sigma^\dagger \right)_3^2
    - B_0 ( m_s - m_d ) \left( \Sigma - \Sigma^\dagger \right)_3^2  \cr
&=&
 -i \frac{ 4 }{ f^2 } \partial_\mu 
    \left[ 
       \frac{ m_s + m_d }{ m_s - m_d } \left( V_\mu \right)_3^2
     - \frac{ m_s - m_d }{ m_s + m_d } \left( A_\mu \right)_3^2           
    \right]  
\label{B-lag} .
\end{eqnarray}
where $V_\mu = ( R_\mu + L_\mu )/2$ and $A_\mu = ( R_\mu - L_\mu )/2$ 
are vector and axial vector currents with 
$L_\mu$ and $R_\mu$ defined in (\ref{def_of_LR}).
The equation of motion for $\Sigma$ is used
to derive the third line from the second line in (\ref{B-lag}).

The counterpart of this operator for QCD can be obtained easily
by $SU(3)_L\otimes SU(3)_R$ and CPS symmetry,
\begin{eqnarray}
Q_{\rm sub} 
&=&    ( m_s + m_d ) \bar{s} d 
     - ( m_s - m_d ) \bar{s}\gamma_5d \cr
&=&  
      \partial_\mu
      \left[
             \frac{ m_s + m_d }{ m_s - m_d } \bar{s} \gamma_\mu d
           - \frac{ m_s - m_d }{ m_s + m_d } \bar{s} \gamma_\mu \gamma_5 d
      \right]
\label{Qsub-def}  ,
\end{eqnarray}
where the equation of motion for $s$ and $d$ quark fields is used.

For physical $K\to\pi\pi$ processes, $Q_{\rm sub}$, and hence ${\bf B}$, does 
not contribute since these operators are total derivative of local operators
and the energy-momentum injected at the weak operator vanishes.
However, for the unphysical processes such as $K\to\pi$ and $K\to 0$ (vacuum) 
which we are to calculate on the lattice, 
the matrix elements of $Q_{\rm sub}$ or ${\bf B}$ does not vanish due to a 
finite energy-momentum transfer for $m_s\not= m_d$. 
Therefore, a mixing between $Q_i$'s and $Q_{\rm sub}$ in $K\to\pi$ 
matrix elements exists which should be removed. 
We should also note that this mixing inevitably arises in the case of 
$m_d=m_s$, as is often chosen in numerical simulations on the lattice, 
since $Q_{\rm sub}$ is not a total divergence for this case.

We assume that there are linear relations in the sense of matrix elements
between the local operators 
$\{Q_i\ (i=1,\dots,6,9,10), Q_{\rm sub}\}$ and 
$\{{\bf A},{\bf B},{\bf C}\}$ which belong to 
the same representations, {\it i.e.} $\{(27_L,1_R),(8_L,1_R)\}$ :
\begin{eqnarray}
Q_i^{(0)}&=&a_i{\bf A}+b_i{\bf B}+c_i^{(0)}{\bf C}^{(0)},\label{rel-ABC}\\
Q_{\rm sub}&=&r{\bf B},\label{rel-B}\\
Q_i^{(2)}&=&c_i^{(2)}{\bf C}^{(2)},\label{rel-C2}
\end{eqnarray}
where the coefficients $a_i, b_i, c^{(I)}_i$ and $r$ are unknown
parameters. 
Taking the matrix elements of the two sides of (\ref{rel-ABC}), 
(\ref{rel-B}) and (\ref{rel-C2})  for 
$K^0\to0$, $K^+\to\pi^+$ and $K^0\to\pi^+\pi^-$, one obtains 
\begin{eqnarray}
   \VEV{0    }{Q_i^{(0)} - \alpha_i Q_{\rm sub}} {K^0} 
&=& 
   0 , \label{Q0sub}  \\
   \VEV{\pi^+}{Q_i^{(0)} - \alpha_i Q_{\rm sub}} {K^+} 
&=& 
     \frac{ 2 p_K \cdot p_\pi }{f^2} ( a_i - c_i^{(0)} ) 
   + {\cal O}(p^4) , \label{kpi0} \\
   \VEV{\pi^+}{Q_i^{(2)}}{K^+}
&=&
   - \frac{ 2p_K\cdot p_\pi}{f^2} c_i^{(2)}
   + {\cal O}(p^4) , \label{kpi2} \\
   \VEV{\pi^+\pi^-}{Q_i^{(0)}}{K^0}
&=&
      \frac{ \sqrt{2}}{f^3}(m_K^2-m_\pi^2) ( a_i - c_i^{(0)} )
   + {\cal O}(p^4) , \label{kpipi0} \\
      \VEV{\pi^+\pi^-}{Q_i^{(2)}}{K^0}
&=& 
    - \frac{ \sqrt{2}}{f^3}(m_K^2-m_\pi^2) c_i^{(2)}
    + {\cal O}(p^4)   \label{kpipi2} ,
\end{eqnarray}
where $\alpha_i\equiv b_i/r$ in (\ref{Q0sub}) and (\ref{kpi0}), 
$p_K$ and $p_\pi$ are the momenta of kaon and pion, respectively,  
and $p$ denotes either of them.
In (\ref{kpipi0}) and (\ref{kpipi2}), $m_K$ and $m_\pi$ are the physical 
meson masses.
After eliminating $a_i-c_i^{(0)}$ from (\ref{kpi0}) and (\ref{kpipi0}), we 
arrive at the relation between $\VEV{\pi^+\pi^-}{Q_i^{(0)}}{K^0}$ and 
$\VEV{\pi^+}{Q_i^{(0)}}{K^+}$ in the $I=0$ case:
\begin{eqnarray}
     \VEV{ \pi^{+} \pi^{-} }{ Q_i^{(0)} }{ K^0 }
&=&  \frac{ ( m_K^2 - m_\pi^2 ) }{ \sqrt{2} f (p_K\cdot p_\pi) }
        \left< \pi^{+} \left | Q_i^{(0)} - \alpha_i Q_{\rm sub} \right | K^{+} \right>
    + {\cal O}\left(p^2\right)   \label{chipt0},\\
\alpha_i&=&\frac{\VEV{0}{Q_i^{(0)}}{K^0}}{\VEV{0}{Q_{\rm sub}}{K^0}}, 
\qquad i=1,\cdots, 6, 9, 10.
\label{alpha-ratio}
\end{eqnarray}
The $K\to 0$ (vacuum) 
matrix elements are used only to determine the $\alpha_i$'s
which govern the subtraction of unphysical contributions originating 
from $Q_{\rm sub}$.
The relation for the $I=2$ case is derived in the same way from 
(\ref{kpi2}) and (\ref{kpipi2}):
\begin{eqnarray}
      \left< \pi^{+} \pi^{-} | Q_i^{(2)} | K^0 \right>
&=&
    \frac{ (m_K^2-m_\pi^2) }{ \sqrt{2} f (p_K\cdot p_\pi)}
        \left< \pi^{+} | Q_i^{(2)} | K^{+} \right>
  + {\cal O}\left(p^2\right),
\qquad i=1,\cdots, 6, 9, 10. \label{chipt2}
\end{eqnarray}

Let us note that the essential point of the reduction method is 
a calculation of the parameters $a_i-c^{(0)}_i$ and $c^{(2)}_i$ from 
$K\to\pi$ three-point correlation functions in
numerical simulations of lattice QCD.
Since these parameters appear in (\ref{kpi0}) and 
(\ref{kpi2}) as the coefficients of $p_K\cdot p_\pi$, 
their values are sensitive to the chiral properties of 
the $K\to\pi$ matrix elements on the left hand side of these equations.
Hence $SU(3)_L\otimes SU(3)_R$ chiral symmetry on the lattice 
is an indispensable requirement for a successful calculation using this method.

\subsection{REDUCTION METHOD FOR ${\bf (8_L,8_R)}$ OPERATORS}\label{Q7Q8}

In order to construct $(8_L,8_R)$ operators in $\chi$PT, 
we observe that $(\Sigma)^i_j(\Sigma^\dagger)^k_l$ transforms as 
$(8_R,8_L)$~\cite{Bij-Wise84,Ciri-Golo00,ishizuka} where
$(j,k)$ and $(l,i)$ correspond to $8_L$ and $8_R$, respectively.
One finds a CPS invariant operator
\begin{eqnarray}
 {\bf D} = 3 \Sigma^1_3(\Sigma^\dagger)^2_1
\end{eqnarray}
as the counterpart of $Y_2$. The decomposition into the $I=0$ and $2$ 
part is given by 
\begin{eqnarray}
 {\bf D}&=& {\bf D}^{(0)}+ {\bf D}^{(2)} ,\\
 {\bf D}^{(0)}&=&2\Sigma^1_3(\Sigma^\dagger)^2_1-\Sigma^2_3
(\Sigma^\dagger)^1_1+\Sigma^2_3(\Sigma^\dagger)^2_2 ,\\
 {\bf D}^{(2)}&=&\Sigma^1_3(\Sigma^\dagger)^2_1+\Sigma^2_3
(\Sigma^\dagger)^1_1-\Sigma^2_3(\Sigma^\dagger)^2_2 .
\end{eqnarray}
Assuming linear relations between $\{Q_7^{(I)},Q_8^{(I)}\}$ and 
${\bf D}^{(I)}$'s,  
\begin{eqnarray}
Q_i^{(0)}=d^{(0)}_i{\bf D}^{(0)}\ ,\ Q_i^{(2)}=d^{(2)}_i{\bf D}^{(2)}\ \ \ \ 
(i=7,8),
\end{eqnarray}
with the unknown parameters $d^{(I)}_i$'s, 
we take the matrix elements of the two sides for 
$K\to\pi\pi$ and $K\to\pi$ to obtain
\begin{eqnarray}
\VEV{\pi^+}{Q_i^{(0)}}{K^+}= 4 d^{(0)}_i/f^2+{\cal O}(p^2) &,& 
\VEV{\pi^+\pi^-}{Q_i^{(0)}}{K^0}= -2\sqrt{2}d^{(0)}_i/ f^3+{\cal O}(p^2)\\
\VEV{\pi^+}{Q_i^{(2)}}{K^+}= 2 d^{(2)}_i/f^2+{\cal O}(p^2) &,& 
\VEV{\pi^+\pi^-}{Q_i^{(2)}}{K^0}= -\sqrt{2}d^{(2)}_i/ f^3+{\cal O}(p^2)
\end{eqnarray}
These relations lead to the reduction formulae for $(8_L,8_R)$ operators, 
namely, 
\begin{eqnarray}
\VEV{\pi^+\pi^-}{Q_i^{(I)}}{K^0} = - \frac{1}{\sqrt{2}f}
    \VEV{\pi^+}{Q_i^{(I)}}{K^+}+{\cal O}(p^2), \qquad i=7,8 \label{88chpt}
\end{eqnarray}
which is common for the $I=0$ and 2 components.


\section{DETAILS OF SIMULATIONS}\label{actions}

\subsection{LATTICE ACTIONS}

The RG-improved gauge action we use is defined by  
\begin{equation}
S_{\rm gluon} = \frac{1}{g^2}\left\{
c_0 \sum_{\rm plaquette} {\rm Tr}\,U_{\rm pl}
+ c_1  \sum_{1\times2\ \rm rectangle} {\rm Tr}\, U_{\rm rtg}\right\} ,
\label{eqn:RG}
\end{equation}
where the coefficients of the plaquette and $1\times 2$ Wilson loop terms 
take the values $c_0=3.648$ and $c_1=-0.331$~\cite{Iwasaki83}.  
This action is expected to lead to a faster approach of physical observables 
to the continuum limit than with the unimproved plaquette gauge action. 

In order to satisfy the requirement of chiral symmetry on the lattice, 
we use the domain-wall formalism~\cite{Kaplan92} for the quark action.
Adopting the Shamir's formulation~\cite{Shamir93,Shamir95}, 
the action is written as 
\begin{eqnarray}
S_F &=& - \sum_{xy,st} \bar{\psi}_{s}(x) D^{DW}_{st}(x,y) \psi_{t}(y) ,\\
D^{DW} &=& D^W+D^5 ,\\
D^W_{st}(x,y) &=&
    \sum_\mu
     \left[
        \frac{ r - \gamma_\mu }{ 2a } U_\mu(x) 
          \delta(x+\hat{\mu}-y)
      + \frac{ r + \gamma_\mu }{ 2a } U^{\dagger}_\mu ( x-\hat{\mu} )
          \delta(x-\hat{\mu}-y)
     \right] \delta_{st} \cr
  & & \qquad\qquad
  + 
    \frac{M-4r}{a} \delta(x-y) \delta_{st} ,\\
D^5_{st}(x,y) &=& 
     \left[   \frac{1-\gamma_5}{ 2a_5 } \delta_{s+1,t} 
            + \frac{1+\gamma_5}{ 2a_5 } \delta_{s-1,t}
     \right] \delta(x-y)  - \frac{1}{a_5}\delta(x-y) \delta_{st},
\end{eqnarray}
where $D^W$ is the ordinary Wilson-Dirac operator in 4 dimensions, 
$M$ is the domain-wall height which has to be adjusted to ensure 
the existence of chiral modes, {\it e.g.,} $0<M<2$ at tree level,  
and $r$ is the Wilson parameter which we choose to be unity. 
The operator $D^5$ is the extended part in the fifth direction 
in which the coordinate is bounded by $1\leq s,t\leq N_5$. 

Using the chirality projection operators 
\begin{eqnarray}
P_L=\frac{1-\gamma_5}{2},\ \ \ P_R=\frac{1+\gamma_5}{2},
\end{eqnarray}
quark fields are defined by
\begin{eqnarray}
      q(x) &=& P_L \psi_1 (x) + P_R \psi_{N_5}(x) , \label{dwquark}\\
\bar{q}(x) &=& \bar{\psi}_{N_5}(x) P_L + \bar{\psi}_{1}(x) P_R , \label{dwbarqua}
\end{eqnarray}
and their mass $m_f$ is introduced as a
parameter in the boundary condition in the 5th direction:
\begin{eqnarray}
\psi_{N_5+1}(x) = m_fa \psi_{1}(x) , \ \ \psi_{0}(x) = m_fa \psi_{N_5}(x) .
\end{eqnarray}
The operators $Q_i$ and $Q_{\rm sub}$ in our numerical simulation
are constructed from $q$ and $\bar q$ only, by identifying
$u$,$d$ and $s$ with $q_u$, $q_d$ and $q_s$.

Axial vector transformations in five dimensions are defined as
\begin{eqnarray}
\delta \psi_{s}(x)       =  i Q(s) \lambda^a \epsilon^a_{s}(x) \psi_{s}(x) ,\ \ \
\delta \bar{\psi}_{s}(x) = -i \bar{\psi}_{s} (x) Q(s)\lambda^a \epsilon^a_{s}(x) ,
\end{eqnarray}
where $Q(s)={\rm sign}(2N_5-s+1)$ and $\epsilon^a_{s}(x)$ is an 
infinitesimal parameter. This definition leads to the variation
\begin{eqnarray}
\delta q(x)       &=& i \gamma_5 \lambda^a \epsilon^a(x) q(x) , \\
\delta \bar{q}(x) &=& i \bar{q}(x) \gamma_5 \lambda^a \epsilon^a(x) ,
\end{eqnarray} 
in terms of quark fields, and the axial-vector current takes the form 
\begin{eqnarray}
A_\mu^a(x) \equiv 
\sum_{s=1}^{N_5}
   Q(s) \frac{1}{2} 
   \left[
          \bar{\psi}_{s}(x)           (1-\gamma_\mu) U_{\mu}(x)         \lambda^a \psi_{s} (x+\hat{\mu}) 
       +  \bar{\psi}_{s}(x+\hat{\mu}) (1+\gamma_\mu) U^\dagger_{\mu}(x) \lambda^a \psi_{s} (x)
   \right].\label{Acurrent}
\end{eqnarray}
Taking the divergence of $A_\mu^a$, one obtains 
\begin{eqnarray} 
\nabla_\mu A_\mu^a(x) = 2 J^a_{5q}(x) + 2 m_fa P^a
\label{axialconserve}
\end{eqnarray} 
with 
\begin{eqnarray} 
J^a_{5q}(x) = 
       \bar{\psi}_{N_5/2}(x)    P_L  \lambda^a  \psi_{N_5/2+1}(x)
     - \bar{\psi}_{N_5/2+1}(x)  P_R  \lambda^a  \psi_{N_5/2}(x)   
\end{eqnarray}
and 
\begin{eqnarray} 
P^a=\bar{q}(x) \lambda^a \gamma_5 q(x).
\end{eqnarray} 
The axial vector current $A_\mu^a$ does not conserve automatically even 
in the chiral limit $m_f\to 0$ due to the first term $J_{5q}$ on the right 
hand side.
Effects of this breaking term, however, 
is expected to vanish as $N_5\to\infty$.
In practice it is necessary to determine the value of $N_5$ 
for a given set of lattice parameters and a type of gluon action,
so that the chiral breaking effect due to this term is acceptably small.

\begin{figure}
 \begin{center}
  \leavevmode
  \includegraphics[width=8.1cm, clip]{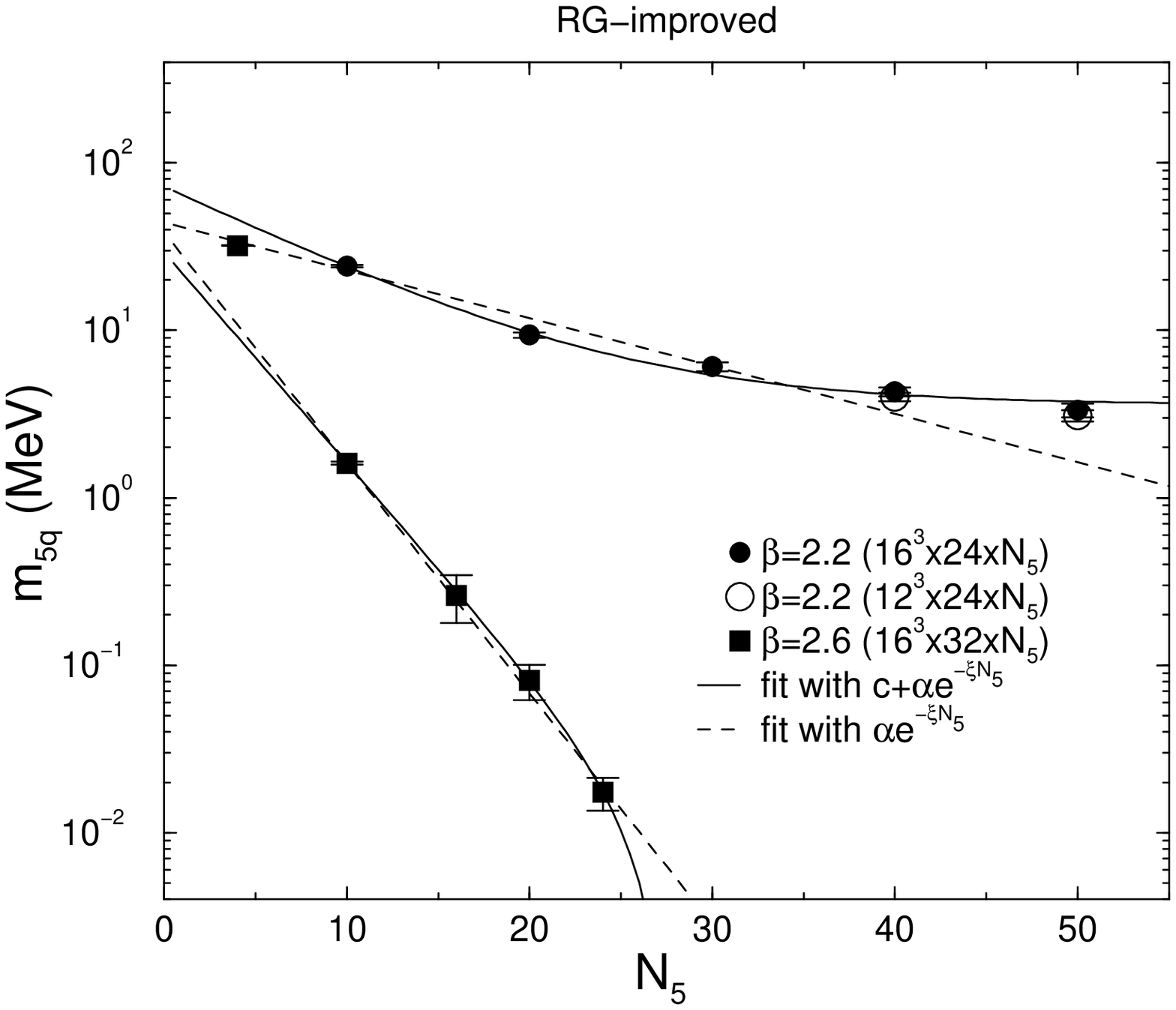}
  \includegraphics[width=8cm, clip]{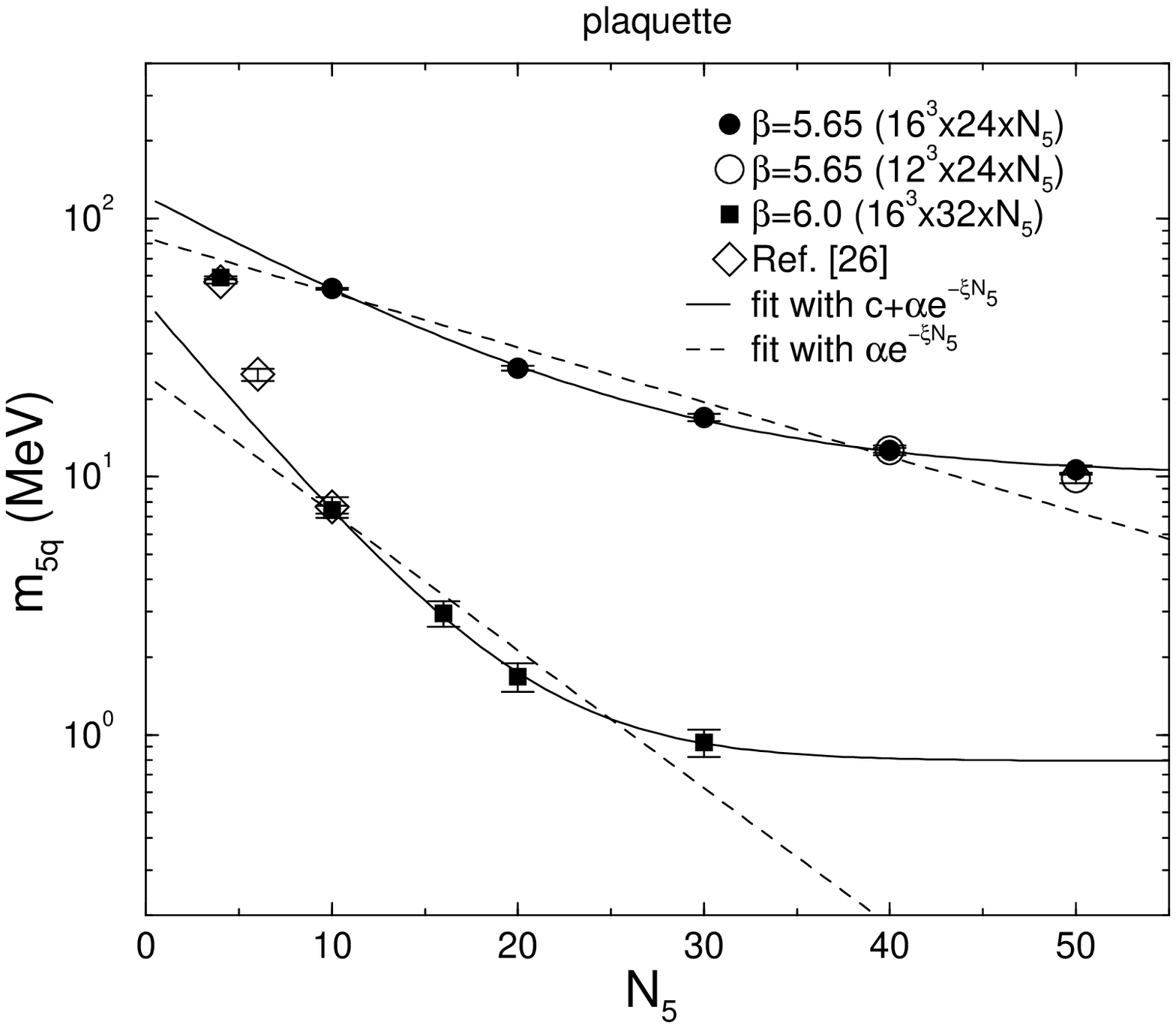}
 \end{center}
 \caption{(Left) Anomalous quark mass $m_{5q}$ as a function of $N_5$ in the
 $m_fa\to 0$ limit for the RG-improved action. Filled (open) circles represent
 data at $(\beta, M)=(2.2,1.7)$ on a $16^3(12^3)\times 24$ lattice. Filled
 squares are those at $(\beta, M)=(2.6,1.8)$ on a $16^3\times 32$ lattice. 
For the latter, data at four larger $N_5$ are used for 
 fits with the functions $\alpha e^{-\xi N_5}$ (dotted line) and 
$c+\alpha e^{-\xi N_5}$ (solid line).
(Right) Same for the plaquette action at $(\beta, M)=$(5.65, 1.7) and
(6.0, 1.8).}
\label{m5qvN}
\end{figure}

In Refs.~\cite{CPPACSdwqcd,RBCcollab}, the chiral property of
domain-wall fermion was investigated in detail in the quenched numerical 
simulation. Defining an anomalous quark mass by~\cite{CPPACSdwqcd} 
\begin{eqnarray}
m_{5q}a\equiv \frac{\VEV{0}{\sum_{\bf x}J^a_{5q}({\bf x},t)P^b(0)}{0}}
{\VEV{0}{\sum_{\bf x}P^a({\bf x},t)P^b(0)}{0}} , 
\end{eqnarray}
the axial Ward-Takahashi identity (\ref{axialconserve}) yields 
\begin{eqnarray}
\nabla_\mu \vev{\sum_{\bf x}A^a_\mu(x)P^b(0)}&=&2a(m_f+m_{5q})
\vev{\sum_{\bf x}P^a(x)P^b(0)}\label{AWTI} .
\end{eqnarray}
In FIG.~\ref{m5qvN}, we quote results of $m_{5q}$ as a function of 
$N_5$ from Ref.~\cite{AIKT00,CPPACSdwqcd}. In the right panel
data from the standard plaquette gluon 
action for $a^{-1}\approx 1$ GeV (circles, $\beta=5.65$) and 
$a^{-1}\approx 2$ GeV (squares, $\beta=6.0$) are summarized with two
types of exponential fits.  The counterparts from the 
RG-improved gluon action are found in the left panel, 
where $\beta=2.2$ and $2.6$ correspond to $a^{-1}\approx 1$ and
$2$ GeV, respectively.
The anomalous quark mass for the RG-improved action is an order of magnitude 
smaller than that for the plaquette action for both 
$a^{-1}\approx 1$ and $2$ GeV. This clearly demonstrates
the advantage of the use of RG-improved gluon action,  
which we therefore adopt in our work.

\subsection{SIMULATION PARAMETERS}
\label{sec:runs}

Our numerical simulations are carried out in the quenched approximation 
at the inverse gauge coupling of $\beta=2.6$. 
From the string tension $\sqrt{\sigma}=440$ MeV ~\cite{IKKY,lat99Kaneko,okamoto}, 
this value of $\beta$ corresponds to 
\begin{equation}
1/a=1.94(7) \ {\rm GeV},
\end{equation}
which we adopt in our analyses. 
If we use other quantities such as the rho meson mass or the pion decay
constant to determine the scale, the lattice spacing is different
from the above value, due to the quenched ambiguity as well as
the scaling violation. We do not include such an ambiguity of $a$
in the systematic uncertainty of our results.

Denoting the five-dimensional lattice size as $N_s^3\times N_t\times N_5$, 
we choose the fifth dimensional length to be $N_5=16$ and the domain wall 
height of the quark action to be $M=1.8$. For these parameter choices
the anomalous quark mass at $\beta=2.6$ is given by 
$m_{5q}=0.283(42)$ MeV \cite{CPPACSdwqcd}.  We expect this magnitude to be 
sufficiently small for viability of the $\chi$PT reduction formulae.
Chiral properties of matrix elements will be discussed in detail in 
Sec.~\ref{chi-prop}.

To investigate the effect of finite spatial volume $V=N_s^3$, two 
sizes of lattice given by $N_s=16$ and $24$ are examined, 
in both cases using the temporal size $N_t=32$.

We work with degenerate quark masses for $u, d$ and $s$ quarks, and 
denote the common bare quark mass as $m_f=m_u=m_d=m_s$.  
Matrix elements are evaluated for the bare quark masses 
$m_fa=0.02, 0.03, 0.04, 0.05$ and 0.06. 
Masses and decay constants of the pseudo-scalar meson 
calculated on the lattice, which are common for pion and kaon, are 
denoted as $m_M$ and $f_M$. 

Gauge configurations are generated by combining one sweep of 
the 5-hit pseudo heat bath algorithm and four overrelaxation sweeps, 
which we call an iteration. We skip 200 iterations between configurations 
for measurements. In TABLE~\ref{Ngconfs}, the numbers of configurations 
used in our analyses are given.  

We emphasize that {\it we generate gauge configurations independently 
for each value of $m_fa$}.  This is practically feasible since most of the 
computer time in our runs is spent in calculating quark propagators. 
A clear advantage is a removal of correlations between data at different 
values of $m_f$, and hence a more reliable control of the chiral extrapolation
as a function of $m_f$ or meson mass squared $m_M^2$ on the basis of 
$\chi^2$ fitting of data.
For error analyses at each $m_f$ a single elimination jackknife estimation is
employed throughout the present work.

TABLE~\ref{mpi2vmf} shows $m_M^2$ for both 
sizes of $16^3\times 32$ and $24^3\times 32$.
The intercepts in $m_f$ and $m_M^2$ are obtained by taking a linear 
extrapolation.
Values of $m_f$ in the limit of $m_M^2\to 0$ 
are 0.95(62) MeV and 1.09(31) MeV on $16^3\times 32$ and $24^3\times 32$
lattices respectively.
These values are larger than the value $m_{5q}=0.283(42)$~MeV
at $m_f=0$. As pointed out in Ref.~\cite{CPPACSdwqcd},
the discrepancy between the direct measurement of $m_{5q}$ and
the estimate from the pion mass is largely explained by finite spatial size 
effects on the pion mass.  
We use $m_M^2$ as a variable in our chiral extrapolation
throughout this paper.  We have checked that our results remain identical 
within estimated statistical errors if $m_f$ is used in chiral fits.

\subsection{CALCULATION OF MATRIX ELEMENTS}

\begin{figure}[h]
 \begin{center}
  \leavevmode
  \includegraphics[width=13.5cm, clip]{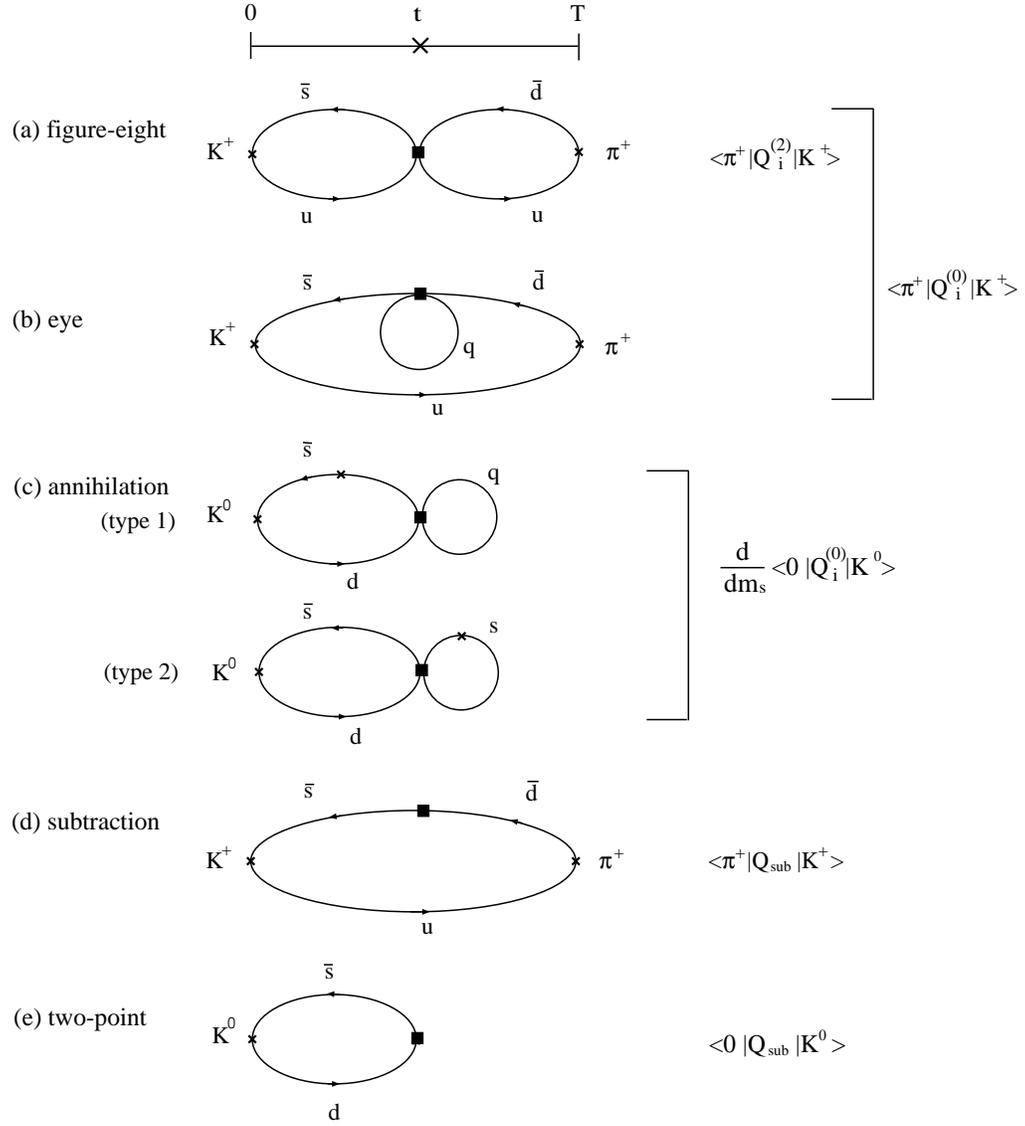}
 \end{center}
  \caption{Types of contractions needed for our calculation.  
Solid lines represent quark propagators on a background gauge field. 
Crosses represent points where meson sources are placed, while filled 
squares  denote four quark operators or the subtraction operator.
(a) ''figure-eight'', 
(b) ''eye'' which contributes only for matrix elements of $Q_i^{(0)}$, 
(c) ''annihilation'' with a quark mass derivative in the external line (type 1) 
      or in the quark loop (type 2),  
(d) ''subtraction'' and (e) ''two-point''.}
\label{allcont}
\end{figure}

In FIG.~\ref{allcont} we display the quark line diagrams of three- and 
two-point correlation functions needed for our simulation.
Filled squares represent the weak operator $Q_i^{(I)}$ or $Q_{\rm sub}$ 
located at the site $({\bf x},t)$.  Crosses are meson operators. 
We fix gauge configurations to the Coulomb gauge.  
A wall source for pion is placed at $t=0$ and that for kaon at 
$t=T\equiv N_t-1$. Quark propagators are solved by the conjugate gradient 
algorithm, imposing the Dirichlet boundary condition in time and 
the periodic boundary condition in space.  The stopping condition is given by 
\begin{equation}
\vert\vert (D+m)\cdot x - b \vert\vert^2 < 10^{-9} \vert\vert b \vert\vert^2 ,
\end{equation}
where $b$ is the source vector,  $x$ is the solution vector and
$D$ is the lattice fermion operator.
With this stopping condition a precision of better than 0.1\% is achieved for 
arbitrary elements of three-point correlation functions.

The three-point correlation functions for $K\to\pi$ matrix elements 
have the contractions of FIG.~\ref{allcont}(a),(b) and (d).
For calculating the $I=0$ amplitudes $\VEV{\pi^+}{Q_i^{(0)}}{K^+}$, 
both the figure-eight contraction of (a) and the eye contraction of (b)  
are needed, while for the $I=2$ amplitudes $\VEV{\pi^+}{Q_i^{(2)}}{K^+}$ 
only the figure-eight contributes. 
Writing ${\cal O}(t)=\frac{1}{V}\sum_{\bf x}{\cal O}({\bf x},t)$, 
we extract the matrix elements from calculation of the ratio of form
\begin{eqnarray}
\frac{ \left< 0 \left| \pi^+(T) Q_i^{(I)}(t) (K^+)^\dagger(0) \right| 0 \right> }
     { \left< 0 \left| \pi^+(T) A_4(t) \right| 0 \right>
       \left< 0 \left| A_4(t) (K^+)^\dagger(0) \right| 0 \right> }
&\stackrel{T\gg t\gg1}
     {\simeq}&\frac{\left<\pi^+\left| Q_i^{(I)}\right|K^+\right>}
     {\left<\pi^+\left|A_4\right|0\right>\left<0\left|A_4\right|K^+\right>}
\label{plateau-2pt}\\
&=& \frac{ 1 }{ 2 m_M^2 f_M^2 } \times \VEV{\pi^+}{Q^{(I)}_i}{K^+}
\label{Kpinorm}.
\end{eqnarray}
We note that a local current $A_\mu(x)=\overline{q}(x)\gamma_\mu\gamma_5q(x)$ 
is employed in the denominator rather than the conserved current given 
in (\ref{Acurrent}) in order to match with the local form of the four-quark 
operator in the numerator.   
 
The contractions in FIG.~\ref{allcont}(c) show the $K^0\to 0$ (vacuum)  
annihilation 
matrix elements from which the parameters $\alpha_i$
in the $\chi$PT reduction formulae (\ref{chipt0}) are obtained.
If $d$ and $s$ quarks are non-degenerate, these parameters are easily 
obtained from the ratio of propagators:
\begin{eqnarray}
\frac{\VEV{0}{Q^{(0)}_i(t)(K^0)^\dagger(0)}{0}}
{\VEV{0}{Q_{\rm sub}(t)(K^0)^\dagger(0)}{0}}\stackrel{t\to\infty}{\simeq}
\frac{\VEV{0}{Q^{(0)}_i}{K^0}}{\VEV{0}{Q_{\rm sub}}{K^0}}
=\alpha_i \label{alnorm}.
\end{eqnarray}
In the limit of degenerate quark masses, which applies to our numerical 
simulation, some care is needed. 
From the definition of $Q_{\rm sub}$ (\ref{Qsub-def}) and the fact 
that CPS symmetry gives $\VEV{0}{Q_i}{K^0}\Bigl|_{m_s = m_d}=0$, we derive
\begin{eqnarray}
\alpha_i&=&
-\lim_{m_s\to m_d}\frac{\VEV{0}{Q_i^{(0)}}{K^0}\Bigl|_{m_s > m_d}}
{(m_s-m_d)\VEV{0}{\bar{s}\gamma_5 d}{K^0}}\label{alpha-lim}\\
&=&-\frac{\frac{d}{dm_s}\VEV{0}{Q_i^{(0)}}{K^0}\Bigl|_{m_s=m_d}}
{\VEV{0}{\bar{s}\gamma_5 d}{K^0}}.\label{alpha-def}
\end{eqnarray}
The derivative acts both on the operator $Q_i^{(0)}$ and on the kaon, 
and hence there are two contributions as shown in FIG.~\ref{allcont}(c).  
The necessary derivative of the quark propagator is obtained through
\begin{eqnarray}
 \frac{dG(x,y)}{dm}=-\sum_z G(x,z)G(z,y) \label{GG}.
\end{eqnarray}

To calculate the quark loops that appear in the eye and
annihilation contractions, we employ the random $U(1)$ noise method. 
We generate  $\zeta^{(j)}(x)=e^{i\theta(x)} (j=1, \cdots, N)$ from a 
uniform random number $\theta(x)$ in the interval $0\leq \theta<2\pi$.
In the limit $N\to \infty$, we have 
\begin{eqnarray}
\frac{1}{N}\sum_{i=1}^N \zeta^{(i)*}(x)\zeta^{(i)}(y)
\stackrel{N\to\infty}{\longrightarrow} \delta(x-y).
\end{eqnarray}
Therefore, calculating quark propagators with $\zeta^{(i)}(x)$ as the source, 
\begin{eqnarray}
\eta^{(i)}(x)&\equiv&\sum_{ x'}(D+m)^{-1}( x, x')\zeta^{(i)}(x'),
\end{eqnarray}
we find
\begin{eqnarray}
\frac{1}{N}\sum_{i=1}^N\eta^{(i)}(x)
\zeta^{(i)*}(x)
\stackrel{N\to\infty}{\longrightarrow}(D+m)^{-1}(x,x) 
\label{noiseloop}
\end{eqnarray}
as the quark loop amplitude for each gauge configuration.

\begin{figure}
 \begin{center}
  \leavevmode
  \includegraphics[width=8cm, clip]{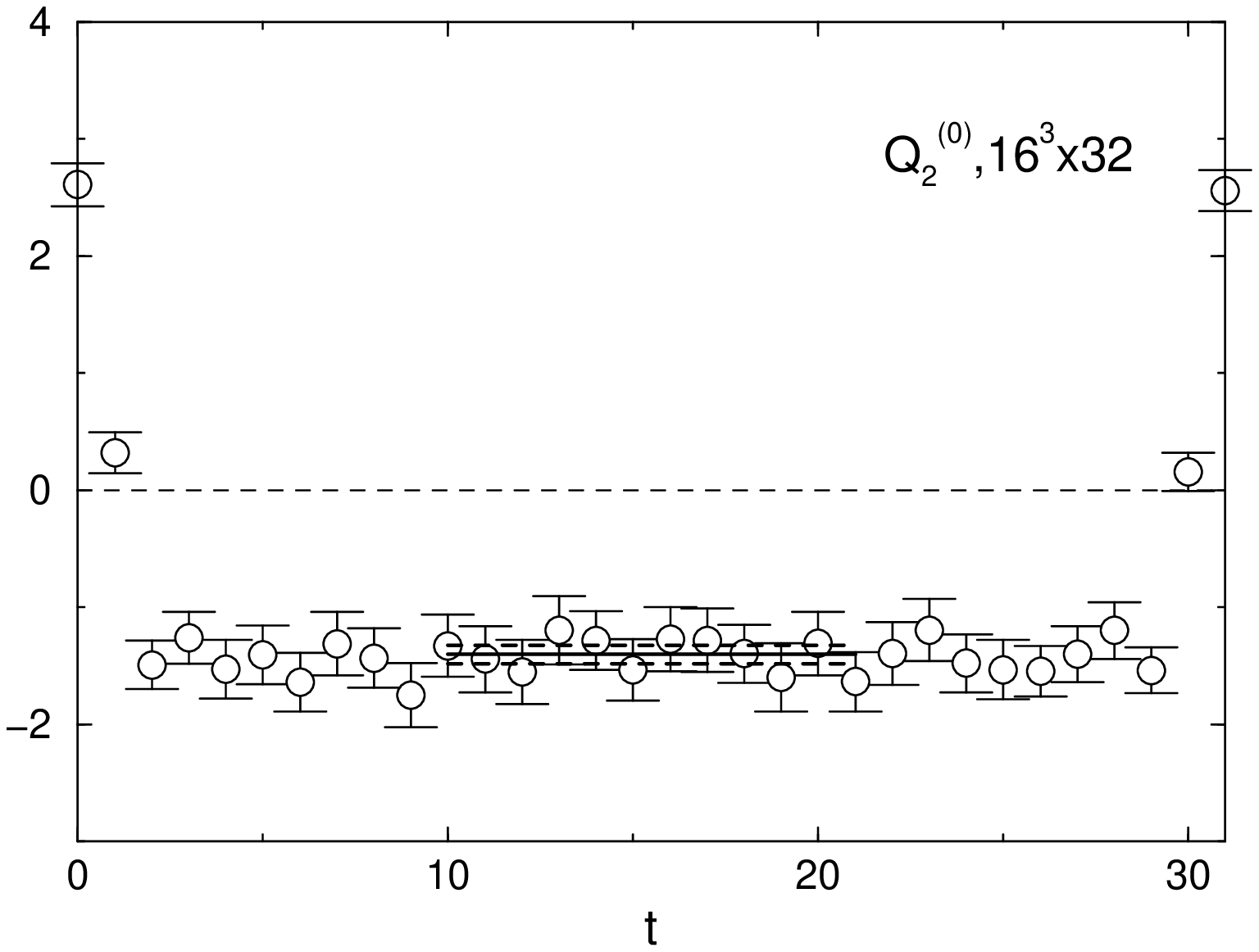}
\hspace*{0.3cm}
  \includegraphics[width=8cm, clip]{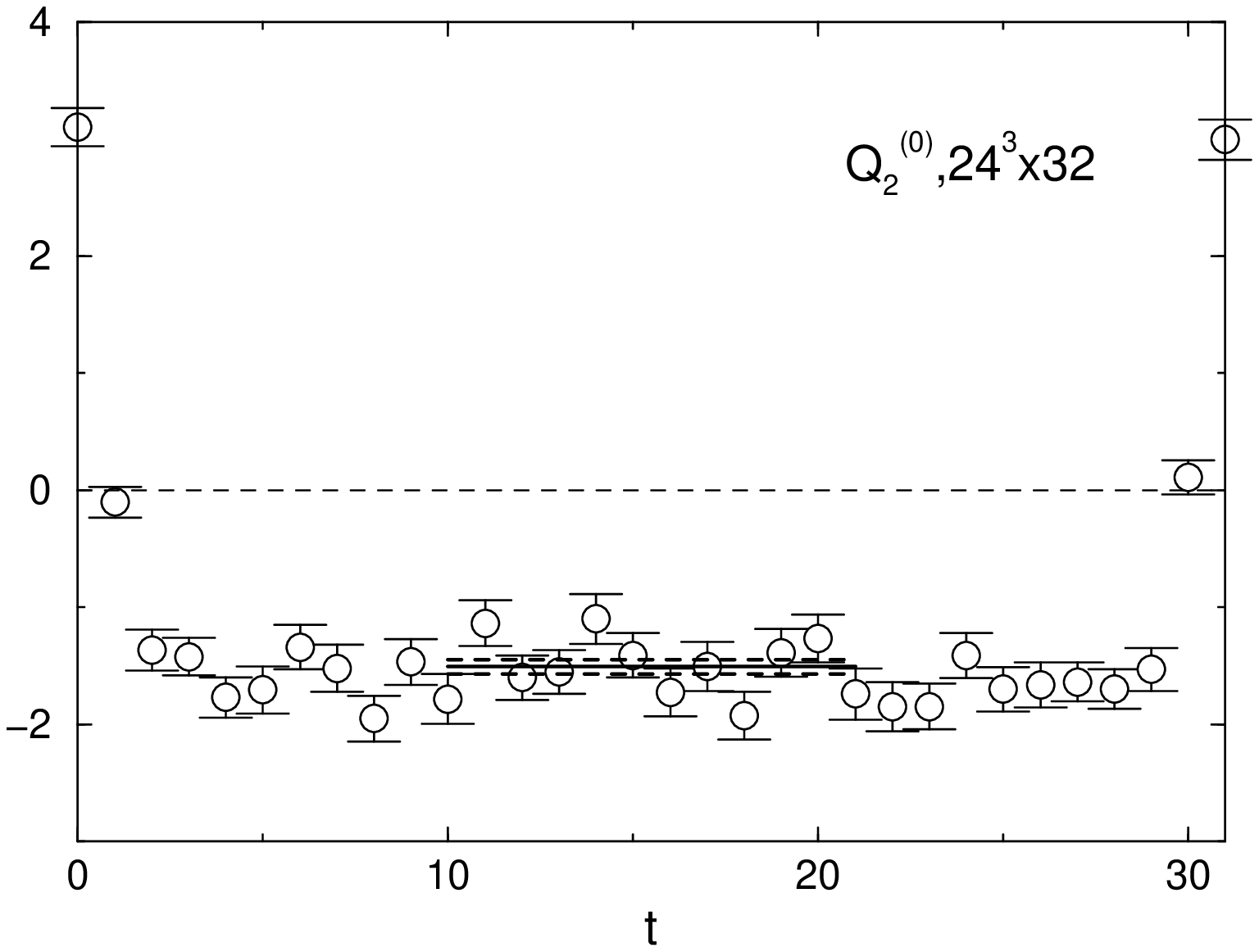}
 \end{center}
 \begin{center}
  \leavevmode
  \includegraphics[width=8cm, clip]{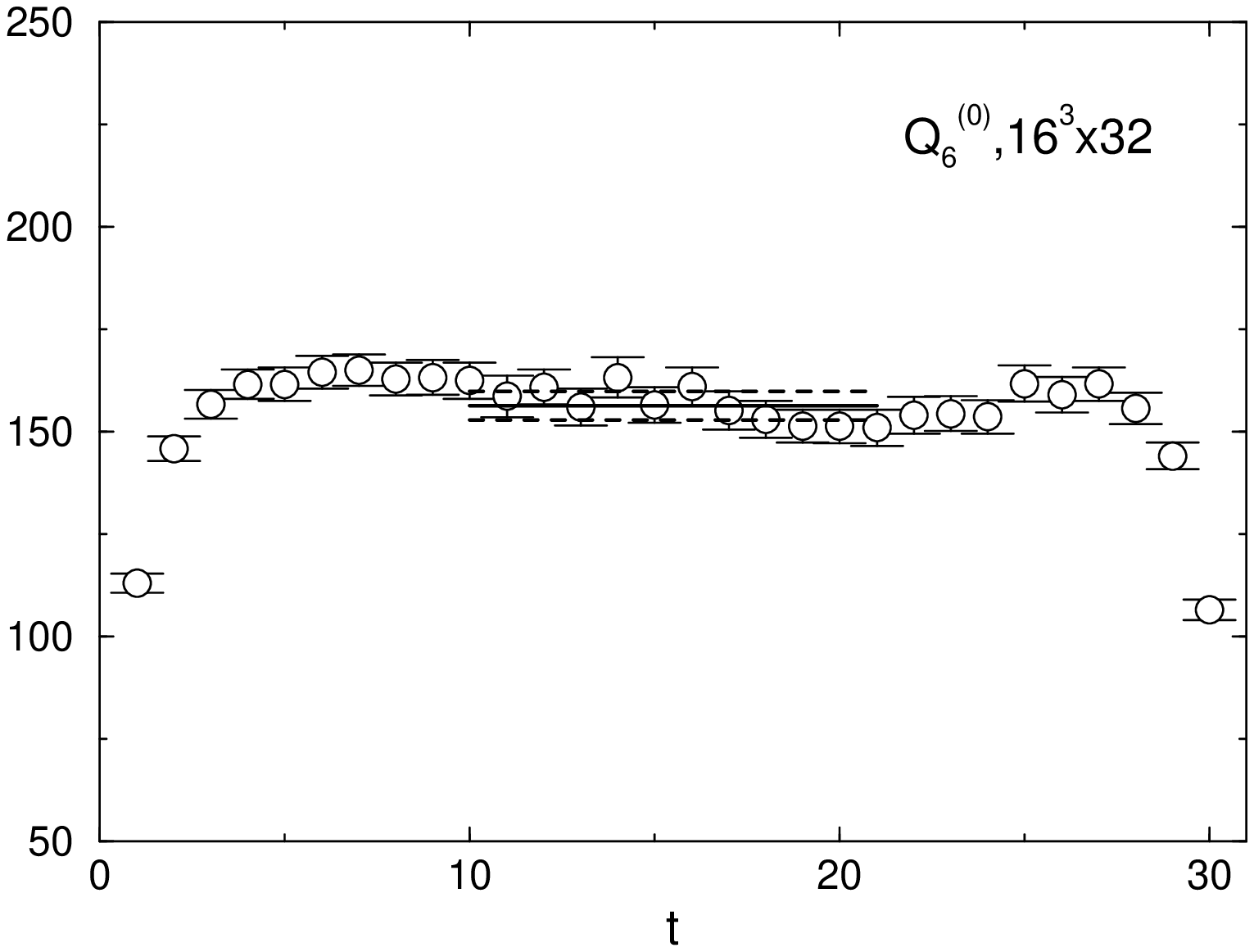}
\hspace*{0.3cm}
  \includegraphics[width=8cm, clip]{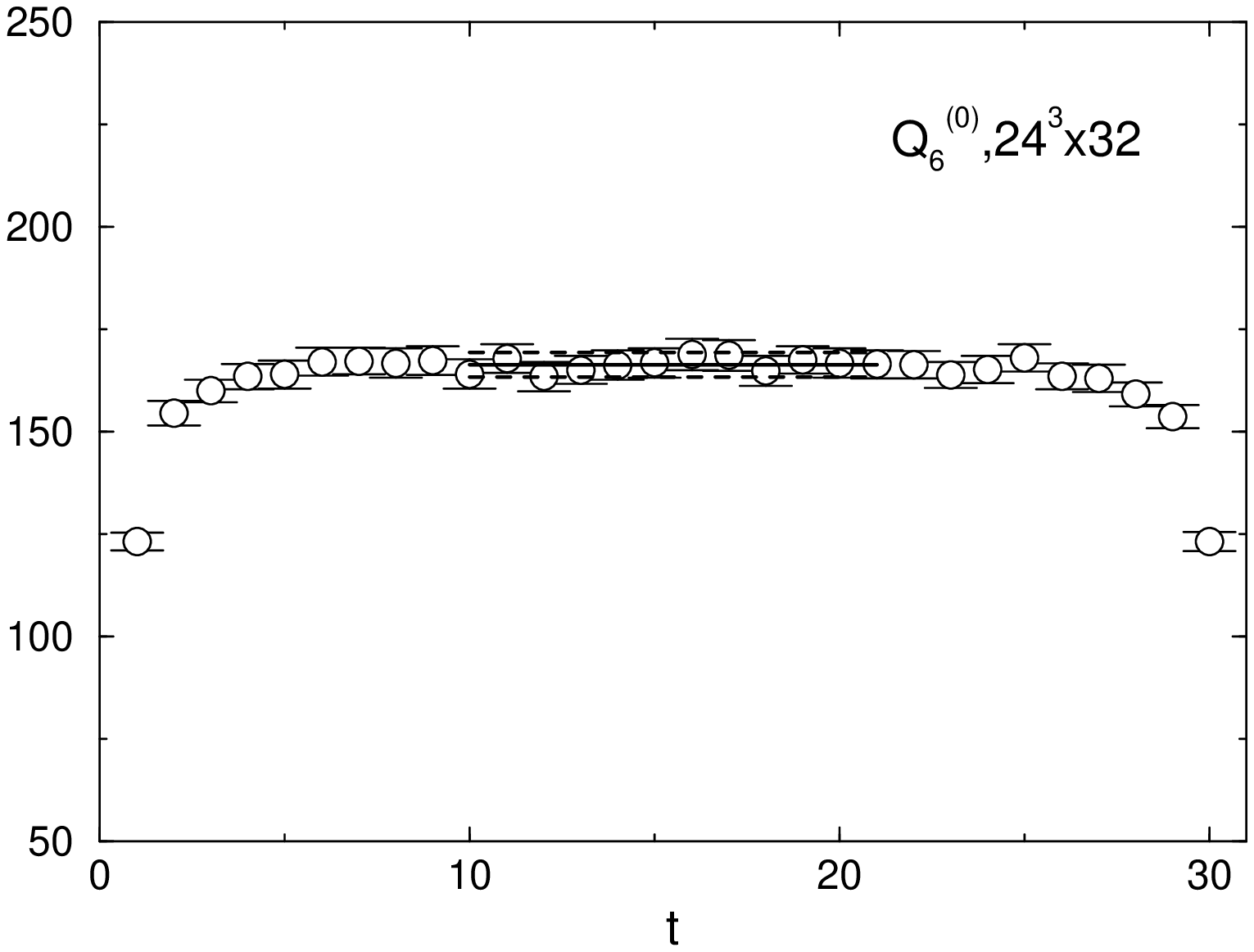}
 \end{center}
 \caption{Time dependence of the propagator ratio defined by 
(\ref{Kpinorm}) for $Q_2^{(0)}$ (upper) and $Q_6^{(0)}$ (lower) 
for $m_fa=0.03$. 
Left and right columns are for the lattice size $16^3\times 32$ and 
$24^3\times 32$ respectively. }
\label{t-depKpi}
\end{figure}

\begin{figure}
 \begin{center}
  \leavevmode
  \includegraphics[width=8cm, clip]{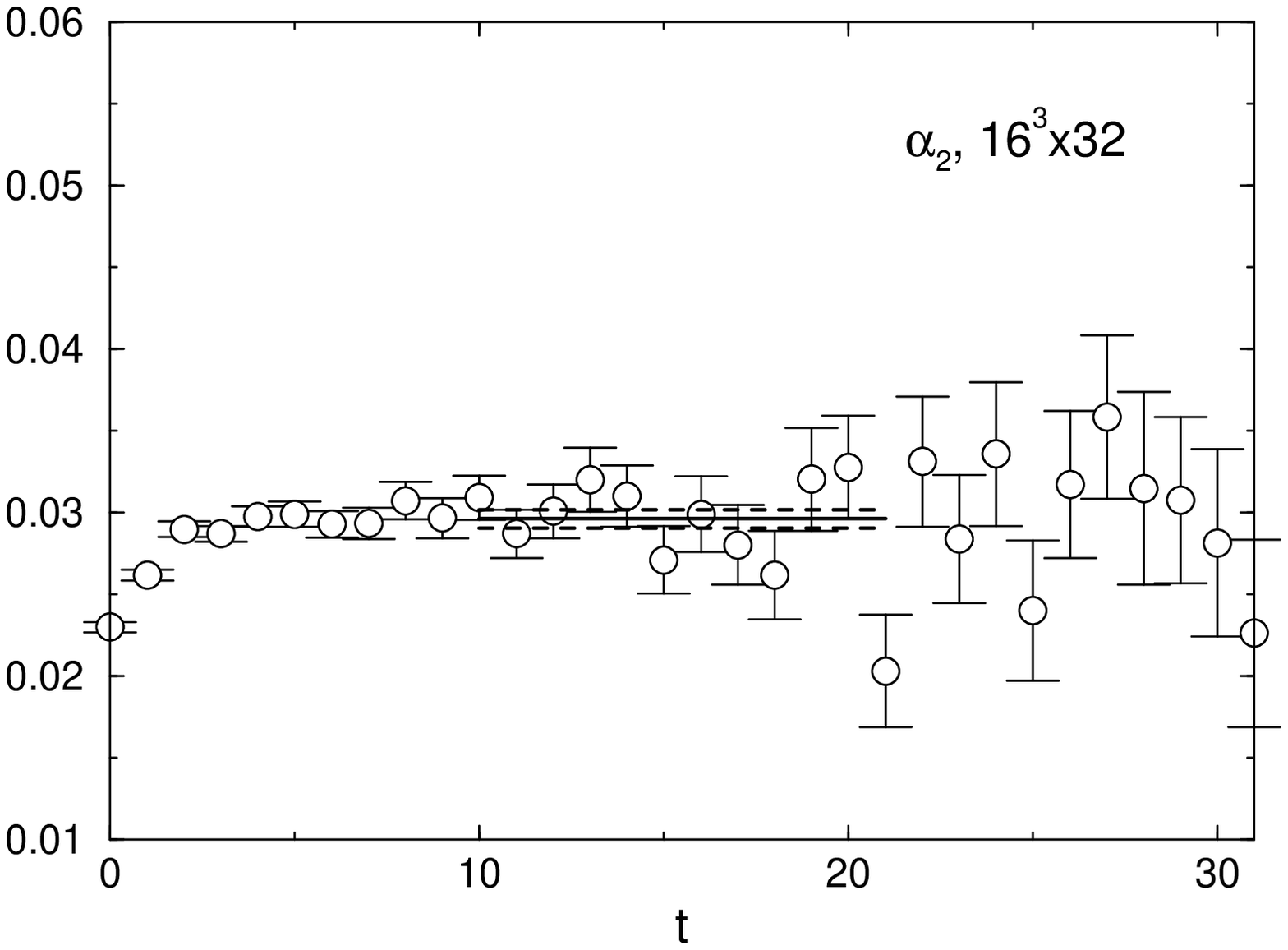}
\hspace*{0.3cm}
  \includegraphics[width=8cm, clip]{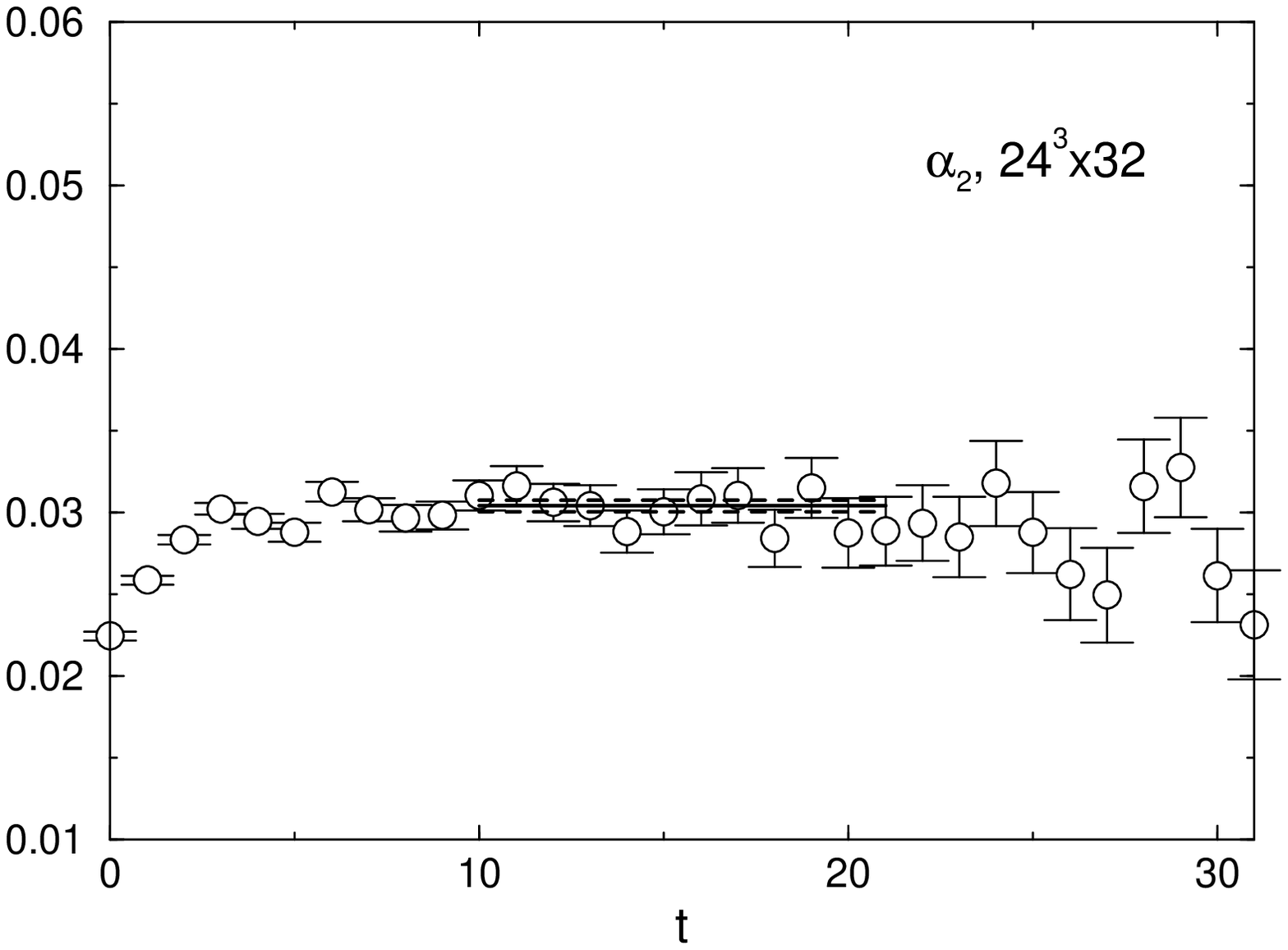}
 \end{center}
 \begin{center}
  \leavevmode
  \includegraphics[width=8cm, clip]{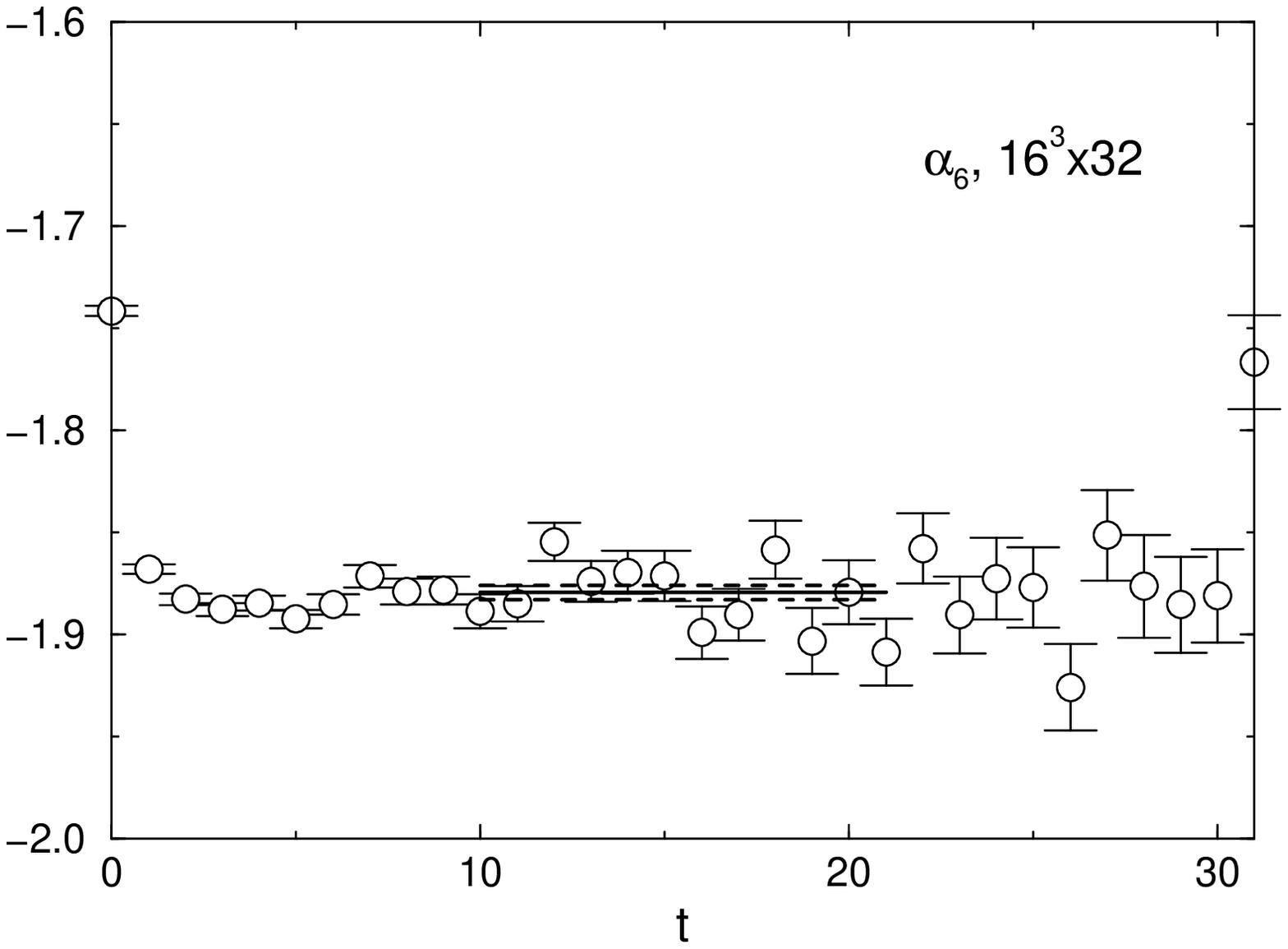}
\hspace*{0.3cm}
  \includegraphics[width=8cm, clip]{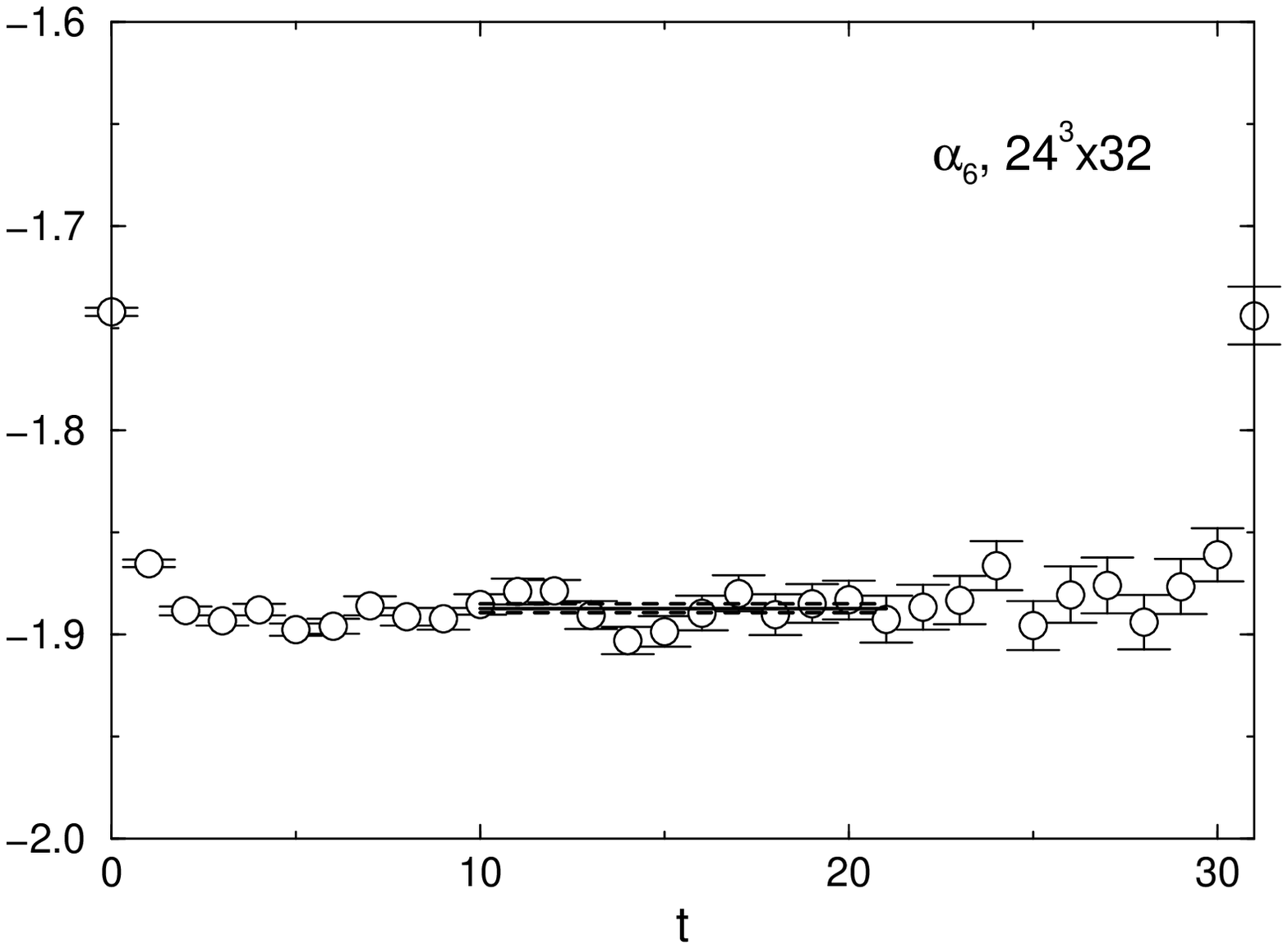}
 \end{center}
 \caption{Time dependence of the propagator ratio defined by
(\ref{alpha-def}) to calculate the parameter $a^2\times\alpha_2$ (upper) and 
$a^2\times\alpha_6$ (lower) at $m_fa=0.03$.  
Left and right columns are for the lattice size $16^3\times 32$ and 
$24^3\times 32$ respectively.}
\label{t-depAlp}
\end{figure}

In our calculation, we generate two noises for {\it each spinor and color 
degree of freedom}, {\it i.e.,} $ 2\times (\#{\rm color})
\times (\#{\rm spinor})=24$ noises for each configuration.  
In FIGs.~\ref{t-depKpi} and \ref{t-depAlp} we show propagator ratios 
for the $Q^{(0)}_2$ and $Q^{(0)}_6$ operators, 
and those for $\alpha_2$ and $\alpha_6$. 
The horizontal lines indicate the values extracted from a constant fit  
over $t=10-21$ and the one standard deviation error band. 
Here correlations between different time slices are not taken into account
for the fit. Instead errors are estimated by the jackknife method.
We observe reasonable signals, which show that 24 noises for 
each configuration we employ is sufficient to evaluate the quark 
loop amplitude. 
From (\ref{Kpinorm}), the $\chi$PT reduction formulae derived in 
Sec.~\ref{almostQs} and \ref{Q7Q8} are converted to the following forms 
at the lowest order of $\chi$PT:
\begin{eqnarray}
{\rm for}\ i=1,\dots,6,9,10\ : \ \ & & \nonumber \\
\left<\pi^+\pi^-\left|Q_i^{(0)}\right|K^0\right>
 &=& \sqrt{2}f_\pi(m_K^2-m_{\pi}^2)\times
\frac{\left<\pi^+\left|Q_i^{(0)}-\alpha_i
Q_{\rm sub}\right|K^+\right>}{\left<\pi^+\left|A_4\right|0\right>
        \left<0\left|A_4\right|K^+\right>} ,
\label{chptred0}\\
\left<\pi^+\pi^-\left|Q_i^{(2)}\right|K^0\right>
 &=&\sqrt{2}f_\pi(m_K^2-m_{\pi}^2)\times
\frac{\left<\pi^+\left|Q_i^{(2)}\right|K^+\right>}
{\left<\pi^+\left|A_4\right|0\right>
        \left<0\left|A_4\right|K^+\right>} ,
 \label{chptred2}\\
{\rm for\ } i=7,8\ \ (I=0,2):\ \ \ & &\nonumber \\
\left<\pi^+\pi^-\left|Q_i^{(I)}\right|K^0\right>
&=& -\sqrt{2}f_\pi m_M^2 \times\frac{\left<\pi^+\left|
Q_i^{(I)}\right|K^+\right>}
{\left<\pi^+\left|A_4\right|0\right>\left<0\left|A_4\right|K^+\right>} , 
\label{chptred88}
\end{eqnarray}
where we set $p_K = ( i m_M , \vec{0} )$ and $p_\pi = ( -i m_M , \vec{0} )$
for $K^+ \to \pi^+$ matrix elements on the right hand side.
We identify $f_M$ with $f$, and assign to it the physical value of 
$f_\pi$, since $f_M$ agrees with $f_\pi$ in the chiral limit.  
On the other hand, the meson masses $m_K^2$ and $m_\pi^2$ in (\ref{chptred0}) 
and (\ref{chptred2}) 
represent the experimental values since they arise from the physical 
$K\to\pi\pi$ matrix elements.
All of the experimental values used in our calculation are summarized in 
appendix~\ref{appendix-param}. 
We emphasize that these formulae are valid to the lowest order in $\chi$PT. 
If higher order corrections are small, 
the right hand sides of eqs. (\ref{chptred0})--(\ref{chptred88}) should
depend only weakly on the lattice meson mass $m_M^2$.

The two-pion states in the isospin basis are decomposed as
\begin{eqnarray}
\ket{(\pi\pi)_0}&=&\sqrt{\frac{2}{3}}\ket{\pi^+\pi^-}+
\sqrt{\frac{1}{3}}\ket{\pi^0\pi^0},\\
\ket{(\pi\pi)_2}&=&\sqrt{\frac{1}{3}}\ket{\pi^+\pi^-}-
\sqrt{\frac{2}{3}}\ket{\pi^0\pi^0}.
\end{eqnarray}
Therefore, matrix elements in this basis are given by 
$\VEV{\pi^+\pi^-}{Q^{(I)}_i}{K^0}$ times constants :
\begin{eqnarray}
\VEV{(\pi\pi)_0}{Q_i}{K^0}&=&\sqrt{\frac{3}{2}}\VEV{\pi^+\pi^-}
{Q_i^{(0)}}{K^0}\\
\VEV{(\pi\pi)_2}{Q_i}{K^0}&=&\sqrt{3}\VEV{\pi^+\pi^-}{Q_i^{(2)}}{K^0},
\label{isospin-basis}
\end{eqnarray}
We use a shorthand notation 
\begin{equation}
\vev{Q_i}_I\equiv\VEV{(\pi\pi)_I}{Q_i}{K^0}, \qquad I=0, 2 
\end{equation}
for the matrix elements in the isospin basis hereafter.

\subsection{SUBTRACTIONS IN $\Delta I=1/2$ MATRIX ELEMENTS}

\begin{figure}
 \begin{center}
  \leavevmode
  \includegraphics[width=8.3cm, clip]{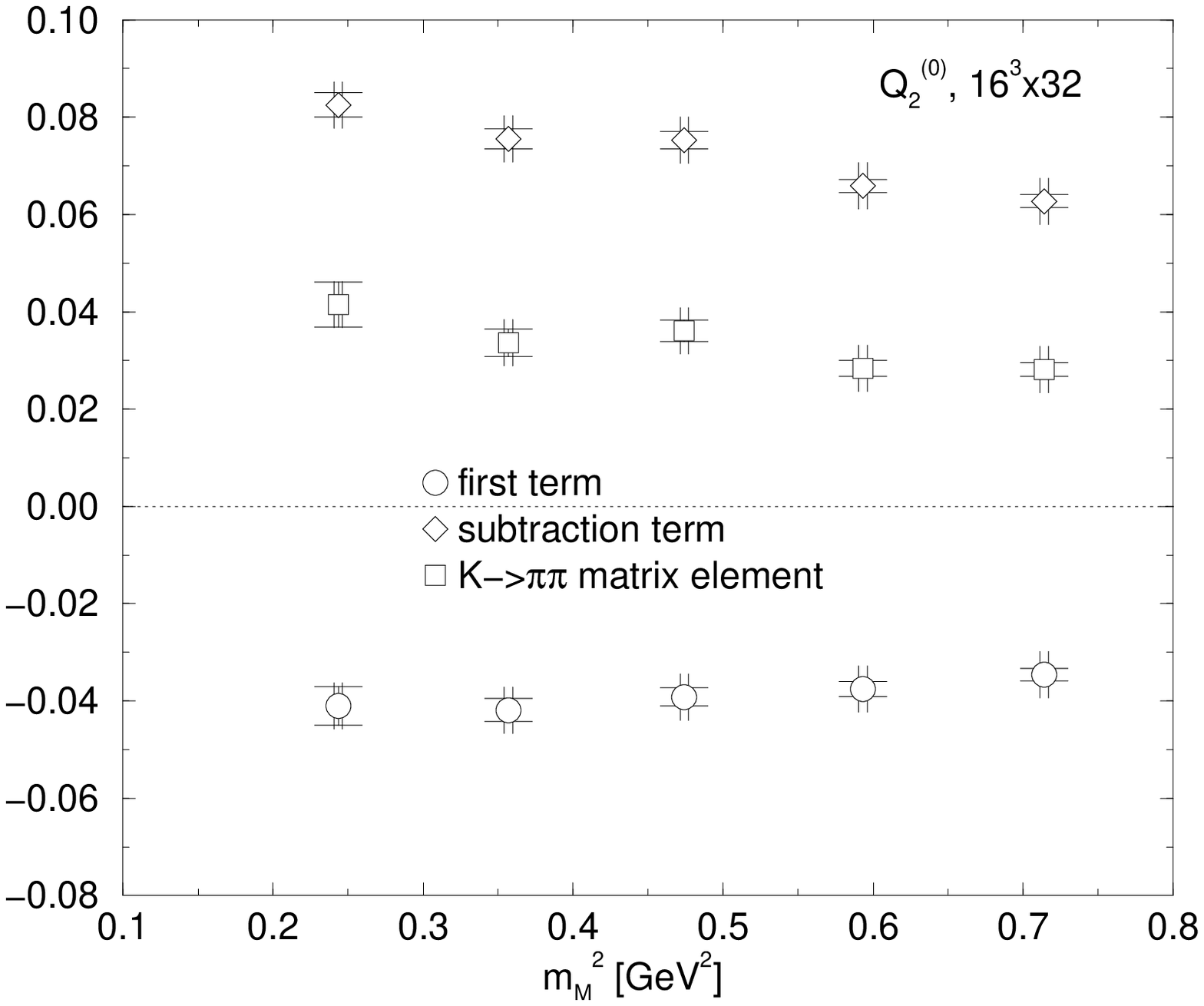}
  \includegraphics[width=8.3cm, clip]{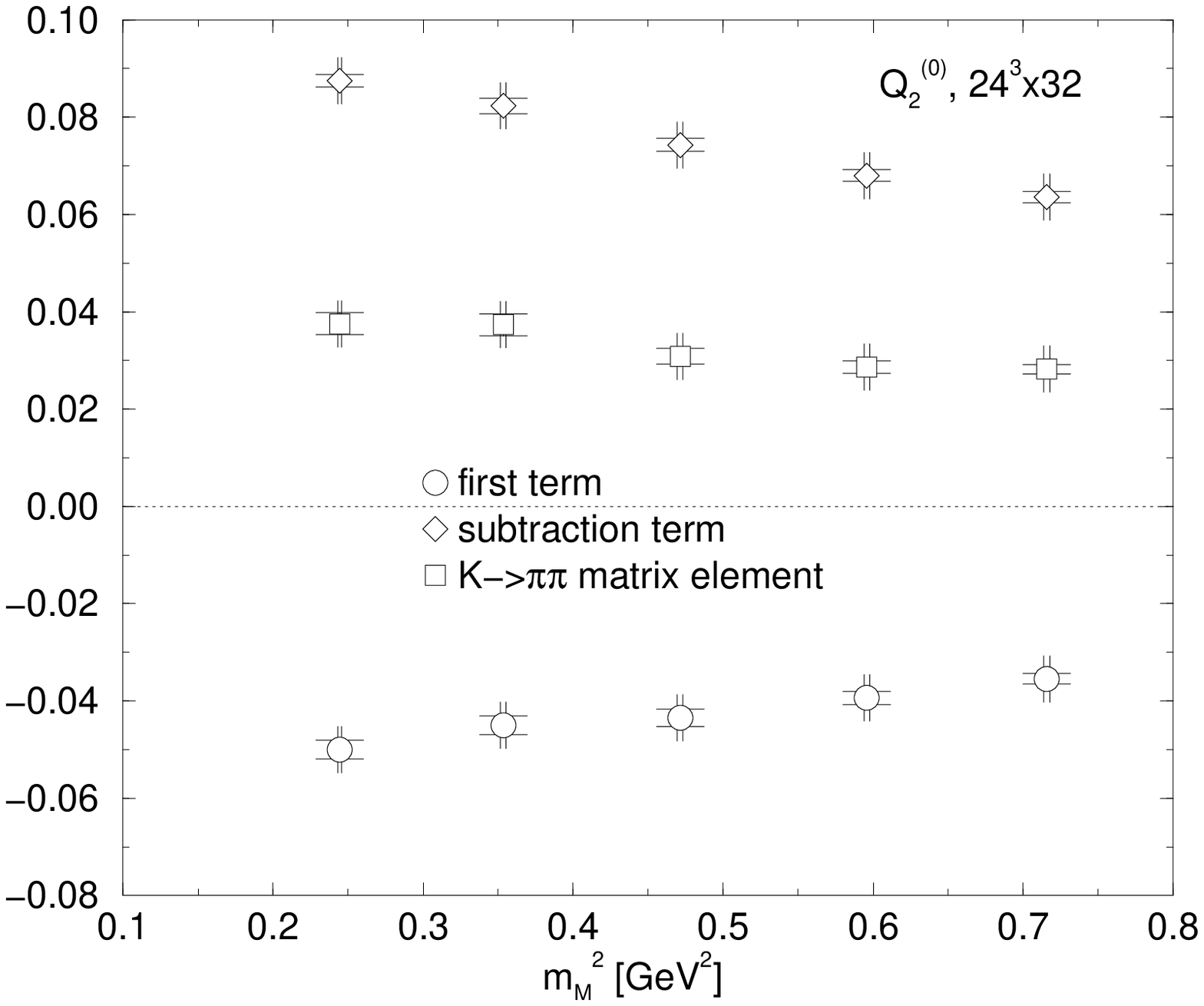}
 \end{center}
 \begin{center}
  \leavevmode
\hspace*{2mm}
  \includegraphics[width=7.9cm, clip]{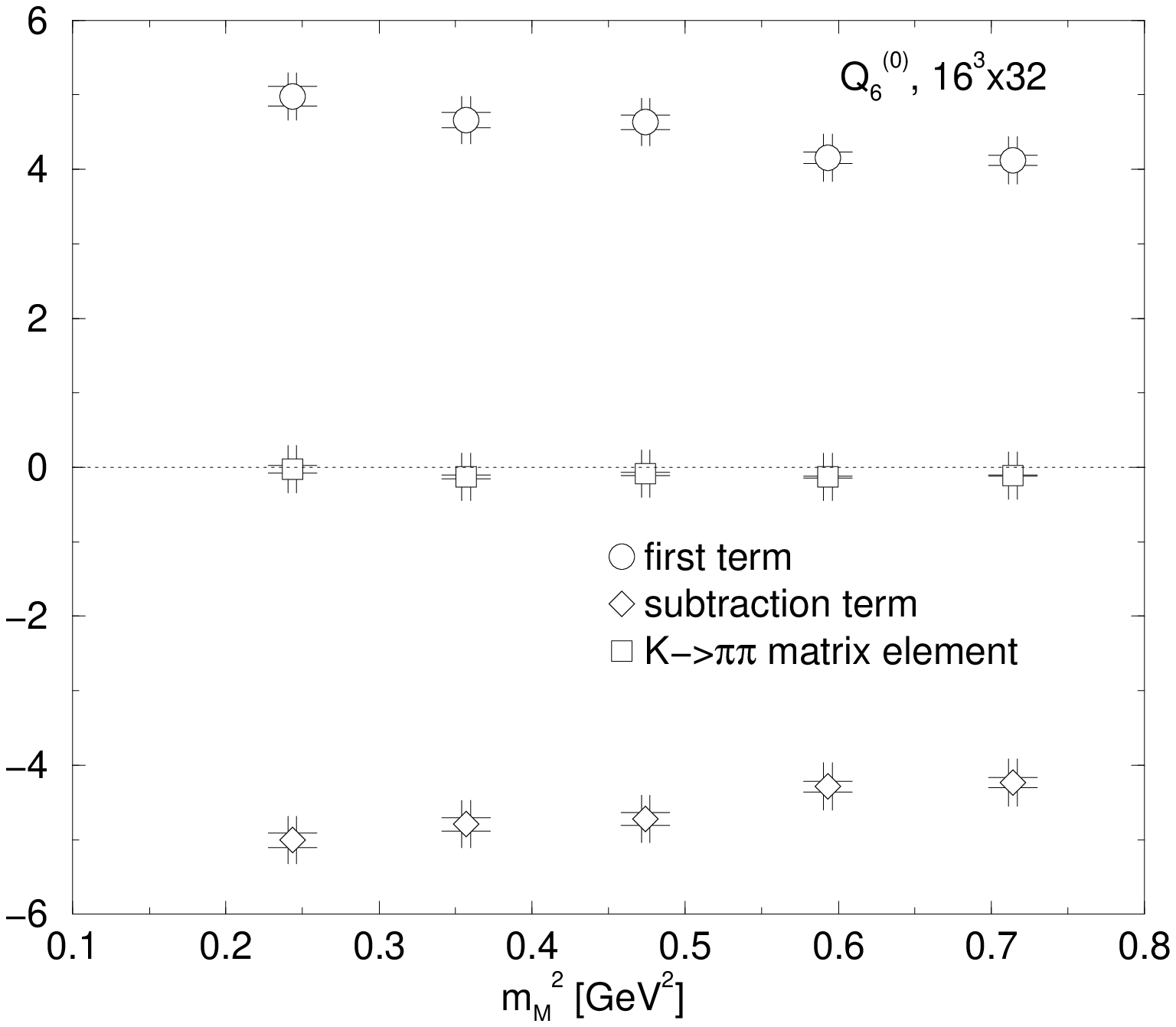}
\hspace{3mm}
  \includegraphics[width=7.9cm, clip]{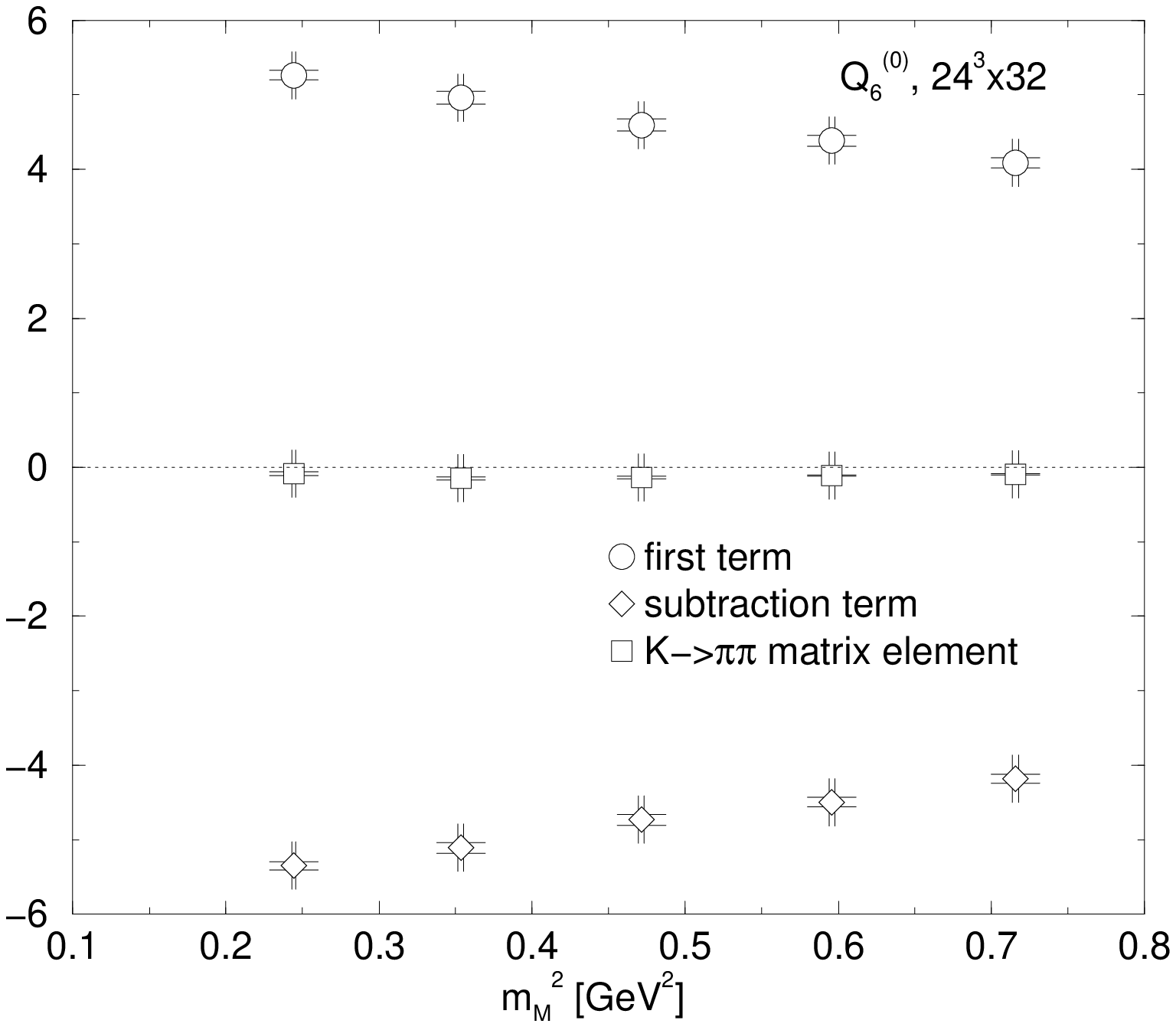}
 \end{center}
 \caption{Effect of subtractions illustrated for $Q_2^{(0)}$ (upper) and 
$Q_6^{(0)}$ (lower) as a function of $m_M^{2}$. 
The original matrix element $\VEV{\pi^+}{Q^{(0)}_i}{K^+}$ (circles) and  
the subtraction term $-\alpha_i\VEV{\pi^+}{Q_{\rm sub}}{K^+}$ 
(diamonds) are added to obtain the physical matrix element (squares). 
Values are multiplied with a factor
$\sqrt{2} f_\pi (m_K^2-m_\pi^2)/\VEV{\pi^+}{A_4}{0}\VEV{0}{A_4}{K^+}$ so that 
the vertical axis has dimension $[{\rm GeV}^3]$. 
Left and right columns are for the lattice sizes $16^3\times 32$ and 
$24^3\times 32$ respectively.}
\label{effect-fig}
\end{figure}

According to (\ref{chptred0}) the contribution of the unphysical 
operator $Q_{\rm sub}$ has to be subtracted for calculating the 
$\Delta I=1/2$ matrix elements.  FIG.~\ref{effect-fig} shows the 
original matrix element $\VEV{\pi^+}{Q^{(0)}_i}{K^+}$ (circles), 
the subtraction term $-\alpha_i\VEV{\pi^+}{Q_{\rm sub}}{K^+}$ 
(diamonds), and their sum (squares), 
multiplied with a factor
$ \sqrt{2} f_\pi ( m_K^2 - m_\pi^2 ) / \VEV{\pi^+}{A_4}{0}\VEV{0}{A_4}{K^+}$
for conversion to the $K\to\pi\pi$ matrix elements (see (\ref{chptred0})).
The left and right columns correspond to the spatial sizes $16^3$ and 
$24^3$ respectively, and the upper and lower rows exhibit the data 
for $Q^{(0)}_2$ and $Q^{(0)}_6$ as typical examples.
These matrix elements play a dominant role in the $\Delta I=1/2$ 
rule and $\varepsilon'/\varepsilon$ as we see in later sections. 
The numerical details of subtractions for all of 
the relevant operators $Q_i^{(0)}$ for $i=1,2,3,5,6$
are collected in TABLE~\ref{effect-tab}.

We observe that the subtraction term represents a crucial contribution 
in the physical matrix element.
In the case of $Q^{(0)}_2$ the subtraction term is twice larger than 
the original matrix element and opposite in sign.  Thus, the physical 
matrix element is similar in magnitude but flipped in sign compared to the 
original matrix element.  

For the case of $Q_6^{(0)}$ the subtraction term 
almost cancels the original matrix element so that 
the physical matrix element is an order of magnitude reduced in size. 
Nonetheless, as one can see from inspection of TABLE~\ref{effect-tab},  
the physical matrix elements are well determined with errors of 10--20\%. 

These results show that the subtraction plays a crucial role in 
calculations with the reduction method.  Numerically this procedure is 
well controlled in our case.

\subsection{RENORMALIZATION AND RG-RUNNING} \label{renorm-running}

Throughout this paper, the renormalization of the operators and the RG-running
of the matrix elements are carried out within the perturbation
theory in $\overline{\rm MS}$ scheme with NDR.

The physical $K\to\pi\pi$ amplitudes in the isospin basis $A_I$ are 
given by 
\begin{eqnarray}
A_I=\frac{G_F}{\sqrt{2}}V_{\rm us}V_{\rm ud}^* \sum_{i=1}^{10}
W_i(\mu)\vev{Q_i}_I^{\overline{\rm MS}}(\mu), \label{AI-ope}
\end{eqnarray}
where we set $\delta_I = 0$ since our calculation at the tree level 
of $\chi$PT does not incorporate the effect of the final state interaction;  
this effect begins from the next to leading order of $\chi$PT.
The Wilson coefficient functions have a form 
\begin{eqnarray}
W_i(\mu)&=&z_i(\mu)+\tau\cdot y_i(\mu)
\end{eqnarray}
where $y_i$ are non-vanishing only for $i=3, \cdots, 10$ and 
$\tau\equiv -(V^*_{\rm ts}V_{\rm td})/(V^*_{\rm us}V_{\rm ud})$ is 
a complex constant. 
With our choice of scale $\mu=m_c=1.3$~GeV, the functions $z_i(m_c)$ 
are negligibly small for $i=3,\cdots, 10$~\cite{Buch-Buras-Lauten95}.

The coefficient functions $y_i(\mu)$ and $z_i(\mu)$ at $m_c=1.3$~GeV 
have been calculated for several values of the QCD parameter 
$\Lambda_{\ovl{\rm MS}}^{(4)}$~\cite{Buch-Buras-Lauten95}.
We employ $\Lambda_{\ovl{\rm MS}}^{(4)}=$ 325 MeV for our main results, and
also consider $\Lambda_{\ovl{\rm MS}}^{(4)}=$ 215 MeV and 435 MeV 
to examine the magnitude of systematic error. 
The choice of the central value is motivated by recent phenomenological  
compilations of the strong coupling constant, {\it e.g.,} Ref.~\cite{Bethke00}
quotes  $\Lambda_{\ovl{\rm MS}}^{(4)}=296^{+46}_{-44}$~MeV corresponding to 
$\alpha_S^{\overline{\rm MS}}(M_{Z^0})=0.1184(31)$. 
We list the values of coefficient functions we use in TABLE~\ref{Wilsoncoeff}.
The experimental parameters are 
summarized in Appendix~\ref{appendix-param}.

To calculate the renormalized matrix elements in the $\overline{\rm MS}$ 
scheme $\vev{Q_i}_I^{\overline{\rm MS}}(\mu)$,  
we first translate the lattice values into the renormalized ones 
at a matching scale $q^*$:
\begin{eqnarray}
\vev{Q_i}^{\ovl{\rm MS}}_I(q^*)={\cal Z}_{ij}(q^*a)\vev{Q_j}^{\rm latt}_I
(1/a). 
\end{eqnarray}
This step is carried out using the renormalization factor 
calculated to one-loop order of perturbation theory 
~\cite{AIKT98,AIKT99,AK00,Taniguchi01}.  
The detailed form of the one-loop terms and explicit numerical values 
for $q^*=1/a$ in quenched QCD,  appropriate for our case, are given in 
Appendix \ref{appendix-ZU}.   

The next step is to evolve the renormalized matrix elements from the scale 
$q^*=1/a$ to $\mu=m_c$ using the renormalization group, and combine them 
with the Wilson coefficient functions $W_i(\mu)$. The RG-evolution of 
the matrix elements $\vev{Q_i}_I^{\overline{\rm MS}}(\mu)$ is inverse to that 
of the coefficient functions $W_i(\mu)$, {\it i.e.,}
\begin{eqnarray}
W_i(\mu_1)&=&U(\mu_1,\mu_2)_{ij}W_j(\mu_2),\\
\vev{Q_i}_I^{\overline{\rm MS}}(\mu_1)&=&[U^{-1}(\mu_1,\mu_2)^T]_{ij}
\vev{Q_j}_I^{\overline{\rm MS}}(\mu_2).
\end{eqnarray}
Perturbative calculations of $U(m_c,q^*)$ at the next-to-leading order 
are available~\cite{Buch-Buras-Lauten95}. 
In Appendix~\ref{appendix-ZU} we adapt the known results to calculate the 
numerical values of the evolution matrix for our case 
in which $\mu_1=m_c=1.3$~GeV and $\mu_2=1/a= 1.94$~GeV.  
The evolution may be made either for quenched QCD or for $N_f=3$ flavors 
corresponding to $u$, $d$ and $s$ quarks, depending on the view if the 
matching at $\mu=1/a$ is made to the quenched theory or to the $N_f=3$ theory 
in the continuum space-time. This is an uncertainty inherent in quenched 
lattice QCD, and we choose the $N_f=3$ evolution in our calculation. 
We have also tested the evolution with quenched QCD, and found that 
the results for hadronic matrix elements do not change beyond a 
10--20\% level. 

For the coupling constant in our $N_f=3$ evolution, 
we employ the two-loop form 
\begin{equation}
\alpha_S^{\overline{\rm MS}}(\mu)=\dfrac{4\pi}{\beta_0 \ln 
\frac{\mu^2}{\Lambda_{\overline{\rm MS}}^2}} \left[
1-\dfrac{\beta_1}{\beta_0^2}\dfrac{\ln\ln 
\frac{\mu^2}{\Lambda_{\overline{\rm MS}}^2}}
{\ln \frac{\mu^2}{\Lambda_{\overline{\rm MS}}^2}}
\right] 
\end{equation}
with $\Lambda_{\overline{\rm MS}}^{(3)}=372$ MeV, which corresponds to
$\Lambda_{\overline{\rm MS}}^{(4)}=325$ MeV.  
In order to check systematic errors associated with this choice,
we also make calculations for  
$\Lambda_{\overline{\rm MS}}^{(3)}=259$~MeV 
($\Lambda_{\overline{\rm MS}}^{(4)}=215$~MeV) and
$\Lambda_{\overline{\rm MS}}^{(3)}=478$~MeV
($\Lambda_{\overline{\rm MS}}^{(4)}=435$~MeV).

\section{RESULTS OF HADRONIC MATRIX ELEMENTS}\label{result-HME}

\subsection{CHIRAL PROPERTIES OF ${\bf K\to\pi}$ MATRIX ELEMENTS }
\label{chi-prop}

As we mentioned in Sec.~\ref{sec:runs}, the RG-improved gauge action 
provides the advantage that the measure of residual chiral symmetry breaking 
$m_{5q}$ due to finite $N_5$ is small at $a^{-1}\simeq 2$ GeV. 
It is nonetheless desirable to check the size of chiral symmetry 
breaking effect directly for the $K\to\pi$ matrix elements.

Explicit chiral symmetry breaking, if present, causes mixing of the 
$I=0$ four-quark operators $Q_i^{(0)}$ with the lower dimensional operator 
$\bar s d$ without quark mass suppression, 
so that $K\to\pi$ matrix elements at $m_d = m_s = m_f$
behave as
\begin{eqnarray}
   \VEV{\pi^+}{Q_i^{(0)} - \alpha_i Q_{\rm sub}} {K^+} 
&=& 
     \frac{ 2 m_M^2 }{f^2} ( a_i - c_i^{(0)} ) 
   + \left(\frac{\beta_i}{a^3} + \frac{\gamma_i}{a^2} m_f \right)
\VEV{\pi^+}{\bar s d} {K^+} 
   + {\cal O}(m_M^4)
\label{eq:sd}
\end{eqnarray}
for $(8_L, 1_R)$  operators, and
\begin{eqnarray}
   \VEV{\pi^+}{Q_i^{(0)}} {K^+} 
&=& 
     \frac{4}{f^2}d_i^{(0)}
   + \frac{\delta_i}{a^3} \VEV{\pi^+}{\bar s d} {K^+} 
   + {\cal O}(m_M^2) 
\end{eqnarray}
for $(8_L, 8_R)$ operators.  
Here $\beta_i, \gamma_i$ and $\delta_i$ 
are dimensionless quantities which represent 
magnitudes of residual chiral symmetry breaking,  and hence are
proportional to $ e^{-c N_5}$ with some constant $c$. 
The matrix element $\VEV{\pi^+}{\bar s d} {K^+}$ stays non-zero 
in the chiral limit.
Motivated by eqs.~(\ref{Qsub-def}) and (\ref{axialconserve}),
one may consider modifications of the subtraction operator such as 
\begin{equation}
Q_{\rm sub} \rightarrow (m_s + m_d + 2 m_{5q} ) \bar s d - (m_s-m_d)
\bar s\gamma_5 d .
\end{equation}
Such modifications, however, will not ensure the complete removal of residual
chiral symmetry breaking from the matrix elements.

The $I=2$ operators $Q_{1,2}^{(2)}$ do not mix with the $\bar s d$ operator.
Their matrix elements can have constant terms in the chiral limit, however, 
due to mixings with dimension 6 operators such as $Q_{7,8}^{(2)}$ 
in the presence of chiral symmetry breaking. 
Hence we also consider the chiral behavior of these matrix elements.

Of the ten operators $Q_i$, we recall that $Q_{4,9,10}$ are dependent 
operators as shown in (\ref{Q4rel}--\ref{Q10rel}).  
Furthermore, there is an identity $Q_1^{(2)}=Q_2^{(2)}$ 
which follows from (\ref{Q1}--\ref{Q2}), and the $I=2$ component is absent  
in the $Q_{3,5,6}$ operators.  Thus we only need to examine 
the matrix elements of $Q_{1,2,3,5,6}^{(0)}$ and $Q_1^{(2)}$. 

\begin{figure}
 \begin{center}
  \leavevmode
  \includegraphics[width=8.3cm, clip]{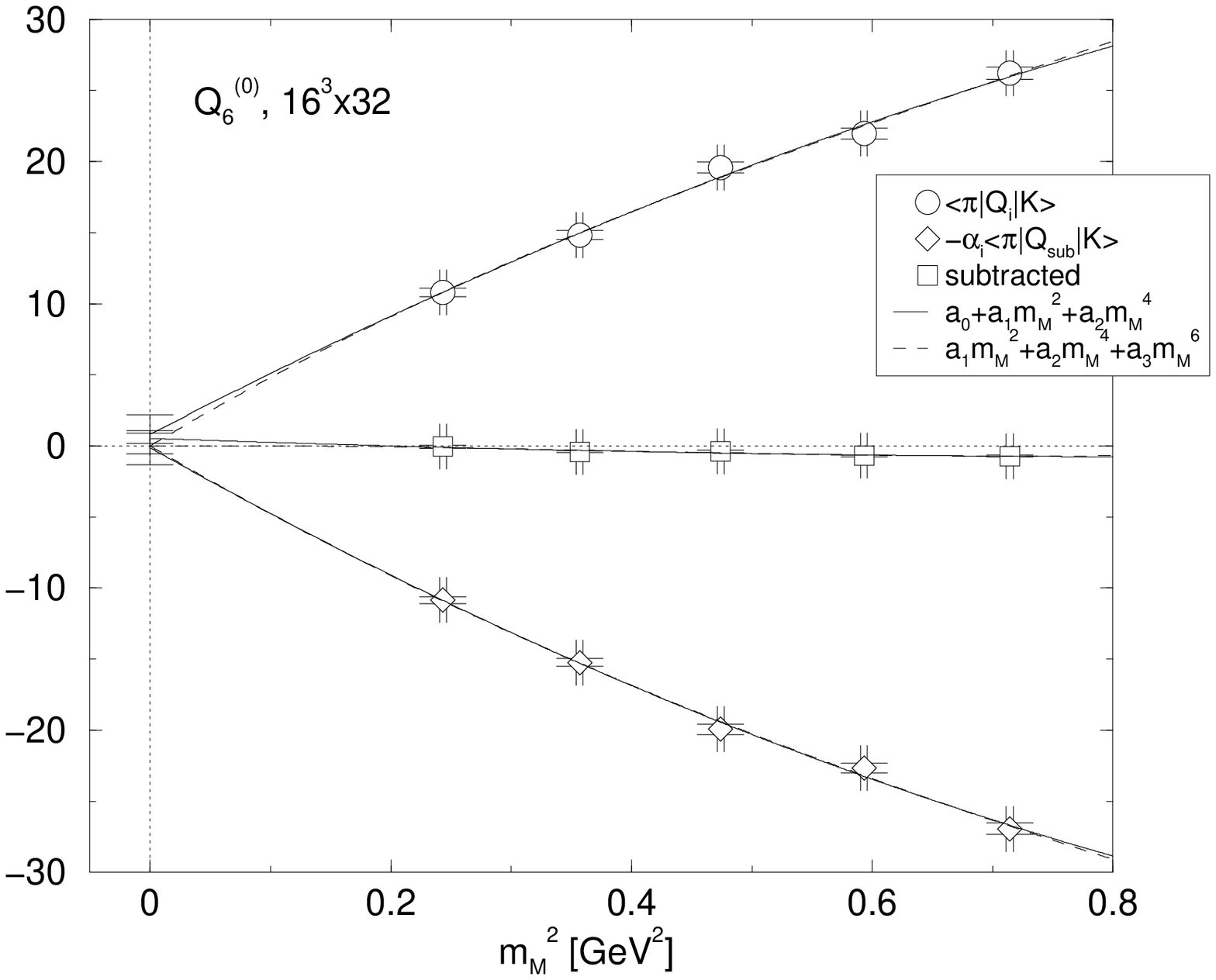}
\hspace{0.1cm}
  \includegraphics[width=8.3cm, clip]{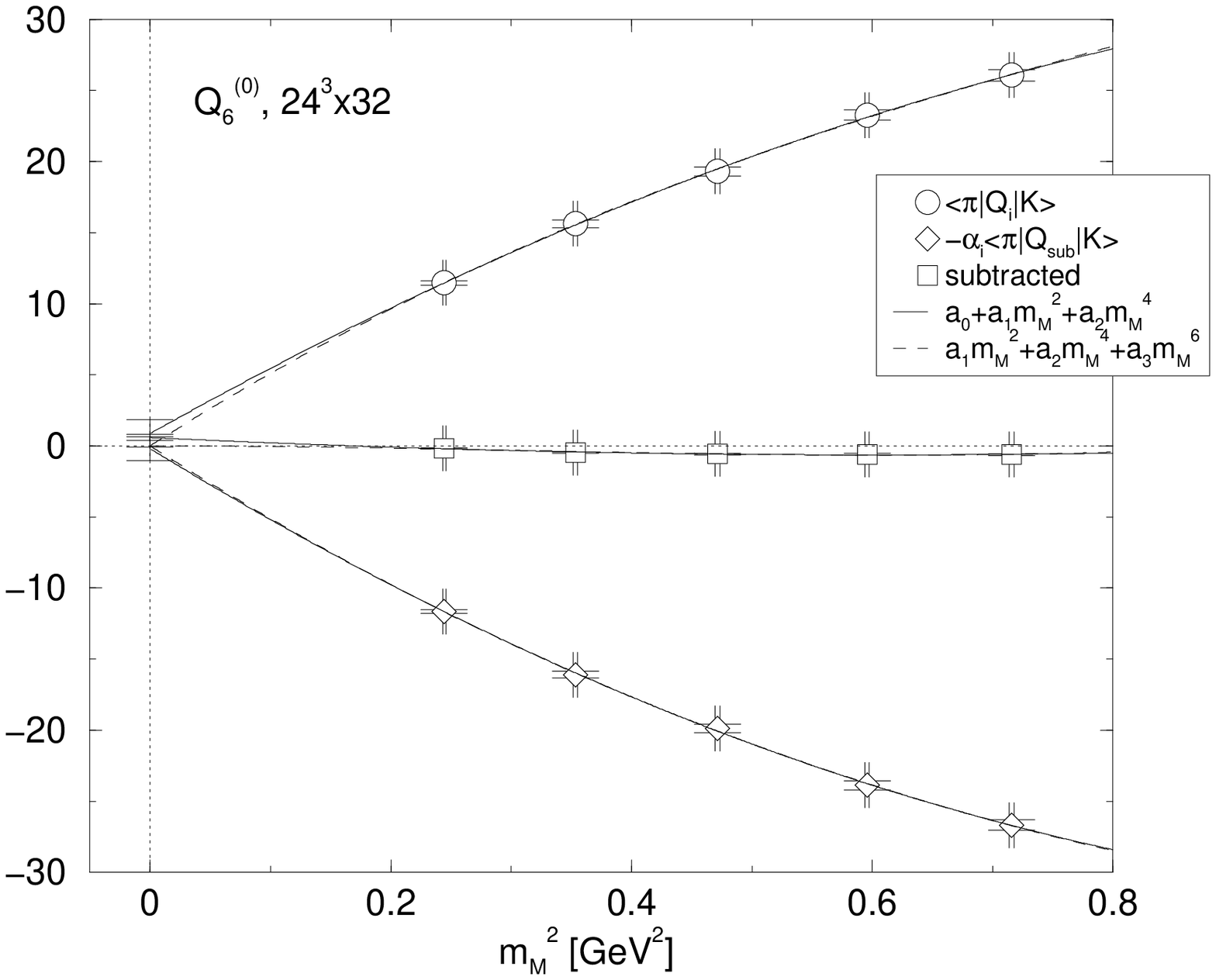}
 \end{center}

 \begin{center}
  \leavevmode
\hspace*{-0.9cm}
  \includegraphics[width=8.0cm, clip]{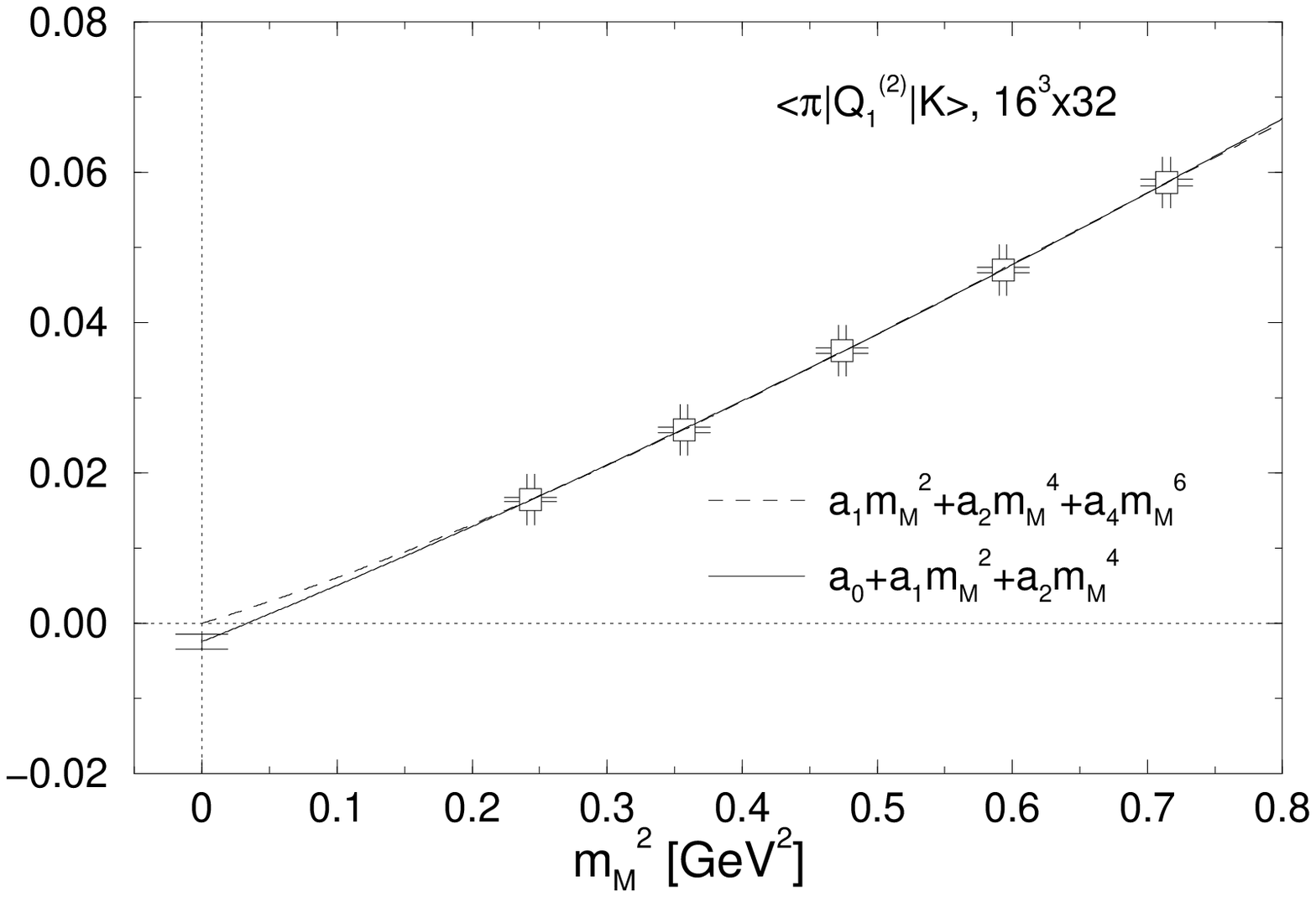}
\hspace{0.4cm}
  \includegraphics[width=8.0cm, clip]{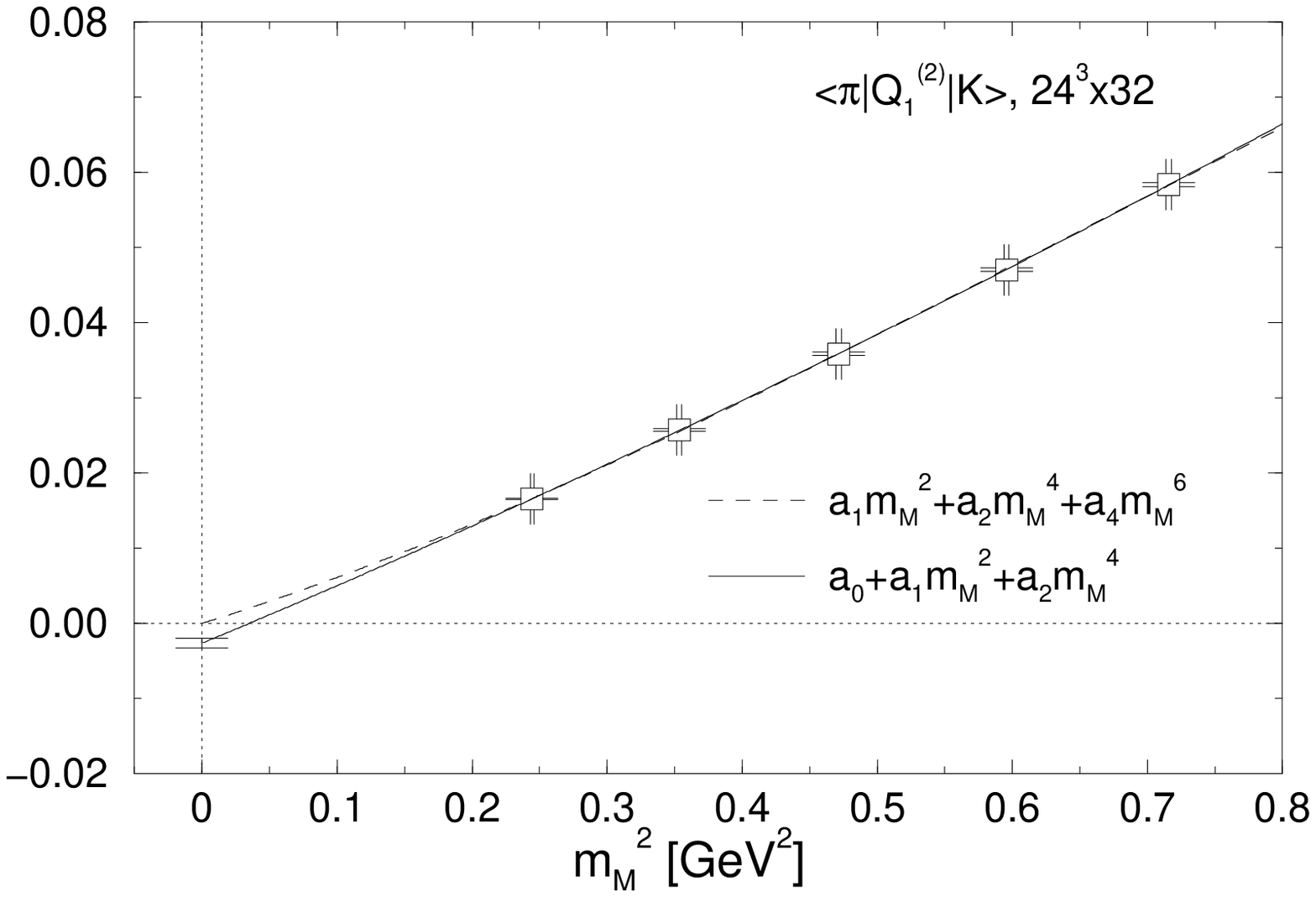}
 \end{center}
 \caption{Ratio of matrix elements 
$\VEV{\pi^+}{X^{(I)}_i}{K^+}/\VEV{\pi^+}{A_4}{0}\VEV{0}{A_4}{K^+}\times 
m_M^2 a^2$ as a function of $m_M^2 [{\rm GeV}^2]$ 
for $i=1,2,3,5,6$ $(I=0)$ and $i=1$ $(I=2)$ from top to bottom. 
Left and right columns are for the lattice sizes 
$16^3\times 32$ and $24^3\times 32$ respectively.
Solid lines represent the chiral extrapolation to $m_M^2\to 0$
with a quadratic function of $m_M^2$, while
dashed lines are with a cubic function as described in the text.}
\label{chiralprops}
\end{figure}

FIG.~\ref{chiralprops} shows these matrix elements
as functions of $m_M^2$ (GeV$^2$) for the two spatial 
volumes $V=16^3$ (left column) and $V=24^3$ (right column), 
adopting the normalization defined by  
\begin{eqnarray}
  \frac{ \VEV{\pi^+}{ X_i^{(I)} }{K^+} }{ \VEV{\pi^+}{A_4}{0}\VEV{0}{A_4}{K^+} }  \times m_M^2 a^2 
= \frac{a^2}{2 f_M^2} \VEV{\pi^+}{X_i^{(I)}}{K^+}
\label{chiral-ratio}
\end{eqnarray}
For the $I=0$ channel, three data sets are plotted, corresponding to the 
original matrix element $X_i^{(I)} = Q_i^{(I)}$ (circles), the subtraction 
term $-\alpha_i Q_{\rm sub}$ (diamonds) and the subtracted matrix element 
$Q_i^{(I)}-\alpha_i Q_{\rm sub}$ (squares).  For the $I=2$ channel, 
subtractions are absent and hence $X_i^{(I)} = Q_i^{(I)}$.

The denominator of (\ref{chiral-ratio}) behaves as 
\begin{eqnarray}
\VEV{\pi^+}{A_4}{0}\VEV{0}{A_4}{K^+} &=& 2 f_M^2 m_M^2,
\end{eqnarray}
irrespective of whether chiral symmetry holds exactly or not. 
The advantage of our normalization is that the coefficient of the $m_M^2$ 
term of the ratio is directly related to the $K^0\to\pi^+\pi^-$ matrix 
elements. 
An alternative normalization is provided by the ratio 
\begin{eqnarray}
\frac{\VEV{\pi^+}{X_i^{(I)}}{K^+}}{\VEV{\pi^+}{P}{0}\VEV{0}{P}{K^+}}
\label{chiral-ratio2}
\end{eqnarray}
where $P=\overline{q}\gamma_5q$ is the pseudo scalar density.  
This method avoids the use of measured values of pion 
mass, but it loses the straightforward relation to the physical matrix 
elements. We use the normalization (\ref{chiral-ratio}) in our analyses. 
We have checked, however, that the conclusion remains unchanged even if 
(\ref{chiral-ratio2}) is employed instead.

For chiral extrapolation we consider an expansion of the form 
\begin{eqnarray}
\frac{ a^2 }{2 f_M^2}\VEV{\pi^+}{X_i^{(I)}}{K^+}=
  a_0 + a_1 m_M^2 + a_2(m_M^2)^2 + a_3(m_M^2)^2 \ln m_M^2 + a_4(m_M^2)^3 + 
\cdots .
\label{KPIexpand}
\end{eqnarray}
Chiral extrapolations using the first three terms are indicated by
the solid line in each panel of FIG.~\ref{chiralprops}.  
The fit parameters are summarized in TABLE~\ref{quadtab}.  
The results for the intercept $a_0$ in the chiral limit are consistent with 
zero within the fitting errors except for 
the $I=2$ operator $Q^{(2)}_1$ for the volumes 
$V=16^3 (1.8\sigma)$ and $24^3 (4\sigma)$,
the $I=0$ subtracted operator 
$Q_6^{(0)}-\alpha_6Q_{\rm sub}$ for $V=16^3 (1.4\sigma)$ and  
$24^3 (2.7\sigma)$,
and the subtraction term for the $i=1$ operator $-\alpha_1 Q_{\rm sub}$ 
for $V=24^3 (2.8\sigma)$.
Since no systematic tendency that the intercepts become larger for 
smaller volume is observed, it is unlikely that the non-zero intercepts of
these matrix elements are caused by the finite spatial size effect.
Indeed even an opposite tendency that the intercept becomes larger
for larger spatial volumes is observed.

The absence of systematic trend in our data suggests the possibility 
that non-zero intercepts observed for some of the matrix elements are
artifacts of the long extrapolation in $m_M^2$.  To test this point, 
we attempt a fit with a cubic polynomial of form 
$a_1m_M^2+a_2(m_M^2)^2+a_4(m_M^2)^3$ 
and a form with chiral logarithm given by 
$a_1m_M^2+a_2(m_M^2)^2+a_3(m_M^2)^2\ln m_M^2$, both having a built-in 
chiral behavior of vanishing at $m_M^2=0$. 
We show the former fit curves by dashed lines in FIG.~\ref{chiralprops} and 
the fitted parameters in TABLE~\ref{cubic-tab}. Numerical results of the 
chiral logarithm fit are given in TABLE~\ref{chlogtab}. The fit curves 
are similar to those of the cubic fit. 
Both functions provide good fit of data with reasonable $\chi^2/$dof.  

Let us try to analyze the chiral behavior of $I=0$ matrix elements 
in terms of mixing with the $\bar{s}d$ operator as given in (\ref{eq:sd}). 
The existence of the constant $\beta_i$ can be detected from the 
chiral limit of the matrix elements.
On the other hand, separating the contribution of $\gamma_i$ and $\delta_i$ 
from the physical ones would require results at different $N_5$. 
We leave such an investigation for future studies, and 
assume that the latter contributions are negligible. 
We also ignore mixings with the dimension 5 operator
$\bar s \sigma_{\mu\nu}F_{\mu\nu} d$ since their contributions are 
subleading in $1/a$. 

We estimate $\beta_i$ from the values of $a_0$ obtained in 
the chiral fit of the matrix elements for the subtracted operator 
$Q_i^{(0)}-\alpha_i Q_{\rm sub}$ given in TABLE~\ref{quadtab}.   
For this purpose, we repeat the calculation of (\ref{chiral-ratio}) 
for $X_i^{(I)}=\bar{s}d$, and extract 
\begin{equation}
\frac{ \VEV{\pi^+}{ \bar{s}d }{K^+} a}{ \VEV{\pi^+}{A_4}{0}\VEV{0}{A_4}{K^+} a^4}
\times m_M^2 a^2 
= \frac{1}{2 f_M^2 a} \VEV{\pi^+}{ \bar{s}d }{K^+}
\label{eq:mixing}
\end{equation}
where powers of $a$ are supplied to absorb dimensions of matrix elements.
We then fit the results to a quadratic polynomial 
$b_0+b_1m_M^2+b_2(m_M^2)^2$. 
The numerical values of (\ref{eq:mixing}) are given in Table~\ref{sbard},
and the result for $b_i$ are given in Table~\ref{quadtab}.
Normalizing with $m_{5q}=0.283$~MeV to take 
into account the $e^{-cN_5}$ dependence expected for $\beta_i$, one 
has 
\begin{equation}
\frac{\beta_i}{m_{5q}a}=\frac{a_0}{b_0}\frac{1}{m_{5q}a} .
\end{equation}
In the case of $V=24^3$, the results are 
$\beta_i/(m_{5q}a)= 0.9(1.1)$ for $i=1$,
                  $-0.91(87)$ for $i=2$,
                  $ 0.8(4.2)$ for $i=3$,
                  $-4.7(4.4)$ for $i=5$,
             and $-21.6(8.1)$ for $i=6$. 
Except for the $i=6$ operator for which the coefficient 
is exceptionally large, we find values consistent with 
zero within the errors. 

The analyses described here do not show strong evidence for the effect of 
residual chiral symmetry breaking in the $K\to\pi$ matrix elements. 
Although more data at smaller quark masses will be needed for the definite
conclusion, we conclude here that our results for the matrix elements 
is consistent with the expected chiral behavior within the statistical 
precision of our data.
Therefore, for the chiral extrapolation in the rest of this article, 
we employ the cubic polynomial without constant term for the central value 
and use the form with chiral logarithm to estimate 
the systematic uncertainty.
Since non-zero intercepts beyond statistical errors cannot be excluded for 
some of the matrix elements,  we examine 
possible effects of the residual chiral symmetry breaking to the
physical matrix elements in Sec.\ref{result-Phys}.

Let us also make a comment on comparison of lattice data with 
predictions of quenched chiral perturbation theory.
For the $I=0$ channel, data for more values of $m_f$ are required for 
such a comparison because of the presence of a number of unknown parameters 
as well as a new term of form $b_1m_M^2\ln m_M^2$ in the predicted matrix 
elements~\cite{Golterman:2000fw}.  
On the other hand, quenched chiral logarithm terms are absent for the $I=2$ 
matrix elements governed by the $(27_L,1_R)$ operator, 
and the ratio $a_3/a_1$ for $Q^{(2)}_1$ is predicted to be 
$a_3/a_1 = -6/(16\pi^2f_\pi^2) = - 2.180$ GeV$^{-2}$.
We observe in TABLE \ref{chlogtab} that the fitted value agrees in sign 
but 3 to 4 times smaller in magnitude than the prediction, 
{\it e.g.}, $a_3/a_1 = -0.58(10)$ GeV$^{-2}$ on a $24^3\times32$ lattice. 

Quenched chiral perturbation theory makes the same prediction for the 
coefficient of the logarithm term of the chiral expansion of $B_K$ 
as it is  governed by the same operator in $\chi$PT.  
For this case, similar discrepancies of lattice results from the prediction 
is  found for the case of the staggered fermion action~\cite{BK-JLQCD} 
as well as for the domain wall fermion action~\cite{BK-CPPACS}.
A possible explanation for these large discrepancies is that  
higher order corrections in (quenched) $\chi$PT are non-negligible
at quark masses employed in the current simulation.
Indeed we have confirmed that data for $Q^{(2)}_1$ can not be fitted by
the form $a_1m_M^2+a_2(m_M^2)^2+a_3(m_M^2)^2\ln m_M^2
+ a_4 (m_M^2)^3$ with $a_3/a_1 = - 2.180$ GeV$^{-2}$ fixed.
The complete form in $\chi$PT to this order,
$a_1m_M^2+a_2(m_M^2)^2+a_3(m_M^2)^2\ln m_M^2 + a_4 (m_M^2)^3
+a_5 (m_M^2)^3\ln m_M^2+a_6 (m_M^2)^3(\ln m_M^2)^2$, unfortunately, 
can not be employed for our data calculated only at five values of 
quark masses.
Understanding the small value of $a_3/a_1$ for $Q^{(2)}_1$ require further 
studies.

\subsection{PHYSICAL VALUES OF HADRONIC MATRIX ELEMENTS}

We tabulate the values of all the $K\to\pi\pi$ matrix elements 
in TABLEs~\ref{HMEmf-16} (for $16^3\times 32$) and \ref{HMEmf-24}
(for $24^3\times 32$). The upper half of each table lists the bare lattice 
values, $\vev{Q_i}^{\rm latt}_I$, and the lower half the physical 
values, $\vev{Q_i}^{\overline{\rm MS}}_I$, obtained through 
matching at the scale $q^*=1/a$ followed by an RG-evolution to 
$\mu=m_c$. Note that $\vev{Q_{3-6}}^{\overline{\rm MS}}_2$
become non-zero due to the RG-evolution which breaks
the isospin symmetry in the presence of the QED interaction.
The two sets of numbers do not differ beyond a 10--20\% level 
except for $\vev{Q_{5,6,7,8}}_0$, for which the difference amounts 
to 30--40\%.  The latter situation arises from a larger magnitude of 
mixing of order 5--10\% among the $Q_{5,6,7,8}^{(0)}$ operators compared 
to the other operators which are typically less than 5\%. 
In the following, the superscript ``$\overline{\rm MS}$'' will be
omitted unless the confusion may arise.

In TABLE~\ref{matching-diff} we illustrate the magnitude of uncertainty 
due to the choice of $q^*$ by comparing the values of physical hadronic 
matrix elements $\vev{Q_i}_I(m_c)$ for the choices  
$q^*=1/a$ and $q^*=\pi/a$ at $m_f=0.02$ on a $24^3$ spatial volume.  

One finds that the difference is at most 20--30\%.  

\begin{figure}
\hspace*{0.8cm}
 \begin{center}
  \includegraphics[width=8.1cm, clip]{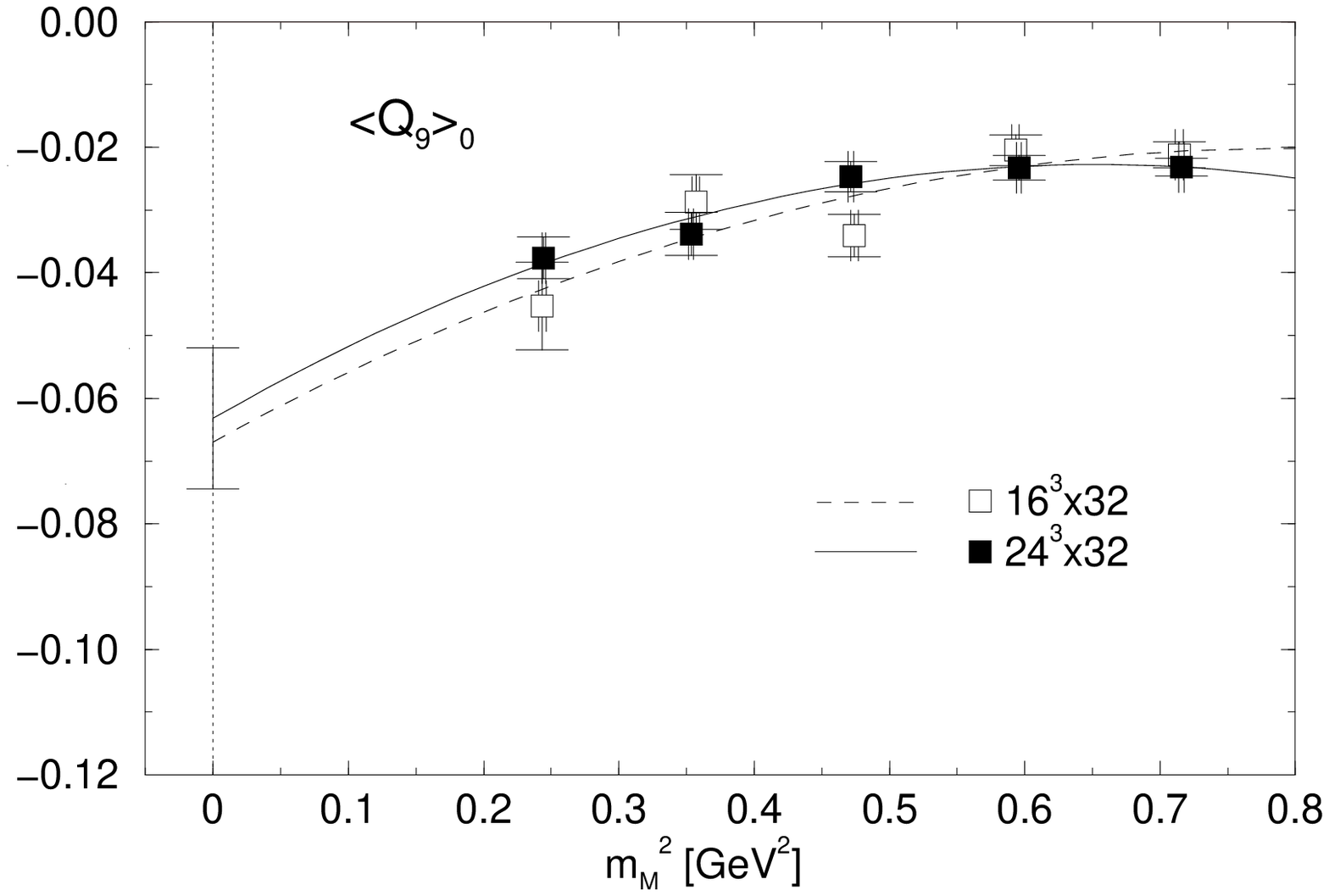}
  \includegraphics[width=8cm, clip]{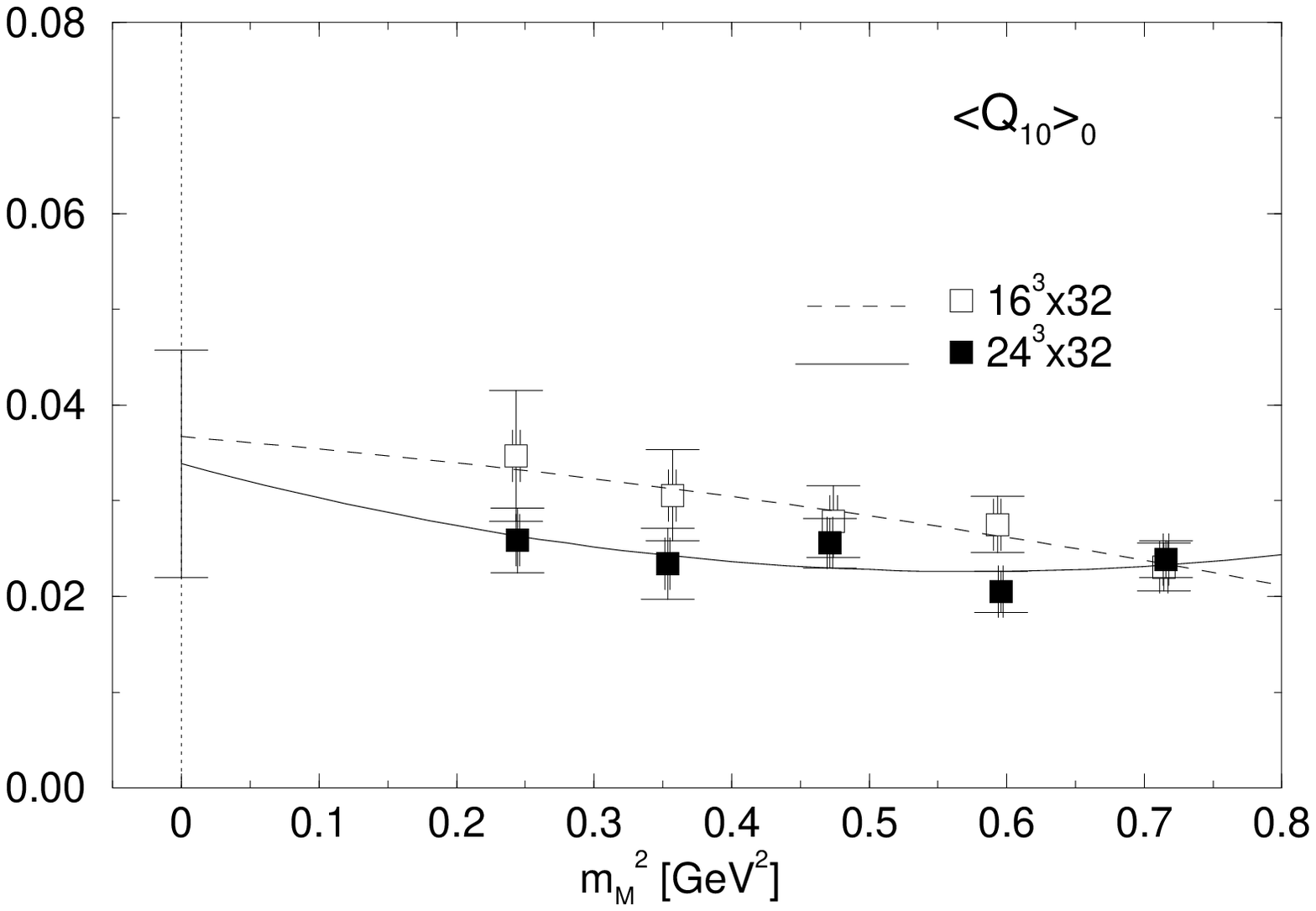}
 \end{center}
\caption{Physical hadronic matrix elements
$\vev{Q_{i}}_0$ for $i=1,2,3,4,5,6,9,10$ 
as a function of $m_M^2$ from top to bottom.  These matrix elements involve 
subtractions of unphysical effects.
Open and filled symbols are from the spatial volume $V=16^3$ and $24^3$ 
respectively. Chiral extrapolations with a quadratic polynomial is shown by 
solid ($V=24^3$) and dashed ($V=16^3$) lines. Fit error in the chiral
limit is added for the former. }
\label{HMEvmp0}
\end{figure}

In FIG.~\ref{HMEvmp0} we plot the physical matrix elements for the
$\Delta I=1/2$ amplitudes $\vev{Q_i}_0\ (i=1,\cdots,6, 9, 10)$ 
as a function of $m_M^2$.  
These eight matrix elements involve the subtraction of unphysical 
effects. The open and filled symbols indicate the data from $V=16^3$ and 
$24^3$ volumes, respectively. Within the statistical errors at each 
$m_f$ and the fluctuation for different values of $m_f$, both of 
which are larger for the smaller spatial size $16^3$, the data from the 
two spatial volumes do not show indications of presence of 
finite size effects. 

\begin{figure}
 \begin{center}
  \includegraphics[width=8.1cm, clip]{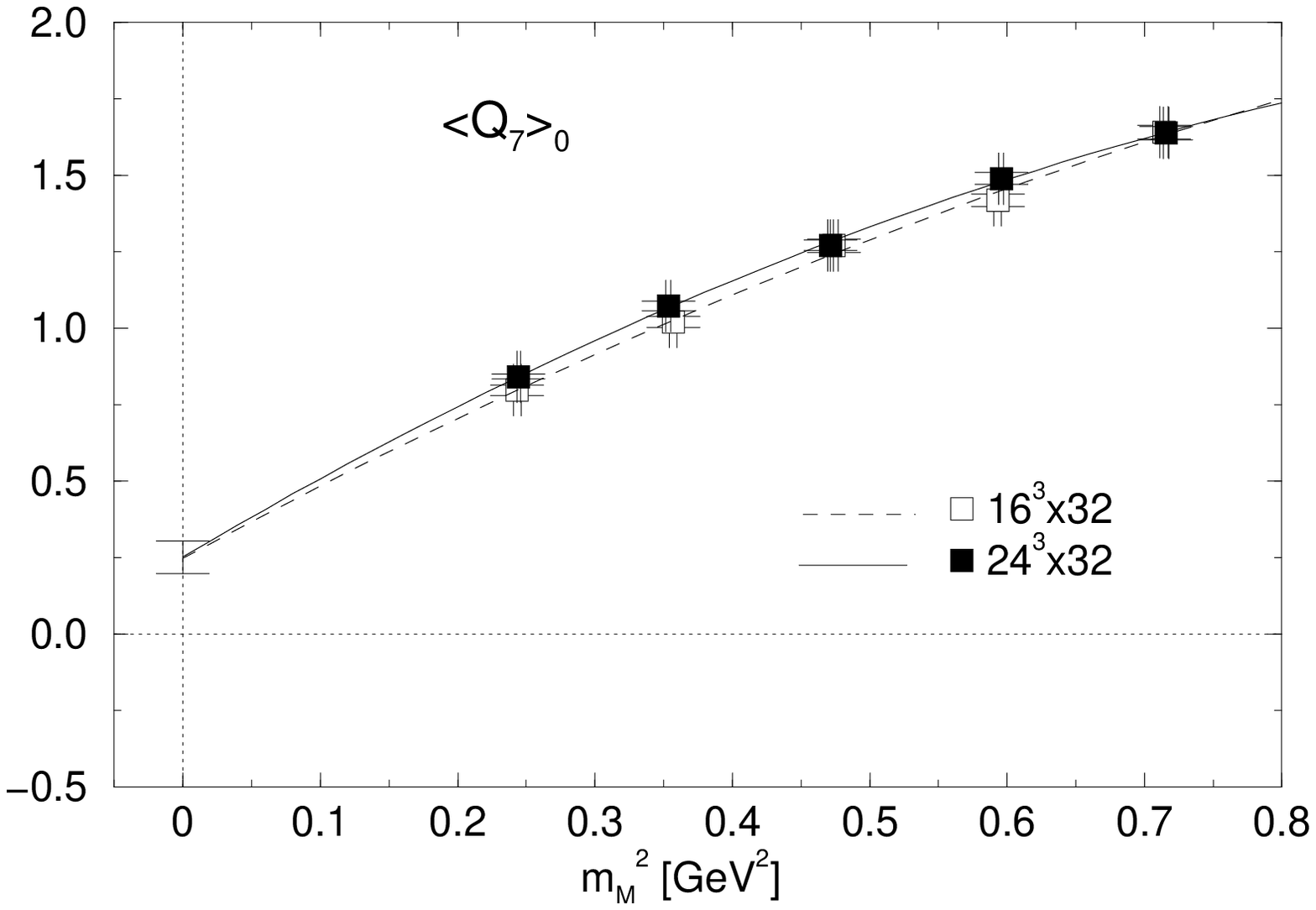}
  \includegraphics[width=8cm, clip]{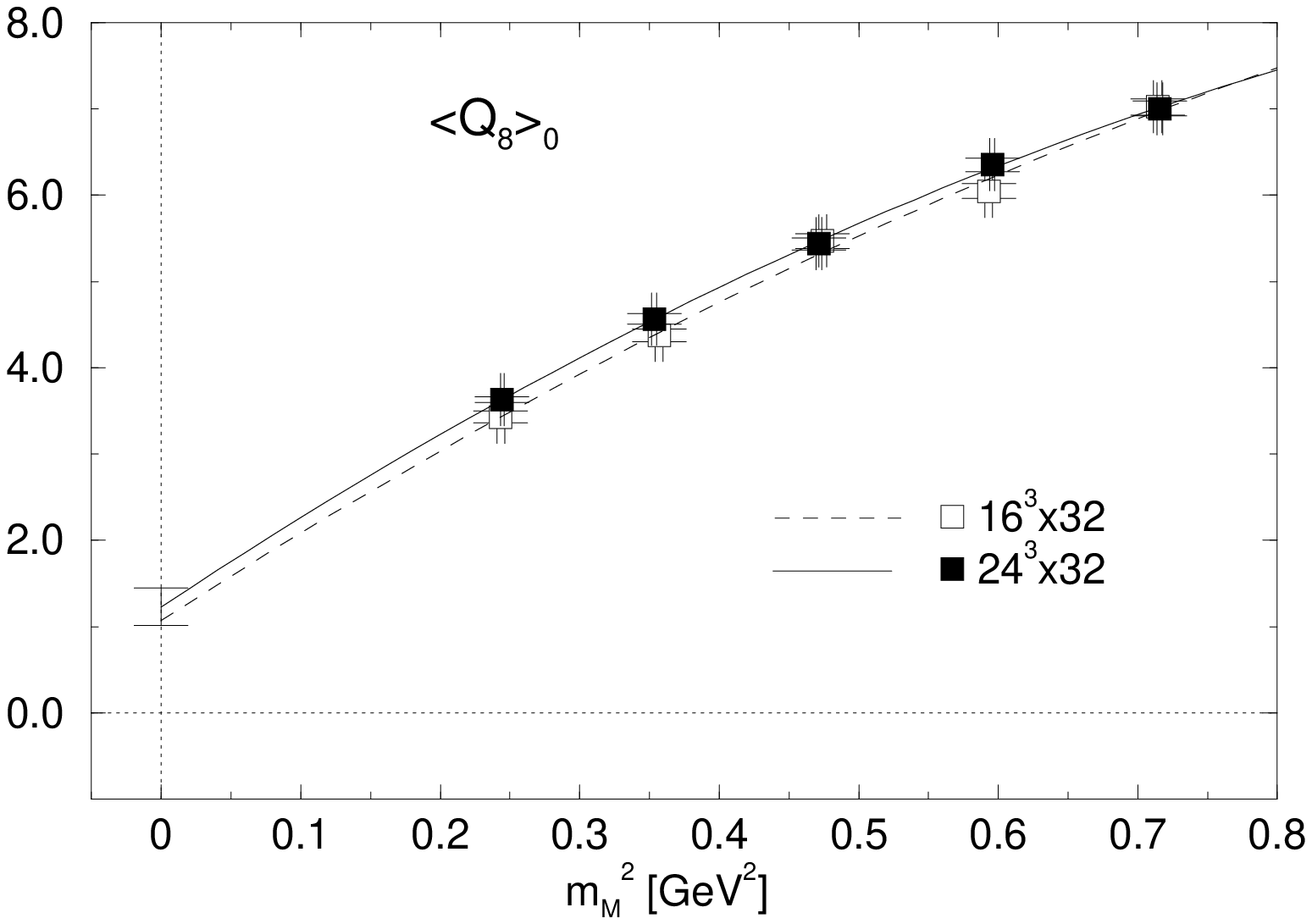}

 \end{center}
\caption{Physical hadronic matrix elements $\vev{Q_{7,8}}_0$ 
as a function of $m_M^2$. The organization of each panel is the same as that
in FIG.~\ref{HMEvmp0}.}
\label{HMEvmp088}
\end{figure}

The remaining matrix elements $\vev{Q_{7,8}}_0$ for the $\Delta I=1/2$ 
amplitude, which do not require the subtraction, are shown in 
FIG.~\ref{HMEvmp088}.  These matrix elements are well determined and 
exhibit a clear $m_M^2$ dependences. 

\begin{figure}
\hspace*{0.5cm}
\begin{center}
  \includegraphics[width=8cm, clip]{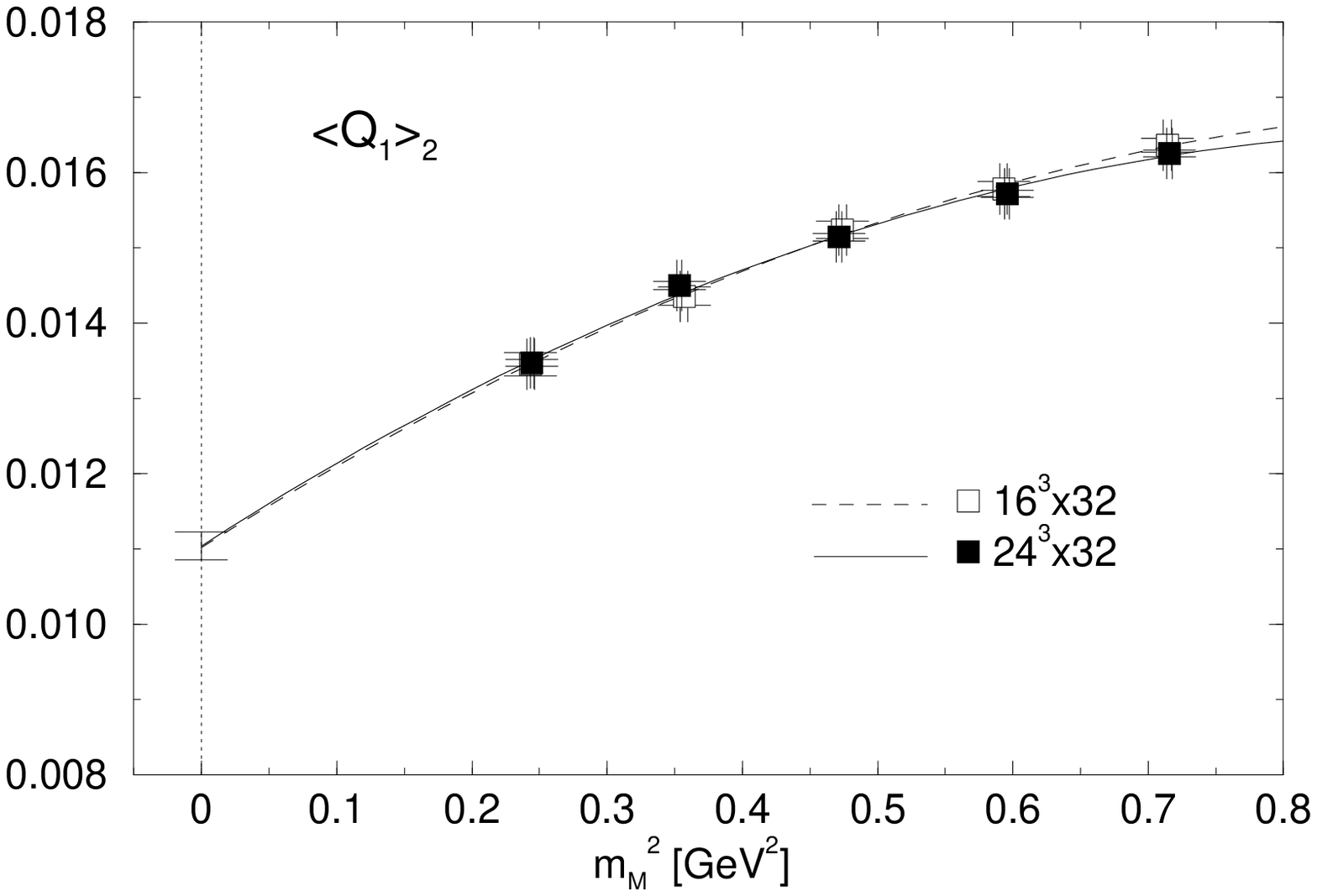}
\end{center}
\hspace*{0.8cm}
\begin{center}
  \includegraphics[width=8cm, clip]{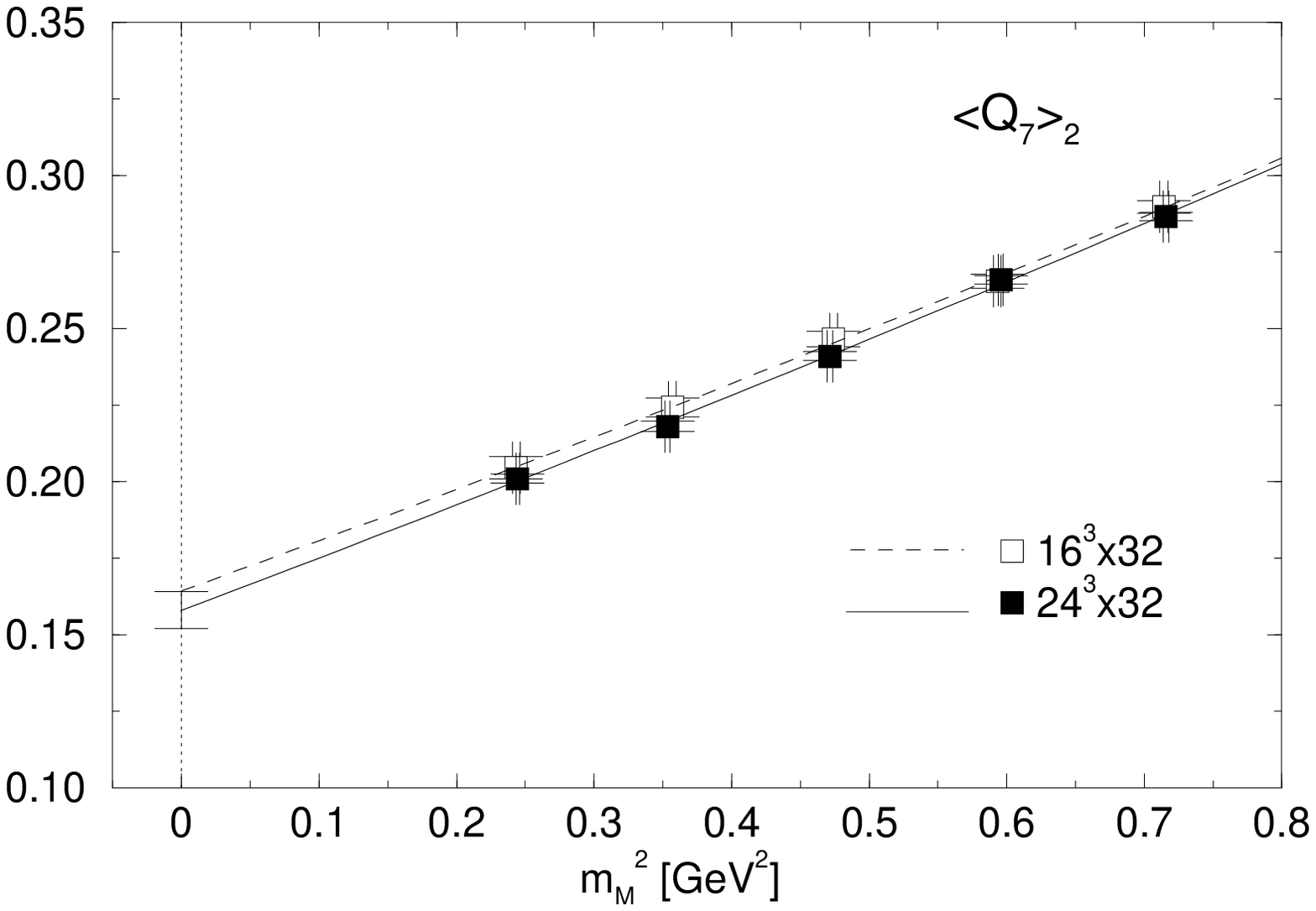}
  \includegraphics[width=8cm, clip]{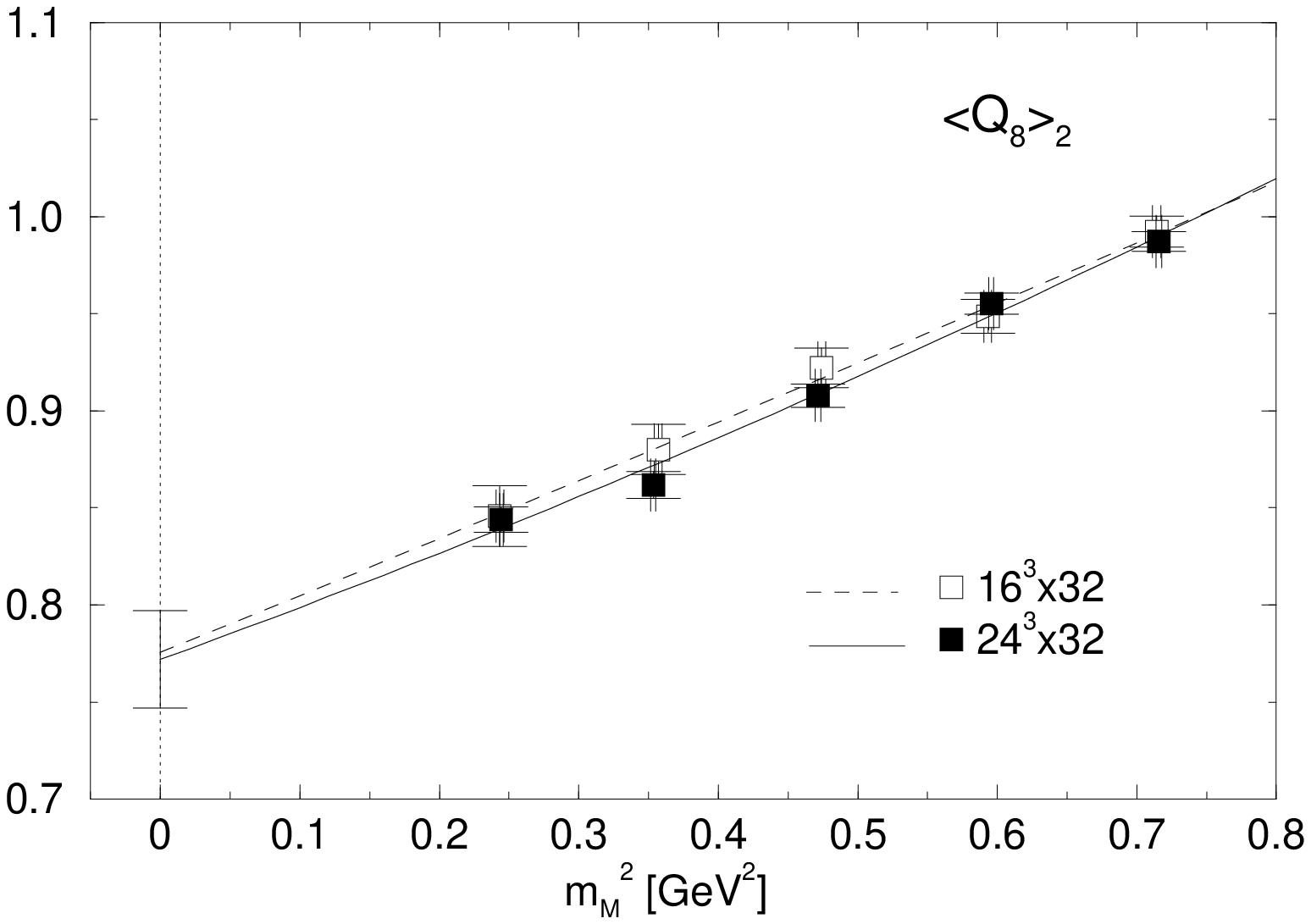}
 \end{center}
\caption{Physical hadronic matrix elements, $\vev{Q_1}_2$
and $\vev{Q_{7,8}}_2$ as a function of $m_M^2$. 
The organization of each panel is the same as that in Fig.~\ref{HMEvmp0}.}
\label{HMEvmp21}
\end{figure}

The matrix elements for the $\Delta I=3/2$ channel given by 
$\vev{Q_1}_2=\vev{Q_2}_2$ and $\vev{Q_{7,8}}_2$ are plotted 
in FIG.~\ref{HMEvmp21}.
Their statistical quality and $m_M^2$ dependence are similar to those 
for $\vev{Q_{7,8}}_0$.  

As discussed in Sec.~\ref{chi-prop},
for extracting the values in the chiral limit,
we adopt a quadratic polynomial form,
\begin{eqnarray}
\vev{Q_i}_I &=& 
\xi_0+\xi_1m_M^2 +\xi_3m_M^4 .
\label{eq:quadra}
\end{eqnarray}
In addition we also employ the chiral logarithm form,
\begin{eqnarray}
\vev{Q_i}_I &=& \xi_0+\xi_1m_M^2 +\xi_2 m_M^2\ln m_M^2 .
\end{eqnarray}

In TABLEs~\ref{HME0-16} and~\ref{HME0-24}, results from these 
chiral extrapolations are summarized with the values of $\chi^2/{\rm dof}$.
The differences between two types of fits should be taken as a measure 
of systematic error.
For $\vev{Q_6}_0$, one observes in FIG.~\ref{HMEvmp0}
an exceptional behavior of the data at $m_f=0.02$. An additional 
chiral extrapolation excluding this quark mass is hence also made
for comparison 
and the fit lines indicated in the figures are obtained.

\subsection{B parameters}

We convert renormalized hadronic matrix elements 
at $\mu =m_c=$ 1.3 GeV into B parameters defined by~\cite{Buch-Buras-Lauten95}
\begin{eqnarray}
B^{(1/2)}_1&=&-\frac{9}{X}\vev{Q_1}_0\ ,\\ 
B^{(1/2)}_2&=&\frac{9}{5X}\vev{Q_2}_0,\\ 
B^{(1/2)}_3&=&\frac{3}{X}\vev{Q_3}_0,\\ 
B^{(1/2)}_5&=&\frac{3}{Y}\vev{Q_5}_0\ ,\\
B^{(1/2)}_6&=&\frac{1}{Y}\vev{Q_6}_0,\\ 
B^{(1/2)}_7&=&-\frac{\vev{Q_7}_0}{\frac{1}{6}Y(\kappa+1)-\frac{1}{2}X}\ ,\\
B^{(1/2)}_8&=&-\frac{\vev{Q_8}_0}{\frac{1}{2}Y(\kappa+1)-\frac{1}{6}X},\\
B^{(3/2)}_1&=&\frac{9}{4\sqrt{2}X}\vev{Q_1}_2\ ,\\
B^{(3/2)}_7&=&-\frac{\vev{Q_7}_2}{\frac{\kappa}{6\sqrt{2}}Y+
\frac{1}{\sqrt{2}}X}\ ,\\
B^{(3/2)}_8&=&-\frac{\vev{Q_8}_2}{\frac{\kappa}{2\sqrt{2}}Y+
\frac{\sqrt{2}}{6}X}\ ,
\end{eqnarray}
where 
\begin{eqnarray}
\kappa&=&\frac{f_\pi}{f_K-f_\pi}\ ,\ X=\sqrt{3}f_\pi(m_K^2-m_\pi^2)\ ,\ 
Y= -4\sqrt{3}\left[\frac{m_K^2}{m_s+m_d}\right]^2\frac{f_\pi}{\kappa} .
\end{eqnarray}
We summarize the values of B parameters in the chiral limit obtained 
by the fit with quadratic polynomial or chiral logarithm 
in TABLE~\ref{Bparams}.
Quark masses and other parameters used in the calculations 
are given in Appendix~\ref{appendix-param}.

Let us compare our values of B parameters with typical ones quoted 
in phenomenology (see {\it e.g.},~\cite{Buch-Buras-Lauten95}).
For the B parameters important for the $\Delta I =1/2$ rule,
the experimental value of $\re A_2$ indicates
$B_{1,NDR}^{(3/2)}(m_c)=0.453$ with $\Lambda_{\overline{\rm MS}}^{(4)}
=$ 325 MeV, with which our value $B_1^{(3/2)}(m_c)\approx 0.4$ -- 0.5 
is consistent. 
On the other hand, our results $B_1^{(1/2)}(m_c)\approx 8$ -- 9 and 
$B_2^{(1/2)}(m_c)\approx 3$ -- 4 are smaller than $B_1^{(1/2)}(m_c) 
\simeq 15$ and 
$B_{2,NDR}^{(1/2)}(m_c)=6.6$ needed to explain the experimental value of 
$\re A_0$.
For the parameter $B_{6}^{(1/2)}$ relevant for the direct CP violation, 
the largest of our estimate $B_{6}^{(1/2)}(m_c)\approx 0.3$ from the 
4-point fit of the data from the $24^3$ spatial volume is still much 
smaller than $B_{6}^{(1/2)}=1$ in the $1/N_c$ approach, while
$B_8^{(3/2)}(m_c)\approx 0.9$ is comparable to $B_8^{(3/2)}=1$ again 
in the $1/N_c$ approach.
In general the B parameters for $I=0$ are smaller than usual estimates.

Previous studies gave 
$B_{7}^{(3/2)}\ ( \mu = 2\ {\rm GeV, NDR})$ = 0.58(7) and
$B_{8}^{(3/2)}( \mu = 2\ {\rm GeV, NDR})$ = 0.81(4)~\cite{Gupta-etal},
$B_{7}^{(3/2)}\ ( \mu = 2\ {\rm GeV, RI(MOM)})$ = 0.38(11)
and $B_{8}^{(3/2)}( \mu = 2\ {\rm GeV, RI(MOM)})$ = 0.77(9)~\cite{Donini-etal99},
$B_{7}^{(3/2)}\ ( \mu = 2\ {\rm GeV, NDR})$ = 0.58(9) and
$B_{8}^{(3/2)}( \mu = 2\ {\rm GeV, NDR})$ = 0.80(9)~\cite{Lell-David},
from quenched lattice QCD,
and
$B_{7}^{(3/2)}( \mu = 2\ {\rm GeV, NDR})$ = 0.55(12) and
$B_{8}^{(3/2)}( \mu = 2\ {\rm GeV, NDR})$ = 1.11(28) from
dispersive sum rules
where $m_s+m_d = 100$ MeV is used~\cite{Donoghue-Golowich}. 
Our values are
$B_{7}^{(3/2)}( \mu = 1.3\ {\rm GeV, NDR})$ = 0.62(3) and
$B_{8}^{(3/2)}( \mu = 1.3\ {\rm GeV, NDR})$ = 0.92(4)
on a $24^3\times 32$ lattice in broad agreement with the above. Note that 
the scale $\mu$ is different between our results and those of other studies.

\section{PHYSICAL RESULTS}\label{result-Phys}

\subsection{$\Delta I=1/2 $ RULE}

The real part of $A_I$ relevant for the $\Delta I=1/2$ rule is written as 
\begin{eqnarray}
\re A_I=\frac{G_F}{\sqrt{2}}|V_{\rm ud}|\cdot|V_{\rm us}|
\left[\sum_{i=1,2}z_i(m_c)\vev{Q_i}_I(m_c)
+(\re\ \tau)\sum_{i=3}^{10}y_i(m_c)\vev{Q_i}_I(m_c)\right],
\label{ampI-OPE2}
\end{eqnarray}
In TABLE~\ref{delI-tab}, we list the values of $\re A_0$, 
$\re A_2$ and $\omega^{-1}=\re A_0/\re A_2$ for each value of $m_f$ and 
spatial volume, and for the three choices of the $\Lambda$ parameter
$\Lambda_{\ovl{\rm MS}}^{(4)}=325$~MeV, 215~MeV and 435~MeV.

\begin{figure}
 \begin{center}
  \leavevmode
  \includegraphics[width=8.1cm, clip]{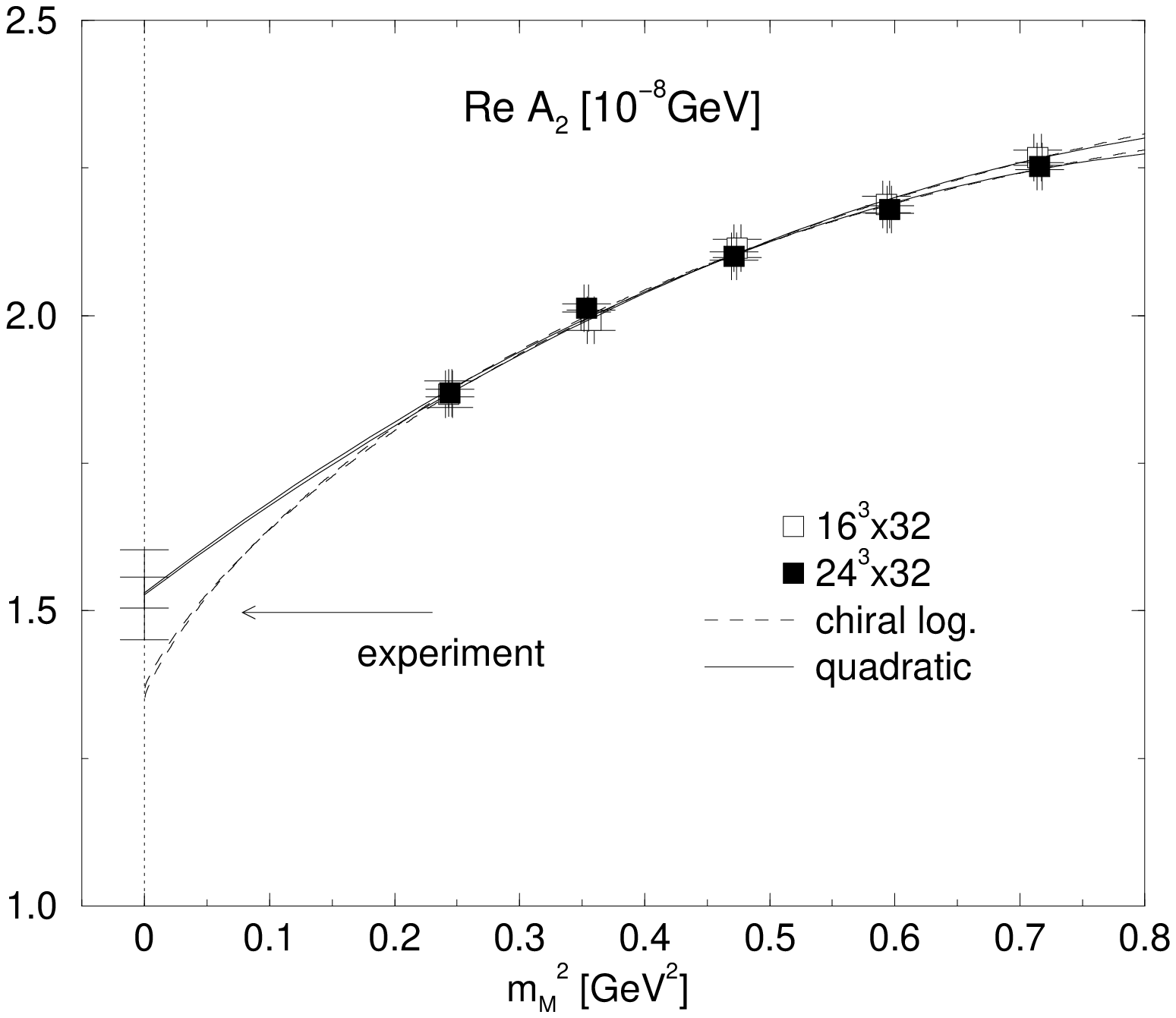}
  \includegraphics[width=8cm, clip]{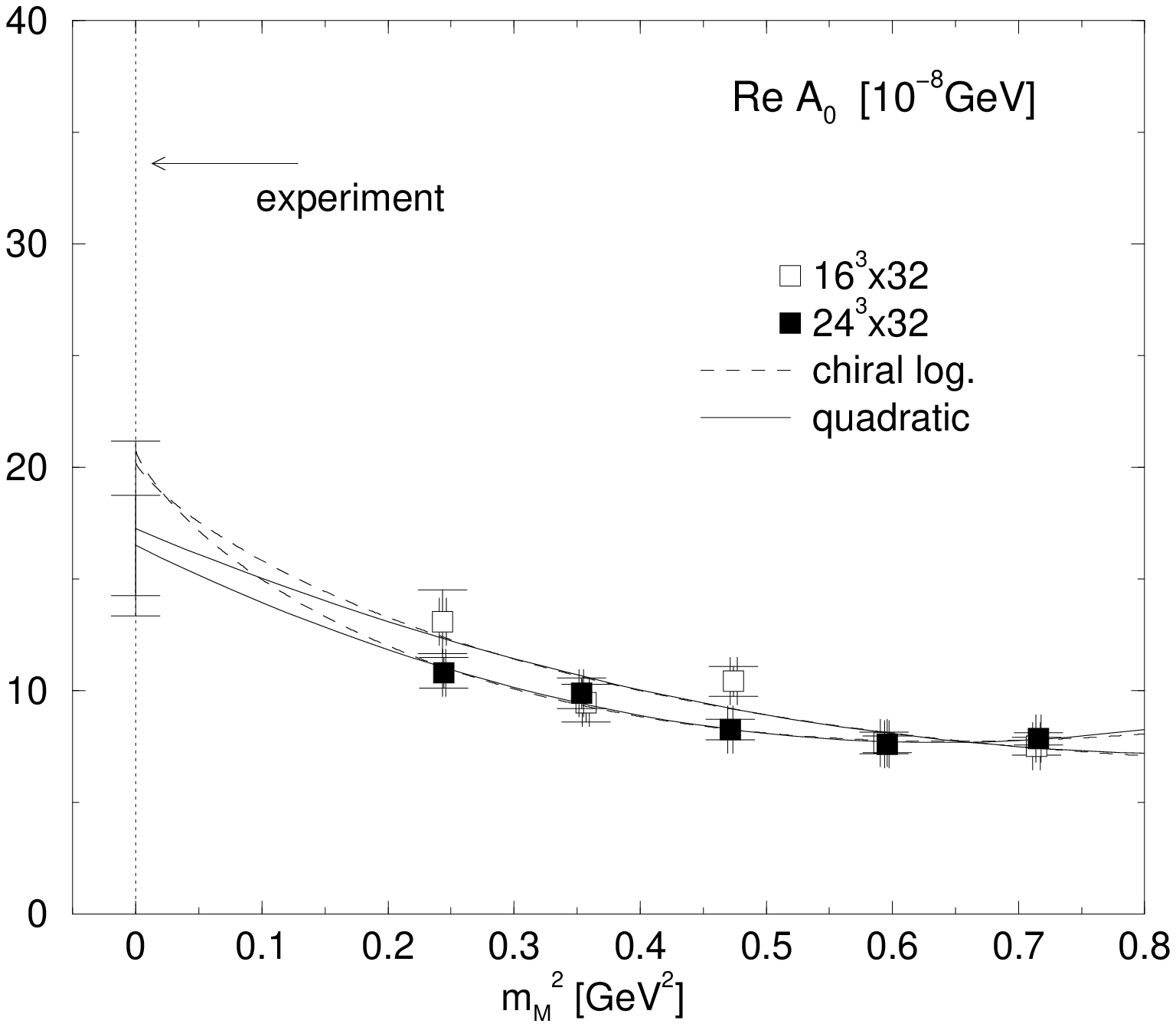}
 \end{center}
\caption{$\re A_2$ (left) and $\re A_0$ (right) in units of GeV as a 
function of $m_M^2$. For chiral extrapolation, quadratic (solid) and 
chiral logarithm (dashed) forms are used.  
For the formar, fit errors are shown in the chiral limit.
Filled and open symbols are for the spatial volume $24^3$ and $16^3$ 
respectively.}
\label{ampI}
\end{figure}

FIG.~\ref{ampI} plots $\re A_2$ (left panel) and $\re A_0$ 
(right panel) as functions of $m_M^2$ for 
$\Lambda_{\ovl{\rm MS}}^{(4)}=325$~MeV.  In both panels, open and filled 
symbols denote the results from the volume $V=16^3$ and $24^3$, respectively.
Signals for $\re A_2$ is quite clean, while those for $\re A_0$ exhibit 
more fluctuations.  Since both amplitudes show a variation with $m_M^2$, 
we need to extrapolate them to the chiral limit to extract the 
physical prediction.  Following the analysis in 
Section~\ref{chi-prop}, we examine two types of fit functions given by
\begin{eqnarray}
\re A_I &=&\left\{
\begin{array}{ll}
 \xi_0+\xi_1m_M^2+\xi_3(m_M^2)^2  &\qquad {\rm (quadratic\ polynomial)}\\
\\
 \xi_0+\xi_1m_M^2+\xi_2m_M^2\ln m_M^2 &\qquad {\rm (chiral\ logarithm)}.
\label{AI-O4}
\end{array}
\right.
\end{eqnarray}
Chiral extrapolations from the quadratic fit are indicated by 
solid lines, and those from the chiral logarithm fit by dashed lines 
in FIG.~\ref{ampI}.

For the $\Delta I=3/2$ amplitude plotted on the left, the extrapolated 
values show good agreement with the experimental value 
$\re A_2=1.50\cdot 10^{-8}$ GeV indicated by the horizontal arrow. 
On the other hand, the $\Delta I=1/2$ amplitude $\re A_0$ is small at 
measured values of quark masses,  
and only amounts to about 50--60\% of the experimental value 
$33.3\cdot 10^{-8}$ GeV even after the chiral extrapolation. 

\begin{figure}
 \begin{center}
  \leavevmode
  \includegraphics[width=7.4cm, clip]{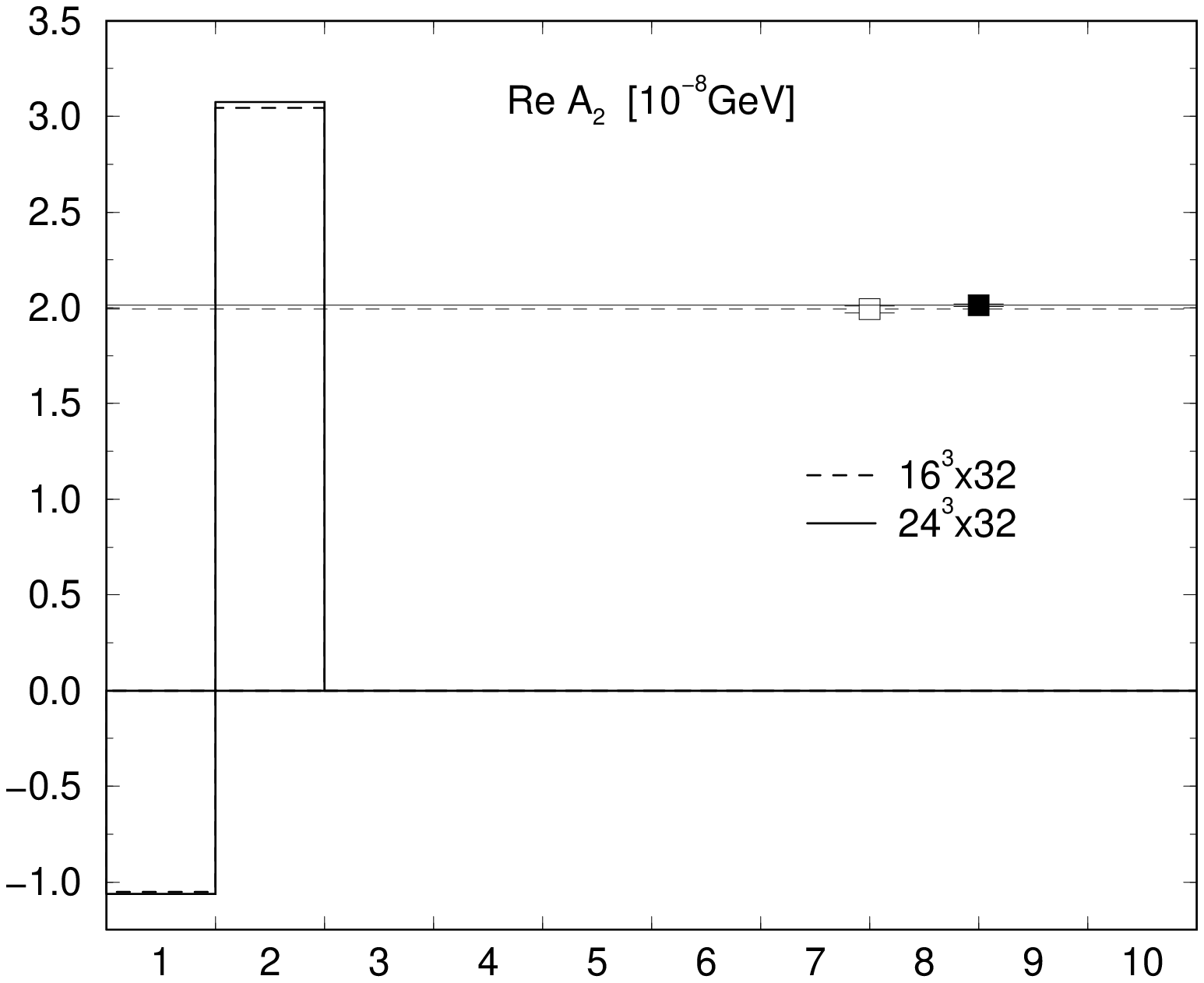}
\hspace{5mm}
  \includegraphics[width=7.5cm, clip]{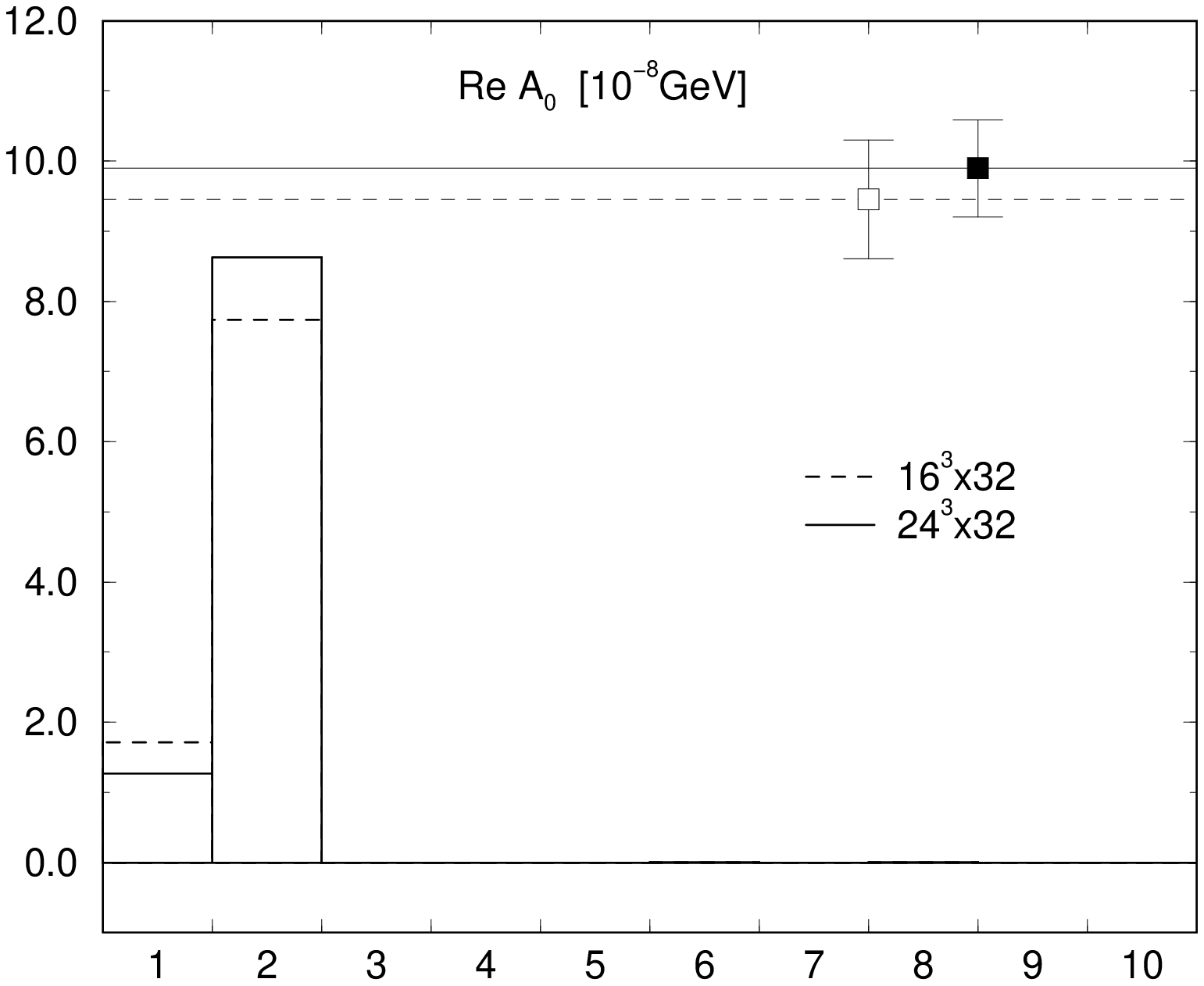}
 \end{center}
 \caption{Breakdown of $\re A_2$ (left) and $\re A_0$ (right) 
 into contributuions from the opertors $Q_i (i=1,\cdots, 10)$ at $m_fa=0.03$. 
Data points placed on horizontal lines show total values and errors.
The solid and dashed lines are 
for the spatial volume $24^3$ and $16^3$ respectively.}
\label{ampI-break}
\end{figure}

A breakdown of the amplitudes into contributions from the ten 
operators $Q_i$ with $i=1,\cdots, 10$ 
is illustrated in FIG.~\ref{ampI-break} for $m_fa=0.03$. 
The histograms for the $V=16^3$ and $24^3$ cases are shown by dashed and 
solid lines, respectively.  
The horizontal lines with statistical errors indicate the total amplitude, 
the dashed and solid lines corresponding to $V=16^3$ and $24^3$.
An apparent absence of contributions from the operators with $i=3, \cdots 10$ 
is due to the small value of the parameter $\re\ \tau\approx 0.002$;  
the real part of the decay amplitudes is determined by the matrix 
elements $\vev{Q_1}_I$ and $\vev{Q_2}_I$, with the latter providing the 
dominant part.

\begin{figure}
 \begin{center}
  \leavevmode
  \includegraphics[width=10cm, clip]{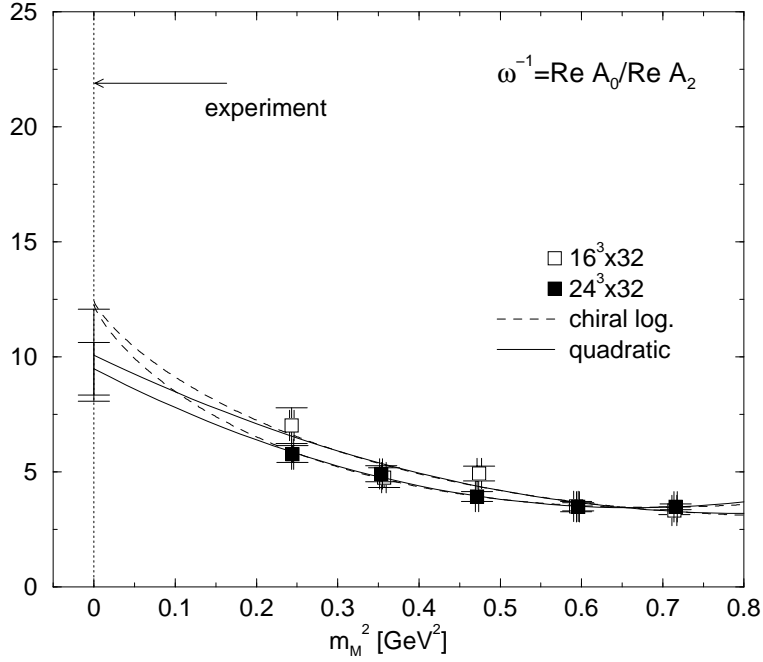}
 \end{center}
 \caption{$\re A_0/\re A_2$ as a function of $m_M^2$. 
For chiral extrapolation, quadratic form (solid line with the fit error
at $m_M^2=0$) and chiral logarithm form (dashed line) are used.
Open and filled symbols are for the spatial volume $16^3$ and $24^3$ 
respectively.}
\label{ominv}
\end{figure}

The ratio $\omega^{-1}=\re A_0/\re A_2$ is shown in FIG.~\ref{ominv}.  
Reflecting 
an insufficient enhancement of the $\Delta I=1/2$ amplitude, it only rises 
to about half of the experimental value $\omega^{-1}\approx 22$. 
The situation hardly changes for $\Lambda_{\ovl{\rm MS}}^{(4)}=215$~MeV 
or 435 MeV,
for which the amplitudes shift by about 5--10\% (see TABLE~\ref{delI-tab}).
We collect chiral fit parameters for the case of larger spatial volume 
$V=24^3$ in TABLE~\ref{fitp-a0a2o24}.

Altogether we find 
\begin{eqnarray}
{\rm Re} A_0 &=&16.5(2.2)(+4.2)(+0.7)\left({}^{+0.8}_{-1.6}\right) 
 \cdot 10^{-8}\ [{\rm GeV}] \\
{\rm Re} A_2 &=&1.531(26)(-178)(-4)\left({}^{+70}_{-38}\right)
\cdot 10^{-8}\ [{\rm GeV}] \\
\omega^{-1} &=& 9.5(1.1)(+2.8)(0.6)\left({}^{+0.7}_{-1.3}\right)
\end{eqnarray}
The central values are taken from the result on a $24^3\times 32$ lattice 
from the quadratic polynomial fit with $\Lambda_{\overline{\rm MS}}^{(4)}=$
325 MeV. 
The first error is statistical, the second one is an estimate of uncertainty 
of chiral extrapolation using the chiral logarithm fit, 
the third one is finite-size variation estimated by the change of value for 
for the $V=16^3$ lattice,
the fourth one, associated with renormalization,
is estimated as the largest variation under changes of 
$\Lambda_{\overline{\rm MS}}^{(4)}$, $q^{*}$ and the RG-running.
If the chiral symmetry breaking term, $\xi_{-1}/m_M^2$, is included in the
chiral fit (\ref{AI-O4}), a non-zero value of $\xi_{-1}$ beyond
the statistical error is obtained only for $\re A_2$, 
resulting in a 60\% increase of the value of $\re A_2$.
The disagreement from experiment becomes worse in this case.
The scaling violation and the quenching error,
which cannot be estimated in our calculation,
are not included in our systematic uncertainty.
In particular, the physical scale of lattice spacing 
set by the string tension in this paper may differ by about 10--20\%
from scales determined by other physical quantities 
due to the quenched approximation.
This uncertainty is not included in the above error estimate.

\subsection{DIRECT CP VIOLATION ($\varepsilon'/\varepsilon$)}

The formula (\ref{epep}) for $\varepsilon'/\varepsilon$ can be rewritten as
\begin{eqnarray}
\varepsilon'/\varepsilon&=&
\im(V_{\rm ts}^*V_{\rm td})\left[P^{(1/2)}-P^{(3/2)}\right],\\
P^{(1/2)}&=& r \sum_i y_i(\mu)\vev{Q_i}_0(\mu)(1-\Omega_{\eta+\eta'}),\\
P^{(3/2)}&=& \frac{r}{\omega}\sum_i y_i(\mu)\vev{Q_i}_2(\mu),
\end{eqnarray}
where
\begin{equation}
r\equiv \frac{G_F\omega}{2|\varepsilon|\re A_0}
\end{equation}
and the parameter $\Omega_{\eta+\eta'}= 0.25(5)$ reflects the 
isospin breaking. 
Since the $\Delta I=1/2$ rule is only partially reproduced with our data, 
we employ the experimental values for $\re A_0$, $\omega$ and $\varepsilon$ 
as input. 

\begin{figure}
 \begin{center}
  \leavevmode
  \includegraphics[width=8cm, clip]{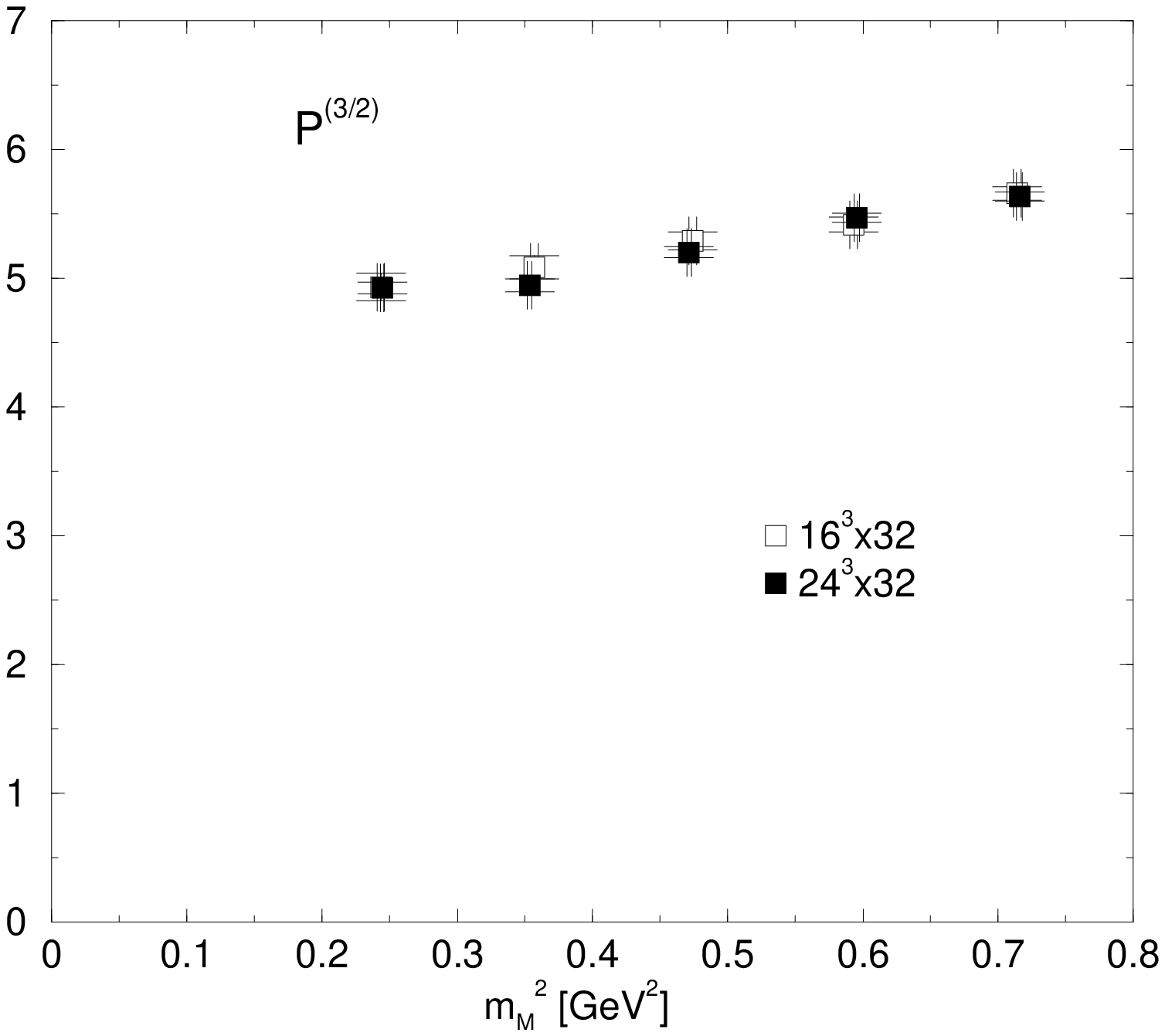}
\hspace{0.2cm}
  \includegraphics[width=8cm, clip]{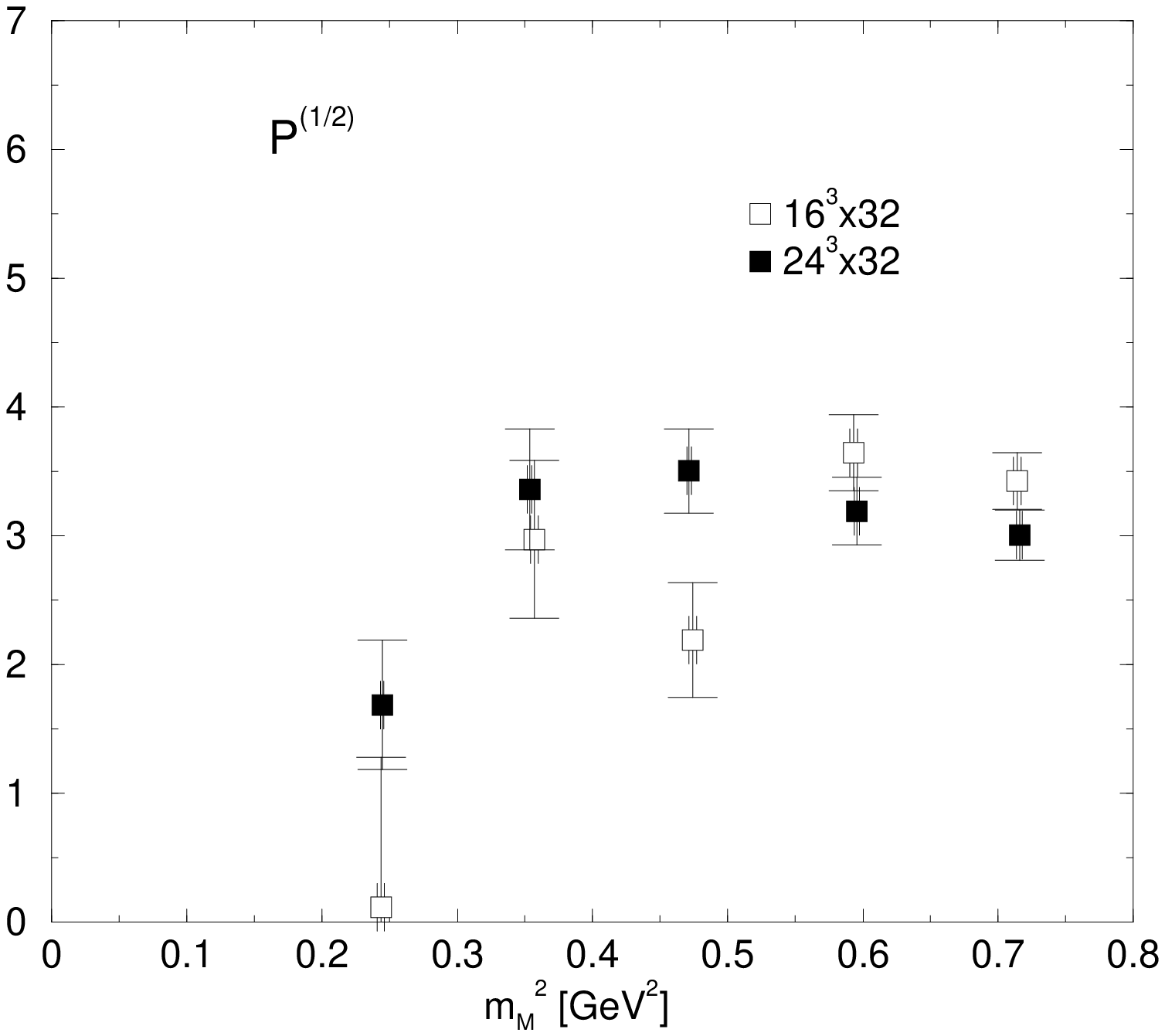}
 \end{center}
 \caption{
$P^{(3/2)}$ (left) and $P^{(1/2)}$ (right) as a function of $m_M^2$. 
Open and filled symbols are for the spatial volume $16^3$ and $24^3$ 
respectively.}
\label{p13hf}
\end{figure}

\begin{figure}
 \begin{center}
  \includegraphics[width=10cm, clip]{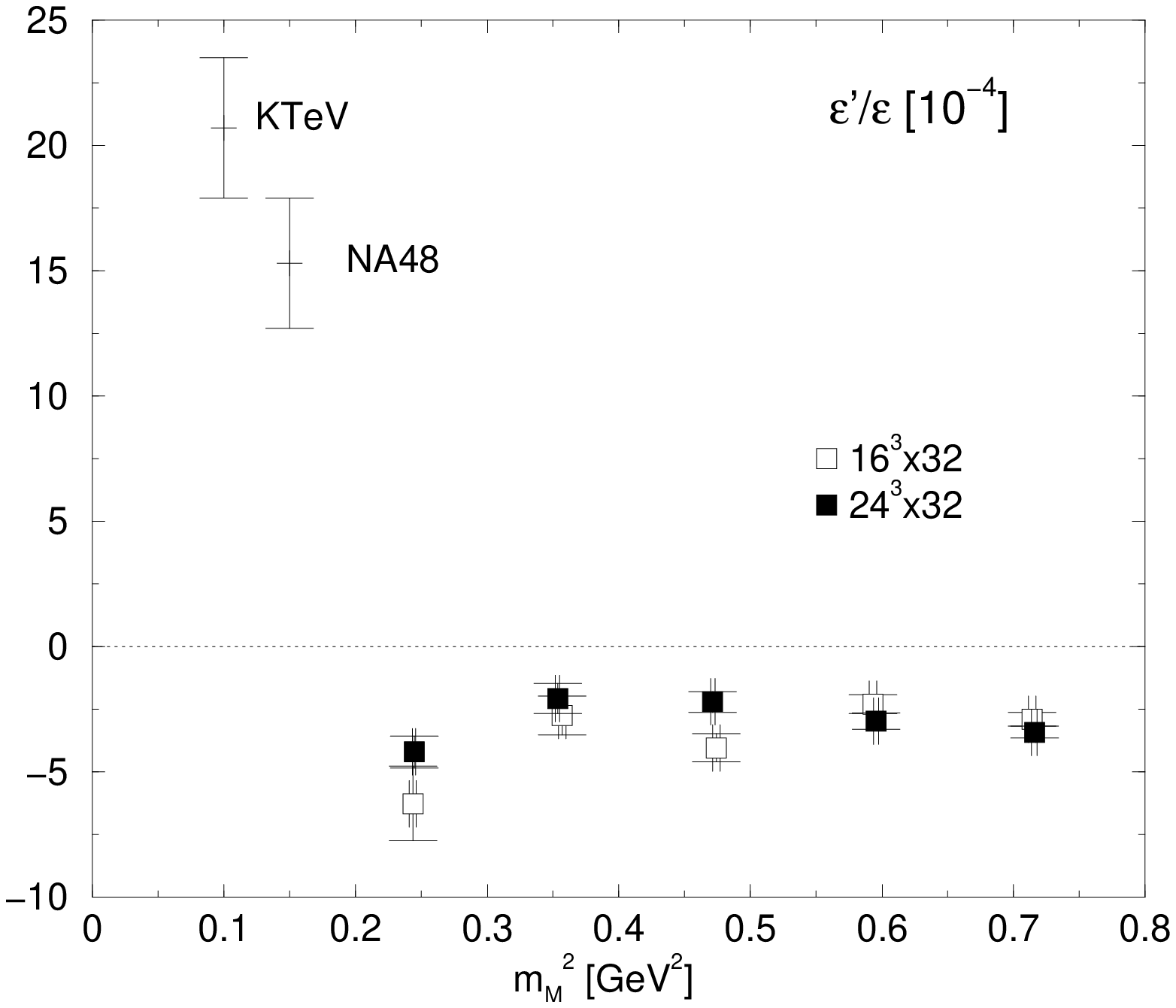}
 \end{center}
 \caption{$\varepsilon'/\varepsilon$ as a function of $m_M^2$.
Open and filled symbols are for the spatial volume $16^3$ and $24^3$ 
respectively. Experimental values quoted in (\ref{epep}) are also shown.}
\label{epep-fig}
\end{figure}

In FIG.~\ref{p13hf} our data for $P^{(3/2)}$ (left panel) and $P^{(1/2)}$ 
(right panel) calculated with $\Lambda_{\ovl{\rm MS}}^{(4)}=325$~MeV 
are plotted as a function of $m_M^2$.  Results for 
$\varepsilon'/\varepsilon$ are shown in FIG.~\ref{epep-fig}. 
Since $P^{(1/2)}$ is smaller than $P^{(3/2)}$ in our data, 
$\varepsilon'/\varepsilon$ tends to be negative.

\begin{figure}
 \begin{center}
  \leavevmode
  \includegraphics[width=4.4cm, clip]{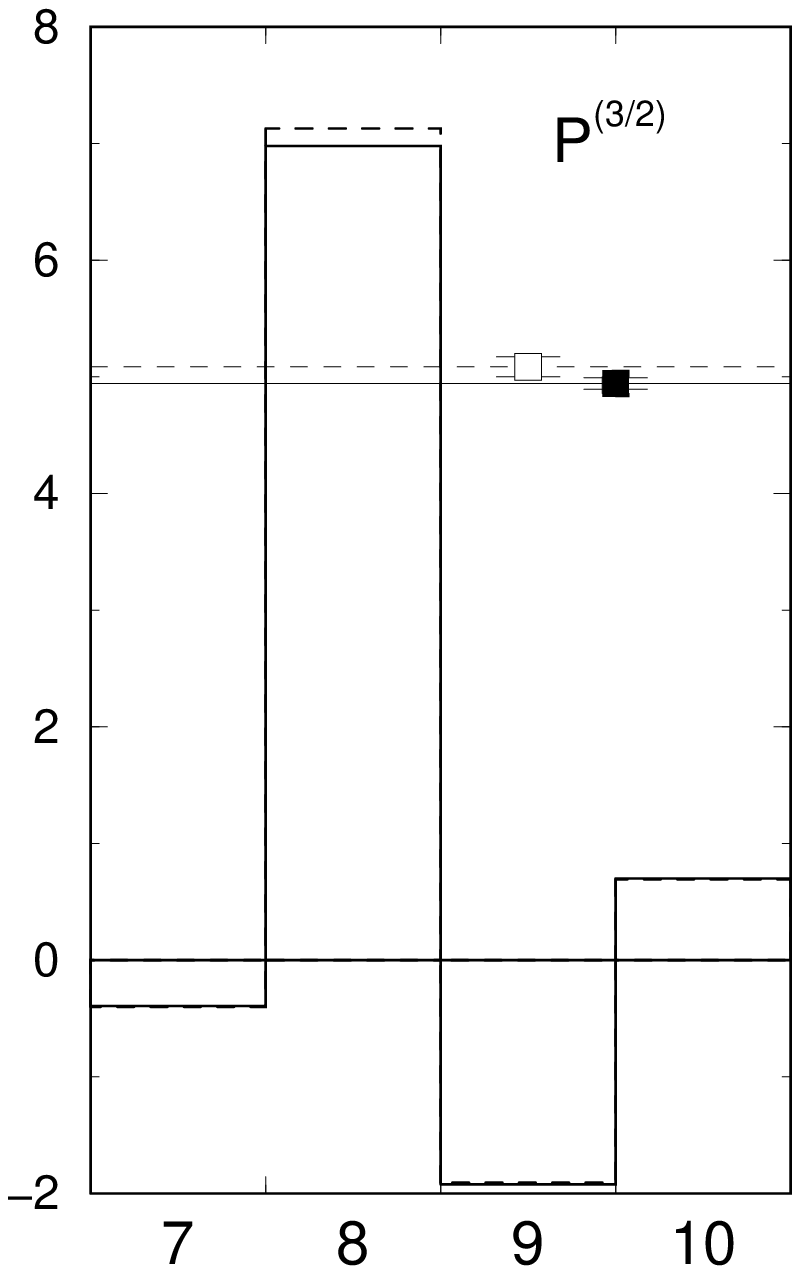}
\hspace{5mm}
  \includegraphics[width=8.1cm, clip]{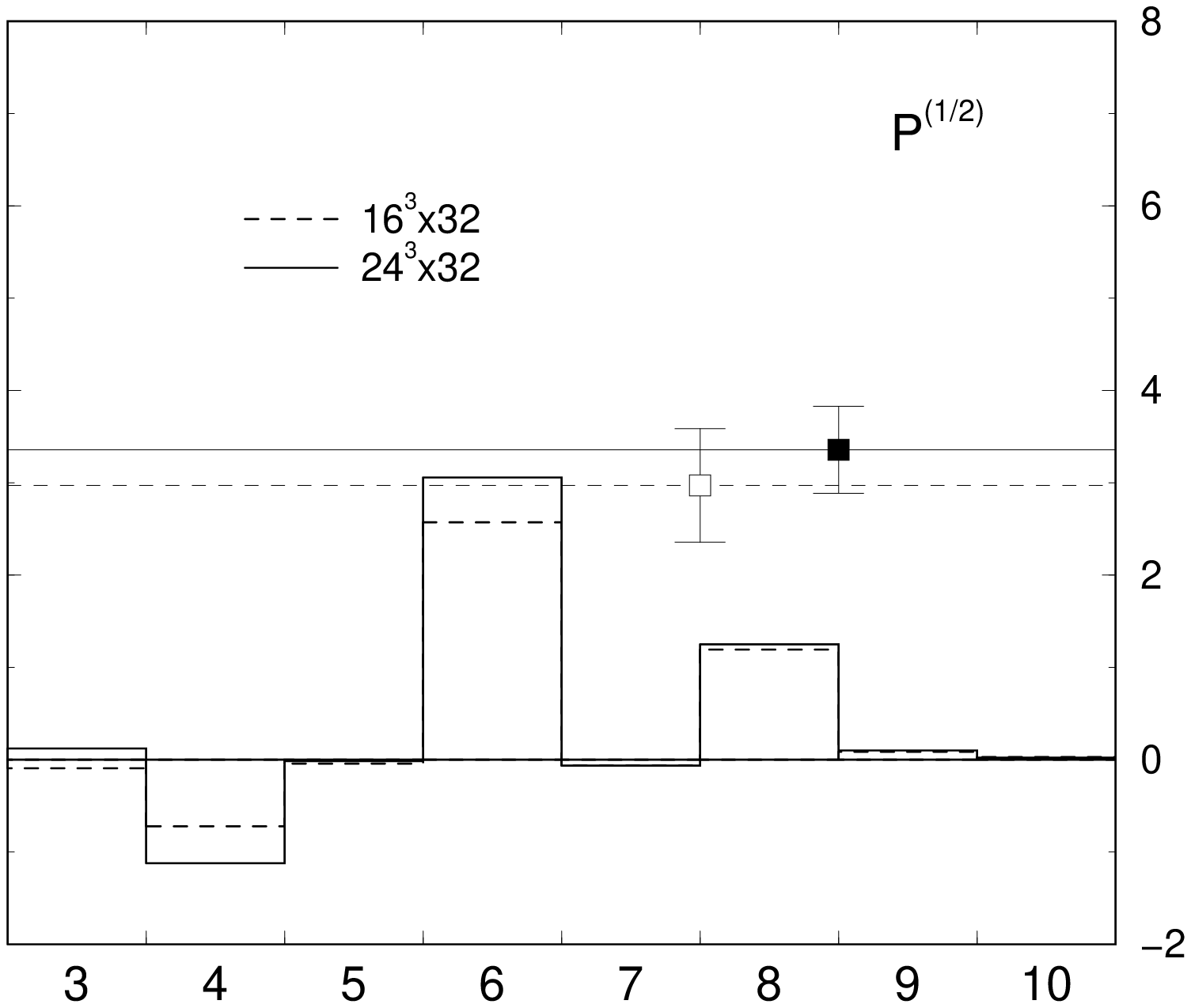}
 \end{center}
 \caption{Breakdown of $P^{(3/2)}$ (left) and $P^{(1/2)}$ (right) 
 into contributuions from the opertors $Q_i (i=3,\cdots, 10)$  at $m_fa=0.03$. 
Data points placed on horizontal lines show total values and errors.
The solid and dashed lines are 
for the spatial volume $24^3$ and $16^3$ respectively.}
\label{p13hf-break}
\end{figure}

A breakdown of $P^{(3/2)}$ and $P^{(1/2)}$ into contributions 
from the operators $Q_i$ ($i=3, \cdots, 10$) is displayed for the case of 
$m_fa=0.03$ in FIG.~\ref{p13hf-break}, where dashed and 
solid lines denote data from $V=16^3$ and $24^3$, respectively. 
This figure demonstrates that $\vev{Q_8}_2$ and $\vev{Q_6}_0$ are 
respectively dominant in $P^{(3/2)}$ and $P^{(1/2)}$ as usually considered. 
However, the matrix element of $\vev{Q_6}_0$ is too small; 
if the experimental value of $\varepsilon'/\varepsilon$ is to be 
reproduced by a change of this matrix element, it has to be increased by 
about a factor of five.   

Numerical values of $P^{(1/2)}, P^{(3/2)}$ and $\varepsilon'/\varepsilon$ 
for each $m_f$ are summarized in TABLE~\ref{epep-tab}.  In addition to the 
features of data discussed above, we observe that changing the $\Lambda$ 
parameter from $\Lambda_{\ovl{\rm MS}}^{(4)}=325$~MeV to 215~MeV   
decreases $P^{(1/2)}$ by 20\% and $P^{(3/2)}$ by 25\%. 
Employing $\Lambda_{\ovl{\rm MS}}^{(4)}=435$~MeV leads to an increase 
by similar percentages for the two functions.  
Therefore, the trend toward a negative value of $\varepsilon'/\varepsilon$ 
is not altered. 

If we make a quadratic chiral extrapolation we find
$\varepsilon'/\varepsilon=-7.7(2.0)\times 10^{-4}$ with 
$\chi^2/\mbox{dof}=1.75$ on a $24^3\times 32$ lattice.  
Including the chiral symmetry breaking term $\xi_{-1}/m_M^2$ in the fit 
changes this value to $+30(20)\times 10^{-4}$ with 
$\chi^2/\mbox{dof}=0.0015$.  
The small $\chi^2$ indicates that more data points, in particular data
at smaller masses, are necessary to constraint the fit parameters well.
The existence of large uncertainties associated with the possible presence 
of the chiral breaking term, and also a subtle quenching effect 
mentioned below, make it difficult to draw a conclusive estimate of 
$\varepsilon'/\varepsilon$. 

Recently, Golterman and Pallante pointed out that the relation between
$K\to \pi$ and $K\to\pi\pi$ matrix elements in chiral perturbation theory
should be modified in the quenched theory~\cite{Golterman:2001}.
We have applied the modified relation to the $Q_{5,6}^{(0)}$ matrix elements, 
and found that the effect is large, ranging between 20\% to 100\% in 
magnitude. For example,
the renormalized $\vev{Q_6}_0 $  on a $24^3\times 32$ lattice increases 
in magnitude to 
$-$0.154(17), $-$0.182(16), $-$0.144(11),$-$0.1238(90), $-$0.0969(72)
at $m_f=0.02$, 0.03, 0.04, 0.05, 0.06, respectively.
(This modification has been tested also in the case of the staggered
fermion~\cite{KS2}, and an increase of $\vev{Q_6}_0 $ of a similar 
magnitude has been observed.)
In terms of $\varepsilon'/\varepsilon$, the modified relation leads to 
$-$1.70(53), $-$0.53(51), $-$1.48(32), $-$2.09(26), $-$2.85(19)
for $\Lambda_{\overline{\rm MS}}^{(4)} =$ 325 MeV.
The modification increases the value of $\varepsilon'/\varepsilon$, 
but it is still negative.
A complete analysis still remains to be made both in the theoretical 
analyses of the relation in quenched chiral perturbation theory and 
in numerical simulations.

\section{CONCLUSIONS}\label{conclusions}

In this article we have presented results of our investigation into 
the reduction method in the framework of chiral perturbation theory 
at the lowest order to calculate the $K\to\pi\pi$ decay amplitudes. 
The $K\to\pi$ and $K\to 0$ hadronic matrix elements of four-quark 
operators were calculated in a quenched numerical 
simulation using domain-wall fermion action for quarks and an RG-improved 
gauge action for gluons to satisfy the requirements of chiral symmetry 
on the lattice.  We have seen that the calculation of quark loop contractions 
which appear in Penguin diagrams by the random noise method works 
successfully.  As a result the $\Delta I=1/2$ amplitudes which require 
subtractions with the quark loop contractions were obtained with a 
statistical accuracy of about 10\%. 
We have investigated the chiral properties required for the $K\to\pi$ matrix 
elements. If we leave aside $Q_6^{(0)}$, 
we have found no strong sign for the existence of the chiral symmetry 
breaking effect within the statistical precision of our data
in the range of quark masses employed in our simulations.
However, $Q_6^{(0)}$ appears to show an exceptionally large chiral 
symmetry breaking effect compared to other channels.  It is not clear to 
us if this is an effect beyond statistical fluctuation. 
For the definite conclusion on this point, 
more data, particularly at smaller quark masses, will be needed.
Matching the lattice matrix elements to those in the continuum 
at $\mu=1/a$ with the perturbative renormalization factor to one loop 
order, and running to the scale $\mu=m_c=1.3$ GeV with the renormalization 
group, we obtained all the matrix elements needed for the decay amplitudes.
Unfortunately the physical amplitudes thus calculated show unsatisfactory 
features. 

One of the pathologies of our results is a poor 
enhancement of the $\Delta I=1/2$ decay amplitude;  
the value of $\re A_0$ is about 50--60\% of the experimental one 
in contrast to $\re A_2$ which reaches the expected value in the chiral limit.
Another deficiency is a small value of the $\Delta I=1/2$ contribution 
to $\varepsilon'/\varepsilon$; if we assume that the $\Delta I=3/2$
contribution has a correct order of magnitude, the $\Delta I=1/2$ contribution
is too small by about a factor of five to explain the experimental 
value $\simeq 2\cdot 10^{-3}$.

The hadronic matrix elements for $\Delta I=1/2$ involve 
significant subtractions.  For some of the matrix elements, this 
results in flips of sign and a reduction in the magnitude.  
Hence insufficient choices of lattice parameters in simulations may 
lead to sizable systematic errors in these matrix elements.  Possible 
origins of the errors are (i) finite fifth dimensional  size $N_5$ of
the domain wall fermion, (ii) finite spatial size $N_s$, (iii) finite lattice 
spacing $a$, (iv) quenching effects, 
and (v) the neglect of the charm quark.
Our use of 
(vi) renormalization factors in one-loop order of perturbation theory is 
another source of error in the renormalized matrix elements. 
Finally (vii) higher order corrections in chiral 
perturbation theory is also a possible source of error.  
It may well be 
that the origin of the deficiency resides in physical phenomena 
such as the effect of $\sigma$ resonance which are difficult to take into
account once the reduction to $K\to\pi$ matrix elements is made. 

\acknowledgements

We thank P. Weisz for informative discussions on the anomalous 
dimension of the four-quark operators. 
JN, SA, YA, SH and TI also thank the Institute for Nuclear Theory
at the University of Washington for its hospitality and the Department
of Energy for partial support during the completion of this work.
This work was supported in part by Grants-in-Aid of the Ministry of
Education (Nos. 10640246, 10640248, 10740107, 11640250, 11640294,
11740162, 12014202, 12304011, 12640253, 12740133, 13640260). VL is
supported by the JSPS Research for the Future Program (No. JSPS-RFTF
97P01102). JN, SE, and KN are JSPS Research Fellows.

\clearpage

\appendix

\section{DECOMPOSITION OF $Q_i$'S INTO $\Delta I=1/2$ AND $\Delta 
I=3/2$ PARTS}\label{app_split}

Four-quark operators which transform under the irreducible representations 
of $SU(3)_L\otimes SU(3)_R$ chiral group and having definite isospin 
$I=0$ or 2 are given by  
\begin{eqnarray}
{\cal X}_{\bf 27,1}^{(2)}&=& (\bar{s}d)_L
\left[(\bar{u}u)_L-(\bar{d}d)_L\right]+(\bar{s}u)_L(\bar{u}d)_L ,\\
{\cal X}_{\bf 27,1}^{(0)}&=& (\bar{s}d)_L\left[
(\bar{u}u)_L+2(\bar{d}d)_L-3(\bar{s}s)_L\right]+(\bar{s}u)_L(\bar{u}d)_L ,\\
{\cal X}_{\bf 8,1}^{(0)}&=& (\bar{s}d)_L(\bar{u}u)_L-
(\bar{s}u)_L(\bar{u}d)_L ,\\
\tilde{\cal X}_{\bf 8,1}^{(0)}&=&(\bar{s}d)_L\left[(\bar{u}u)_L
+2(\bar{d}d)_L+2(\bar{s}s)_L \right]+(\bar{s}u)_L(\bar{u}d)_L ,\\
{\cal Y}_{\bf 8,1}^{(0)}&=&(\bar{s}d)_L\left[
(\bar{u}u)_R+(\bar{d}d)_R+(\bar{s}s)_R\right],\ {\cal Y}_{\bf 8,1}^{(0)\ c} ,\\
{\cal Y}_{\bf 8,8}^{(0)}&=&(\bar{s}d)_L\left[
(\bar{u}u)_R-(\bar{s}s)_R\right]-(\bar{s}u)_L(\bar{u}d)_R, \ \ 
{\cal Y}_{\bf 8,8}^{(0)\ c} ,\\
{\cal Y}_{\bf 8,8}^{(2)}&=&(\bar{s}d)_L\left[
(\bar{u}u)_R-(\bar{d}d)_R\right]+(\bar{s}u)_L(\bar{u}d)_R, \ \ 
{\cal Y}_{\bf 8,8}^{(2)\ c},
\end{eqnarray}
where we use the notation of ${\cal X}$'s and ${\cal Y}$'s
for the Lorentz structure $L\otimes L$ and $L\otimes R$. 
The subscripts ``${\bf i,j}$'' stand for the representation
$(i_L,j_R)$ of the operator and the superscript $(0)$ or $(2)$ denotes
the isospin. A shorthand notation, {\it e.g.}, 
$(\bar{s}d)_L=\bar{s}\gamma_\mu(1-\gamma_5)d$, is employed as in eqs. 
(\ref{CPSop-x1})--(\ref{CPSop-y2}), and  
${\cal Y}_{\bf i,j}^{(I)\ c}$ equals ${\cal Y}_{\bf i,j}^{(I)}$ 
with its color summation changed to cross the two currents.
In terms of these operators the independent local operators are 
rewritten as 
\begin{eqnarray}
Q_1&=&\frac{1}{2}{\cal X}_{\bf 8,1}^{(0)}+\frac{1}{10}
\tilde{\cal X}_{\bf 8,1}^{(0)}+\frac{1}{15}{\cal X}_{\bf 27,1}^{(0)}+
\frac{1}{3}{\cal X}_{\bf 27,1}^{(2)} ,\\
Q_2&=&-\frac{1}{2}{\cal X}_{\bf 8,1}^{(0)}+\frac{1}{10}
\tilde{\cal X}_{\bf 8,1}^{(0)}+\frac{1}{15}{\cal X}_{\bf 27,1}^{(0)}+
\frac{1}{3}{\cal X}_{\bf 27,1}^{(2)} ,\\
Q_3&=&\frac{1}{2}{\cal X}_{\bf 8,1}^{(0)}+\frac{1}{10}
\tilde{\cal X}_{\bf 8,1}^{(0)} ,\\
Q_5&=&{\cal Y}_{\bf 8,1}^{(0)} ,\\ 
Q_6&=&{\cal Y}_{\bf 8,1}^{(0)\ c} ,\\
Q_7&=&\frac{1}{2}\left[{\cal Y}_{\bf 8,8}^{(0)}+
{\cal Y}_{\bf 8,8}^{(2)}\right] ,\\
Q_8&=&\frac{1}{2}\left[{\cal Y}_{\bf 8,8}^{(0)\ c}+
{\cal Y}_{\bf 8,8}^{(2)\ c}\right].
\end{eqnarray}
Therefore, the decomposition of the local operators into $\Delta I=1/2$ 
and $\Delta I=3/2$ parts is summarized as follows:\\

\noindent
$\Delta I=1/2$: 
\begin{eqnarray}
 Q_1^{(0)}&=&\frac{1}{3}[-(\bar{s}_ad_b)_L(\bar{u}_b{u}_a)_L
 +2(\bar{s}_au_b)_L(\bar{u}_bd_a)_L+(\bar{s}_ad_b)_L
(\bar{d}_bd_a)_L]\label{Q0-1} ,\\
 Q_2^{(0)}&=&\frac{1}{3}[-(\bar{s}d)_L(\bar{u}{u})_L
  +2(\bar{s}u)_L(\bar{u}{d})_L+(\bar{s}d)_L(\bar{d}{d})_L]
\label{Q0-2} ,\\
 Q_3^{(0)}&=&(\bar{s}d)_L[(\bar{u}u)_L+(\bar{d}d)_L
  +(\bar{s}{s})_L] \label{Q0-3} ,\\
 Q_4^{(0)}&=&(\bar{s}_ad_b)_L[(\bar{u}_bu_a)_L+(\bar{d}_ad_b)_L
  +(\bar{s}_bs_a)_L] \label{Q0-4} ,\\
 Q_5^{(0)}&=&(\bar{s}d)_L[(\bar{u}u)_R+(\bar{d}d)_R
  +(\bar{s}{s})_R] \label{Q0-5} ,\\
 Q_6^{(0)}&=&(\bar{s}_ad_b)_L[(\bar{u}_bu_a)_R+(\bar{d}_ad_b)_R
  +(\bar{s}_b{s}_a)_R] \label{Q0-6} ,\\
 Q_7^{(0)}&=&\frac{1}{2}[(\bar{s}d)_L(\bar{u}{u})_R
  -(\bar{s}u)_L(\bar{u}{d})_R-(\bar{s}d)_L(\bar{s}{s})_R] 
\label{Q0-7} ,\\
 Q_8^{(0)}&=&\frac{1}{2}[(\bar{s}_ad_b)_L(\bar{u}_b{u}_a)_R
  -(\bar{s}_au_b)_L(\bar{u}_bd_a)_R-(\bar{s}_ad_b)_L
(\bar{s}_bs_a)_R] 
\label{Q0-8} ,\\
 Q_9^{(0)}&=&\frac{1}{2}[(\bar{s}d)_L(\bar{u}{u})_L
  -(\bar{s}u)_L(\bar{u}{d})_L-(\bar{s}d)_L(\bar{s}{s})_L] 
\label{Q0-9} ,\\
 Q_{10}^{(0)}&=&\frac{1}{2}[(\bar{s}_ad_b)_L(\bar{u}_bu_a)_L
  -(\bar{s}_au_b)_L(\bar{u}_bd_a)_L-(\bar{s}_ad_b)_L
(\bar{s}_bs_a)_L] , \label{Q0-10}
\end{eqnarray}
\vspace{2mm}

$\Delta I=3/2$: 
\begin{eqnarray}
 Q_1^{(2)}&=& Q_2^{(2)}=\frac{1}{3}[(\bar{s}d)_L(\bar{u}{u})_L
  +(\bar{s}u)_L(\bar{u}{d})_L-(\bar{s}d)_L(\bar{d}{d})_L]
\label{Q2-2} ,\\
 Q_3^{(2)}&=& Q_4^{(2)}= Q_5^{(2)}= Q_6^{(2)}=0 ,\\
 Q_7^{(2)}&=&\frac{1}{2}[(\bar{s}d)_L(\bar{u}{u})_R
  +(\bar{s}u)_L(\bar{u}{d})_R-(\bar{s}d)_L(\bar{d}{d})_R] 
\label{Q2-7} ,\\
 Q_8^{(2)}&=&\frac{1}{2}[(\bar{s}_ad_b)_L(\bar{u}_bu_a)_R
  +(\bar{s}_au_b)_L(\bar{u}_bd_a)_R-(\bar{s}_ad_b)_L
(\bar{d}_bd_a)_R] \label{Q2-8} ,\\
 Q_9^{(2)}&=& Q_{10}^{(2)}=\frac{3}{2}Q_1^{(2)}\label{Q2-9},
\end{eqnarray}
where color indices are understood within each current 
in the operators with two color traces.
The equivalence between $Q^{(2)}_1$ and $Q^{(2)}_2$ is valid due to
Fierz rearrangement, hence $Q^{(2)}_9=Q^{(2)}_{10}$ follows.

\section{EXPERIMENTAL INPUT PARAMETERS}\label{appendix-param}
\setcounter{equation}{0}

We collect the input parameters which were used in our numerical
calculation~\cite{PDG,Bosch-etal99}. 
\begin{eqnarray}
{\rm Quark\ mass}:& &m_u= 5 {\rm\ MeV}\ , \ \ m_d= 8 {\rm\ MeV},\\
& &m_s = 120 {\rm\ MeV}\ , \ \ m_c= 1.3 {\rm\ GeV},\\
& &m_b = 4.2 {\rm\ GeV}\ , \ \ m_t= 170 {\rm\ GeV}\\
{\rm Meson\ mass:}& &m_\pi=139.6 {\rm\ MeV}\ ,\ \ m_K=497.7 {\rm\ MeV}\\
{\rm Decay\ constant:}& &f_\pi=92.4 {\rm\ MeV}\ ,\ \ f_K =113.1 {\rm\ MeV} \\
{\rm Coupling\ constant:}& &\alpha\equiv e^2/(4\pi)= 1/129 \ \
({\rm at\ }\mu=m_W),\\
& & G_F\equiv\frac{\sqrt{2}g_2{}^2}{8m_W^2}=1.166\cdot10^{-5}\ 
{\rm GeV}^{-2}\\
& & \ \ ( m_W = 80.2\ {\rm GeV} )\nonumber \\
{\rm Quantities\ relevant\ \ \ \ }& & \nonumber \\
{\rm to\ Kaon\ decays:}& & \re A_0 = 33.3\cdot 10^{-8}\ {\rm GeV}\\
& & \re A_2 = 1.50\cdot 10^{-8}\ {\rm GeV}\\
& & |\omega|= 0.045\\
& & \Omega_{\eta+\eta'}=0.25 \\
& & |\varepsilon|= 2.280\cdot 10^{-3}\\
 {\rm CKM\ elements}:
& & |V_{\rm us}|=0.22\ ,\ \ |V_{\rm ud}|=0.974\\
& &\im (V_{\rm ts}^*V_{\rm td})=1.3\cdot 10^{-4}\\
& &\re\ \!\tau=
-\re\left(\frac{V^*_{\rm ts}V_{\rm td}}{V^*_{\rm us}V_{\rm ud}}\right)=0.002.
\end{eqnarray}

\section{RENORMALIZATION FACTORS AND RG-EVOLUTION MATRIX}\label{appendix-ZU}

In this appendix, we summarize the renormalization factors and 
the RG-evolution matrix, 
and calculate their numerical values for our choice of parameters.
Throughout this paper, we employ the perturbative calculation
in $\overline{\rm MS}$ scheme with NDR.

The renormalization formula has the form,
\begin{eqnarray}
\vev{Q_i}^{\ovl{\rm MS}}(q^*)={\cal Z}^g_{ij}(q^*a)\vev{Q_j^{\rm latt}}(1/a)
+{\cal Z}^{\rm pen}_i(q^*a)\vev{Q_{\rm pen}^{\rm latt}}(1/a),
\end{eqnarray}
where,
\begin{eqnarray}
 Q_{\rm pen}^{\rm latt}\equiv Q_4+Q_6- \frac{(Q_3+Q_5)}{N_c}, 
\ \ (N_c=3:\ {\rm \#color}) 
\end{eqnarray}
is the sum of contributions from penguin operators. 
Since our matrix elements are obtained in the form of propagator ratios, 
${\cal Z}^g$ and ${\cal Z}^{\rm pen}$ are also ratios of 
the renormalization factors $Z^g_{ij}$ and $Z_i^{\rm pen}$ 
calculated from corresponding vertex functions and that of the local axial 
current $Z_A$~\cite{AIKT98}:
\begin{eqnarray}
{\cal Z}^g_{ij}= \frac{Z^g_{ij}}{Z_A^2}\ ,\ \
{\cal Z}^{\rm pen}_i= \frac{Z^{\rm pen}_i}{Z_A^2}. \label{Z-redef}
\end{eqnarray}

The diagonal parts $Z^g_{ii}$ are given by  
\begin{eqnarray}
Z^g_{ii}=\left\{
\begin{array}{cc}
1+\dfrac{g^2}{16\pi^2}\left[\dfrac{3}{N_c}\ln (q^* a)^2+\dfrac{z_++z_-}{2} 
\right] ,& i=1,2,3,4,9,10, \\
1+\dfrac{g^2}{16\pi^2}\left[-\dfrac{3}{N_c}\ln (q^* a)^2 +z_1-v_{21} 
\right] ,& i=5,7, \\
1+\dfrac{g^2}{16\pi^2}\left[\dfrac{3(N_c^2-1)}{N_c}\ln (q^* a)^2+ z_2+v_{21} 
\right] ,& i=6,8,
\end{array}
\right.
\end{eqnarray}
while for off-diagonal parts, one has
\begin{eqnarray}
Z^g_{ij}=\left\{
\begin{array}{cc}
\dfrac{g^2}{16\pi^2}\left[-\dfrac{3}{N_c}\ln (q^* a)^2
+\dfrac{z_+-z_-}{2}\right] ,& (i,j)=(1,2),(2,1),(3,4),\\ 
&\hspace*{1.5cm}(4,3),(9,10),(10,9),\\
\dfrac{g^2}{16\pi^2}\left[3\ln (q^* a)^2 +\dfrac{z_2-z_1+v_{21}-v_{12}}{N_c} 
\right] ,& i=(5,6),(7,8), \\
-\dfrac{g^2}{16\pi^2}N_cv_{21} ,& i=(6,5),(8,7),\\
0 & {\rm others}
\end{array}
\right.
\end{eqnarray}
Similarly the contributions from the penguin operators~\cite{AK00} are 
given by
\begin{eqnarray}
Z^{\rm pen}_i=\frac{g^2}{16\pi^2}\frac{C_i}{3}[-\ln (q^* a)^2
+z_i^{\rm pen}],
\end{eqnarray}
where $C_2=1,C_3=2,C_4=C_6=N_f,C_8=C_{10}=N_u-N_d/2,C_9=-1$ and
$C_i=0$ for other $i$ with $N_f,N_u,N_d$ being the number of flavors, 
up-like quarks and down-like quarks in $Q_i$'s,
and $z_i^{\rm pen}$ are constants. 
In our calculation, we should set $N_f=3, N_u=1, N_d=2$. 
Finally the axial vector renormalization constant has the form, 
\begin{eqnarray}
Z_A=1+\frac{g^2}{12\pi^2}z_A\ , 
\end{eqnarray}

In the above $z_\pm, z_1,z_2,v_{12},v_{21}$ and $z_A$ are constants 
depending on the choices of simulation parameters and renormalization scheme.
With the use of mean field improvement at one-loop level, we
obtain the following values~\cite{Taniguchi01} at $\beta=2.6$ and $M=1.8$ 
for the RG-improved gauge action:
\begin{eqnarray}
& &g^2\equiv g^2_{\ovl{\rm MS}}(1/a)= 2.273 \label{eq:gg}\\
& &\tilde{M}=1.41979\\
& &z_A=-4.6930\\
& &z_+=-13.612\ ,\ \ z_-=-10.319\\
& &z_1=-10.063\ ,\ \ z_2=-16.125\\
& &v_{12}=8\ ,\ \ v_{21}=1\\
& &z_i^{\rm pen}=\left\{
\begin{array}{l}
    4.494\ ({\rm for}\ i=2,3,5,7,9)\\
    3.494\ ({\rm for}\ i=4,6,8,10)
\end{array}
\right.
\end{eqnarray}
From the definition of $Q^{\rm pen}$, $Z^{\rm pen}_i$ can be written 
in the form of a $10\times 10$ matrix $\hat{Z}^{\rm pen}$, defined as
$\hat{Z}^{\rm pen}_{i3}=\hat{Z}^{\rm pen}_{i5}=-z^{\rm pen}_i/N_c,
\hat{Z}^{\rm pen}_{i4}=\hat{Z}^{\rm pen}_{i6}=z^{\rm pen}_i,$ 
and $\hat{Z}^{\rm pen}_{ij}=0$ for other $j$. 
The renormalization factor can then be summarized as a $10\times 10$
matrix given by 
\begin{eqnarray}
& &{\cal Z}^g+\hat{\cal Z}^{\rm pen}=\nn \\
& &\left[
\begin{array}{llllllllll}
\ \ 0.9997 &\! -0.0350&\ \ 0     &\ \ 0      &\ \ 0      &\ \ 0      &\ \ 0     &\ \ 0     &\ \ 0     &\ \ 0    \\
\! -0.0350 &\ \ 0.9997&\!-0.0106 &\ \ 0.0318 & \!-0.0106 &\ \ 0.0318 &\ \ 0     &\ \ 0     &\ \ 0     &\ \ 0    \\
\ \ 0      &\ \ 0     &\ \ 0.9785&\ \ 0.0287 & \!-0.0212 &\ \ 0.0636 &\ \ 0     &\ \ 0     &\ \ 0     &\ \ 0    \\
\ \ 0      &\ \ 0     &\!-0.0597 &\ \ 1.0739 & \!-0.0247 &\ \ 0.0742 &\ \ 0     &\ \ 0     &\ \ 0     &\ \ 0    \\
\ \ 0      &\ \ 0     &\ \ 0     &\ \ 0      &\ \ 1.0154 &\!-0.0924  &\ \ 0     &\ \ 0     &\ \ 0     &\ \ 0    \\
\ \ 0      &\ \ 0     &\!-0.0247 &\ \ 0.0742 & \!-0.0884 &\ \ 1.0190 &\ \ 0     &\ \ 0     &\ \ 0     &\ \ 0    \\
\ \ 0      &\ \ 0     &\ \ 0     &\ \ 0      &\ \ 0      &\ \ 0      &\ \ 1.0154&\!-0.0924 &\ \ 0     &\ \ 0    \\
\ \ 0      &\ \ 0     &\ \ 0     &\ \ 0      &\ \ 0      &\ \ 0      &\!-0.0637 &\ \ 0.9448&\ \ 0     &\ \ 0    \\
\ \ 0      &\ \ 0     &\ \ 0.0106& \!-0.0318 &\ \ 0.0106 & \!-0.0318 &\ \ 0     &\ \ 0     &\ \ 0.9997&\!-0.0350\\
\ \ 0      &\ \ 0     &\ \ 0     &\ \ 0      &\ \ 0      &\ \ 0      &\ \ 0     &\ \ 0     &\!-0.0350 &\ \ 0.9997\\
\end{array}
\right].
\end{eqnarray}

For the derivation of the RG-evolution matrix,
we start with constructing the renormalization group equation (RGE) 
of $W_i(\mu)$'s, and hence of $U(\mu,1/a)$'s. If we write the renormalization 
of $Q_i$ as $Q_i^{(0)}=Z_{ij}Q_j$  where the superscript (0) indicates the 
value at tree level, RGE for $Q_i$'s are readily obtained as
\begin{eqnarray}
\frac{d}{d\ln \mu} Q_i=-\gamma_{ij}Q_j\ ,\ \ 
\gamma\equiv\left(Z^{-1}\frac{d}{d\ln\mu}Z\right). \label{RGE-Q}
\end{eqnarray}
On the other hand, interpreting $W_i$'s as coupling constants in the
effective Hamiltonian, renormalization of 
$W_i$'s is possible, $W^{(0)}_i=Z^c_{ij}W_j$, in place of that of $Q_i$'s.
From the equivalence of these renormalizations, $Z^c=(Z^{-1})^T$ follows.
Therefore, using (\ref{RGE-Q}), we obtain 
\begin{eqnarray}
\frac{d}{d\ln \mu} W_i=\gamma_{ij}^TW_j\ ,\ \ {\rm hence}\ \
\frac{d}{d\ln \mu} U_{ij}(\mu,1/a)=(\gamma^T)_{ik}U_{kj}(\mu,1/a).
\label{RGE-W}
\end{eqnarray}
Using the $10\times 10$ anomalous dimension matrix $\gamma$,
defined in (\ref{RGE-Q}),
the RGE for $U(\mu,1/a)$ has been solved for the QCD $\beta$ function and 
anomalous dimension $\gamma$
calculated at next to leading order~\cite{Martinelli-etal,Buras-etal}:
\begin{eqnarray}
\beta(g)&=&-\beta_0\frac{g^3}{16\pi^2}-\beta_1\frac{g^5}{(16\pi^2)^2},\\
\beta_0&=&\frac{11N_c-2N_f}{3}\ ,\ \ \beta_1=\frac{34}{3}N_c^2
-\frac{10}{3}N_cN_f-2C_FN_f,\\
\gamma(\alpha_S,\alpha)&=&\gamma_S(g^2)+\frac{\alpha}{4\pi}\Gamma(g^2),\\
\gamma_S(g^2)&=&\gamma_S^{(0)}\frac{\alpha_S}{4\pi}+\gamma_S^{(1)}
\left(\frac{\alpha_S}{4\pi}\right)^2 ,\\
\Gamma(g^2)&=&\gamma_e^{(0)}+\frac{\alpha_S}{4\pi}\gamma_{\rm se}^{(1)}.
\end{eqnarray}
The solution at this order is written as
\begin{eqnarray}
U(\mu_1,\mu_2,\alpha)&=&U(\mu_1,\mu_2)+\frac{\alpha}{4\pi}R(\mu_1,\mu_2).
\label{evolv-grand}
\end{eqnarray}
Using the matrix $V$ that diagonalize the $\gamma_S^{(0)T}$,
we obtain ${\rm diag}[\gamma^{(0)}_{Di}]=V^{-1}\gamma^{(0)T}V$ and 
$ G=V^{-1}\gamma^{(1)T}V$. Then,
\begin{eqnarray}
U(\mu_1,\mu_2)&=&U^{(0)}(\mu_1,\mu_2)
+\frac{\alpha_S(\mu_1)}{4\pi}J U^{(0)}(\mu_1,\mu_2)
-U^{(0)}(\mu_1,\mu_2)\frac{\alpha_S(\mu_2)}{4\pi}J,\\
U^{(0)}(\mu_1,\mu_2)&=&V\left(\frac{\alpha_S(\mu_2)}{\alpha_S(\mu_1)}
\right)^{\gamma_D^{(0)}/2\beta_0}V^{-1},\\
J&=&VHV^{-1},\\
\ H_{ij}&=&\delta_{ij}\gamma_{Di}^{(0)}\frac{\beta_1}{2\beta^2_0}
-\frac{G_{ij}}{2\beta_0+\gamma_{Di}^{(0)}-\gamma_{Dj}^{(0)}}.
\end{eqnarray}
Moreover, with $M^{(0)}\equiv V^{-1}\gamma_e^{(0)T}V$,
\begin{eqnarray}
R(\mu_1,\mu_2)&=&-\frac{2\pi}{\beta_0}V\left[K^{(0)}(\mu_1,\mu_2)
+\frac{1}{4\pi}\sum_{i=1}^3K^{(1)}_i(\mu_1,\mu_2)\right]V^{-1},\\
K^{(0)}(\mu_1,\mu_2)_{ij}&=&\frac{2\beta_0M^{(0)}_{ij}}
{\gamma^{(0)}_{Di}-\gamma^{(0)}_{Dj}-2\beta_0}\nonumber \\
& &\times\Biggl[\left(\frac{\alpha_S(\mu_2)}{\alpha_S(\mu_1)}
\right)^{\gamma_{Dj}^{(0)}/2\beta_0}\frac{1}{\alpha_S(\mu_1)}
-\left(\frac{\alpha_S(\mu_2)}{\alpha_S(\mu_1)}
\right)^{\gamma_{Di}^{(0)}/2\beta_0}\frac{1}{\alpha_S(\mu_2)}\Biggr],\\
K^{(1)}_1(\mu_1,\mu_2)_{ij}&=&\frac{2\beta_0M^{(1)}_{ij}}
{\gamma^{(0)}_{Di}-\gamma^{(0)}_{Dj}}\Biggl[\left(\frac{\alpha_S(\mu_2)}
{\alpha_S(\mu_1)}\right)^{\gamma_{Dj}^{(0)}/2\beta_0}
-\left(\frac{\alpha_S(\mu_2)}{\alpha_S(\mu_1)}\right)^{
\gamma_{Di}^{(0)}/2\beta_0}\Biggr],\\
M^{(1)}&=&V^{-1}\left(\gamma^{(1)T}_{\rm se}-\frac{\beta_1}{\beta_0}
\gamma_e^{(0)T}+[\gamma_e^{(0)T},J]\right)V,\\
K_2^{(1)}(\mu_1,\mu_2)&=&-\alpha_S(\mu_2)K^{(0)}(\mu_1,\mu_2)H,\\
K_3^{(1)}(\mu_1,\mu_2)&=&\alpha_S(\mu_1)HK^{(0)}(\mu_1,\mu_2),
\end{eqnarray}
where, $\mu_1=\mu_c=1.3$ GeV, $\mu_2=1/a$.

Using the value of the strong coupling constant 
$\alpha_S^{\ovl{\rm MS}}(1/a)=0.30171$ and $\alpha_S^{\ovl{\rm MS}}
(1.3\ {\rm GeV})=0.39601$ with 
$\Lambda_{\overline{\rm MS}}^{(3)} = 372$ MeV,
together with
$\gamma$-functions presented in Ref.~\cite{Buch-Buras-Lauten95},
we obtain the matrix $U(m_c, 1/a, \alpha)$ given in (\ref{evolv-grand})
and the RG-evolution matrix: 

\begin{eqnarray}
& &[U^{-1}(m_c,1/a)]^T= \nn \\
& &\left[
\begin{array}{llllllllll}
0.9738& 0.0730 &\ \ 0.0035 & \!-0.0003 &\!-0.0033  & \!-0.0002 &\ \ 0.0005 &\ \ 0      &\ \ 0.0005 &\ \ 0.0001 \\
0.0731& 0.9736 & \!-0.0024 &\ \ 0.0149 &\!-0.0053  &\ \ 0.0116 &\ \ 0.0002 &\ \ 0      &\ \ 0.0001 &\ \ 0.0001 \\
    0 & 0      &\ \ 0.9794 &\ \ 0.1043 &\!-0.0212  &\ \ 0.0247 & \!-0.0002 &\ \ 0      & \!-0.0006 & \!-0.0001 \\
    0 & 0      &\ \ 0.0731 &\ \ 1.0105 &\!-0.0186  &\ \ 0.0306 & \!-0.0005 &\ \ 0      & \!-0.0004 &\ \ 0\\
    0 & 0      & \!-0.0083 & \!-0.0065 &\ \ 1.0465 & \!-0.0996 &\ \ 0.0005 &\ \ 0      &\ \ 0      &\ \ 0 \\
    0 & 0      & \!-0.0090 &\ \ 0.0228 & \!-0.0421 &\ \ 0.7878 &\ \ 0      &\ \ 0.0007 &\ \ 0      &\ \ 0\\
    0 & 0      &\ \ 0      &\ \ 0      &\ \ 0.0002 &\ \ 0      &\ \ 1.0349 & \!-0.0929 &\ \ 0.0008 &\ \ 0 \\
    0 & 0      &\ \ 0      &\ \ 0      &\ \ 0      &\ \ 0.0004 & \!-0.0367 &\ \ 0.7602 &\ \ 0.0001 &\!-0.0001\\
    0 & 0      &\ \ 0.0021 &\!-0.0149  &\ \ 0.0053 & \!-0.0116 &\ \ 0.0009 &\ \ 0.0001 &\ \ 0.9750 &\ \ 0.0731 \\
    0 & 0      & \!-0.0035 &\!-0.0001  &\ \ 0.0032 &\ \ 0.0002 &\ \ 0.0006 &\ \ 0      &\ \ 0.0736 &\ \ 0.9740\\
\end{array}
\right].
\end{eqnarray}

In order to check the systematic error associated with the matching
procedure above, we also employ an alternative procedure in which  
the RG-evolution is carried out in the {\it quenched} theory from 
$\mu_2=q^*$ to $\mu_1=\mu_c=$1.3 GeV where matching to the $N_f=3$ theory 
is made. 
For the quenched RG-evolution, the two-loop anomalous dimension matrix 
$\gamma_S^{(1)}$ is modified according to~\cite{Buras-etal}
\begin{eqnarray}
[\gamma_S^{(1)}]_{\rm quenched} &=&[\gamma_S^{(1)}]_{\rm full} -
\Delta \gamma_S^{(1)} ,
\end{eqnarray}
where $\Delta \gamma_S^{(1)}= {\rm diag}
[\Gamma_1,\Gamma_2,\Gamma_3,\Gamma_4,\Gamma_5]$ with the 
$2\times 2$ matrices $\Gamma_i$, which are given by 
\begin{eqnarray}
\Gamma_1& = &\Gamma_2=\Gamma_5=
\left[
\begin{array}{cc}
-\frac{2N_f}{3 N_c} & \frac{2N_f}{3} \\
\frac{2N_f}{3 } & -\frac{2N_f}{3N_c} \\
\end{array}
\right] \\
\Gamma_3& = &\Gamma_4 =
\left[
\begin{array}{cc}
-\frac{22N_f}{3 N_c} & \frac{22N_f}{3} \\
4N_f & \frac{20 C_F N_f}{3}-\frac{4N_f}{N_c} \\
\end{array}
\right]. 
\end{eqnarray}
Note that $N_f = 3$ in this case.
For the gauge coupling in the quenched theory,
we employ $\alpha_S^{\ovl{\rm MS}}(1/a)=0.180891$ from (\ref{eq:gg}),
and $\alpha_S^{\ovl{\rm MS}} (1.3\ {\rm GeV})=0.20439$
obtained by the 2-loop running with $N_f=0$. 

\newcommand{\J}[4]{{#1} {\bf #2} (#3) #4}
\newcommand{\RMP}{Rev.~Mod.~Phys.}
\newcommand{\MPL}{Mod.~Phys.~Lett.}
\newcommand{\IJMP}{Int.~J.~Mod.~Phys.}
\newcommand{\NP}{Nucl.~Phys.}
\newcommand{\NPSup}{Nucl.~Phys.~{\bf B} (Proc.~Suppl.)}
\newcommand{\PL}{Phys.~Lett.}
\newcommand{\PR}{Phys.~Rev.}
\newcommand{\PRL}{Phys.~Rev.~Lett.}
\newcommand{\AP}{Ann.~Phys.}
\newcommand{\CMP}{Commun.~Math.~Phys.}
\newcommand{\PTP}{Prog. Theor. Phys.}
\newcommand{\Suppl}{Prog. Theor. Phys. Suppl.}

\clearpage 
%

\begin{table}
\begin{center}
\begin{tabular}{ccc}
\hline
\ \ \ \ $m_fa$&\hspace{0.5cm} $16^3\times 32$&\hspace{0.5cm} $24^3\times 32$\\
\hline
\hspace{0.5cm}0.02 &\hspace{0.5cm} 407 &\hspace{0.5cm} 432 \hspace{0.5cm}\\
\hspace{0.5cm}0.03 &\hspace{0.5cm} 406 &\hspace{0.5cm} 200 \hspace{0.5cm}\\
\hspace{0.5cm}0.04 &\hspace{0.5cm} 406 &\hspace{0.5cm} 200 \hspace{0.5cm}\\
\hspace{0.5cm}0.05 &\hspace{0.5cm} 432 &\hspace{0.5cm} 200 \hspace{0.5cm}\\
\hspace{0.5cm}0.06 &\hspace{0.5cm} 435 &\hspace{0.5cm} 200 \hspace{0.5cm}\\
\hline
\end{tabular}
\caption{Number of gauge configurations, independently generated for
each value of $m_fa$, in our numerical simulation.}
\label{Ngconfs}
\end{center}
\end{table}

\begin{table}
\begin{center}
\begin{tabular}{lll|lll}
\hline
\multicolumn{3}{c|}{$16^3\times 32$}& \multicolumn{3}{c}{$24^3\times 32$}\\
\ \ $m_fa$ & $m_M^2 [{\rm GeV}^2]$& & &\ \ $m_fa$ & $m_M^2 [{\rm GeV}^2]$ \\
\hline\hline
$-$0.00049(32)\hspace*{0.5cm}& 0.00 & & &  $-$0.00056(16)\hspace*{0.5cm}&0.00  \\
\ \ 0.00& 0.0059(37) & & &\ \ 0.00& 0.0066(19) \\
\hline
\ \ 0.02& 0.2434(26) & & &\ \ 0.02 & 0.2445(11) \\
\ \ 0.03& 0.3568(29) & & &\ \ 0.03 & 0.3534(17) \\
\ \ 0.04& 0.4741(28) & & &\ \ 0.04 & 0.4714(19) \\
\ \ 0.05& 0.5932(29) & & &\ \ 0.05 & 0.5957(19) \\
\ \ 0.06& 0.7134(30) & & &\ \ 0.06 & 0.7158(20) \\
\hline
\end{tabular}
\end{center}
\caption{Lattice pseudo scalar meson mass squared $m_M^2\ [{\rm GeV}^2]$ 
at each $m_fa$. The x- and y-intercepts are obtained through a linear 
chiral extrapolation.  Physical scale of lattice spacing equals 
$1/a=1.94$~GeV determined by $\sqrt{\sigma}=440$~MeV. }
\label{mpi2vmf}
\end{table}

\begin{table}
\begin{center}
\begin{tabular}{cclllc|lll}
\hline 
 \multicolumn{6}{c|}{$16^3\times 32$}&
 \multicolumn{3}{c}{$24^3\times 32$}\\
& $m_fa$&\multicolumn{1}{c}{first}
 &\multicolumn{1}{c}{subtraction}&\multicolumn{1}{c}{total}&
&\multicolumn{1}{c}{first}
 &\multicolumn{1}{c}{subtraction}&\multicolumn{1}{c}{total}\\
\hline
$Q_1^{(0)}$ & 0.02 &\ $\!-$0.0135(44) & $\!-$0.0134(25) & $\!-$0.0269(56)&
                   &  $\!-$0.0028(23) & $\!-$0.0164(12) & $\!-$0.0192(28)\\
            & 0.03 &\ $\!-$0.0084(29) & $\!-$0.0133(20) & $\!-$0.0217(39)&
                   &  $\!-$0.0052(24) & $\!-$0.0115(14) & $\!-$0.0167(30)\\
            & 0.04 &\ $\!-$0.0091(21) & $\!-$0.0107(16) & $\!-$0.0198(30)&
                   &  $\!-$0.0082(14) & $\!-$0.0092(10) & $\!-$0.0174(21)\\
            & 0.05 &\ $\!-$0.0096(16) & $\!-$0.0085(13) & $\!-$0.0181(24)&
                   &  $\!-$0.0056(12) & $\!-$0.0076(10) & $\!-$0.0131(18)\\
            & 0.06 &\ $\!-$0.0071(14) & $\!-$0.0073(12) & $\!-$0.0144(20)&
                   &  $\!-$0.0085(11) & $\!-$0.00752(82)& $\!-$0.0160(16)\\
\hline
$Q_2^{(0)}$ & 0.02 &\ $\!-$0.0410(40)&\ \ 0.0825(25)&\ \ 0.0415(46)&
                   &  $\!-$0.0500(19)&\ \ 0.0875(13)&\ \ 0.0375(23)\\
            & 0.03 &\ $\!-$0.0419(24)&\ \ 0.0755(21)&\ \ 0.0336(29)&
                   &  $\!-$0.0450(19)&\ \ 0.0823(16)&\ \ 0.0373(22)\\
            & 0.04 &\ $\!-$0.0392(18)&\ \ 0.0752(18)&\ \ 0.0361(22)&
                   &  $\!-$0.0434(18)&\ \ 0.0743(14)&\ \ 0.0309(16)\\
            & 0.05 &\ $\!-$0.0375(15)&\ \ 0.0659(14)&\ \ 0.0284(16)&
                   &  $\!-$0.0394(13)&\ \ 0.0680(12)&\ \ 0.0286(13)\\
            & 0.06 &\ $\!-$0.0346(13)&\ \ 0.0627(13)&\ \ 0.0281(14)&
                   &  $\!-$0.0354(11)&\ \ 0.0636(11)&\ \ 0.02821(94)\\
\hline
$Q_3^{(0)}$ & 0.02 &\ $\!-$0.130(17) &\ \ 0.1253(90)& $\!-$0.005(21)&
                   &  $\!-$0.1151(81)&\ \ 0.1256(46)&\ \ 0.010(10)\\
            & 0.03 &\ $\!-$0.118(10) &\ \ 0.1107(81)& $\!-$0.007(14)&
                   &  $\!-$0.1140(84)&\ \ 0.1311(53)&\ \ 0.017(11)\\
            & 0.04 &\ $\!-$0.1137(76)&\ \ 0.1179(61)&\ \ 0.004(11)&
                   &  $\!-$0.1198(61)&\ \ 0.1206(42)&\ \ 0.0008(80)\\
            & 0.05 &\ $\!-$0.1132(65)&\ \ 0.1055(50)& $\!-$0.0077(86)&
                   &  $\!-$0.1052(46)&\ \ 0.1146(43)&\ \ 0.0094(66)\\
            & 0.06 &\ $\!-$0.1000(50)&\ \ 0.1031(47)&\ \ 0.0032(75)&
                   &  $\!-$0.1049(44)&\ \ 0.1051(35)&\ \ 0.0002(55)\\
\hline
$Q_5^{(0)}$ & 0.02 &\ \ \ 1.719(45)& $\!-$1.743(36)& $\!-$0.024(24)&
                   &  \ \ 1.832(24)& $\!-$1.853(185) & $\!-$0.022(11)\\
            & 0.03 &\ \ \ 1.608(37)& $\!-$1.657(32)& $\!-$0.048(15)&
                   &  \ \ 1.731(31)& $\!-$1.768(261) & $\!-$0.036(11)\\
            & 0.04 &\ \ \ 1.591(33)& $\!-$1.633(30)& $\!-$0.042(11)&
                   &  \ \ 1.593(27)& $\!-$1.635(249) & $\!-$0.0420(80)\\
            & 0.05 &\ \ \ 1.438(26)& $\!-$1.482(25)& $\!-$0.0444(82)&
                   &  \ \ 1.521(25)& $\!-$1.553(224) & $\!-$0.0321(67)\\
            & 0.06 &\ \ \ 1.430(26)& $\!-$1.465(23)& $\!-$0.0359(71)&
                   &  \ \ 1.412(23)& $\!-$1.448(205) & $\!-$0.0361(53)\\
\hline
$Q_6^{(0)}$ & 0.02 &\ \ \ 4.98(13)  & $\!-$5.01(10) & $\!-$0.025(51)&
                   &  \ \ 5.264(67) & $\!-$5.350(54)& $\!-$0.086(22)\\
            & 0.03 &\ \ \ 4.66(10)  & $\!-$4.792(91)& $\!-$0.129(26)&
                   &  \ \ 4.960(88) & $\!-$5.110(76)& $\!-$0.150(20)\\
            & 0.04 &\ \ \ 4.632(97) & $\!-$4.721(86)& $\!-$0.089(19)&
                   &  \ \ 4.595(80) & $\!-$4.732(71)& $\!-$0.137(14)\\
            & 0.05 &\ \ \ 4.155(78) & $\!-$4.287(71)& $\!-$0.132(12)&
                   &  \ \ 4.385(72) & $\!-$4.496(65)& $\!-$0.111(11)\\
            & 0.06 &\ \ \ 4.121(73) & $\!-$4.234(67)& $\!-$0.1129(95)&
                   &  \ \ 4.087(65) & $\!-$4.183(60)& $\!-$0.0957(88)\\
\hline
\end{tabular}
\end{center}
\caption{Subtraction in $K\to\pi$ matrix element 
$\VEV{\pi^+}{Q_i^{(0)}}{K^+}$ for $i=1,2,3,5,6$ multiplied with 
a factor $\sqrt{2} f_\pi (m_K^2-m_\pi^2)/\VEV{\pi^+}{A_4}{0}\VEV{0}{A_4}{K^+}$. 
The values of the $K^+\to\pi^+$ matrix element (first), 
the subtraction term $-\alpha_i\VEV{\pi^+}{Q_{\rm sub}}{K^+}$ (subtraction) 
and their sum (total) are given in units of ${\rm GeV}^{3}$. }
\label{effect-tab}
\end{table}

\begin{table}
\begin{center}
\begin{tabular}{ccccccccccc}
\hline
$\Lambda_{\overline{\rm MS}}^{(4)}$ & $z_1$ & $z_2$ & $y_3$ & $y_4$ &
$y_5$ & $y_6$ &$y_7/\alpha$ & $y_8/\alpha$ & $y_9/\alpha$ & $y_{10}/\alpha$ \\
\hline
215~MeV &\ \ $\!-$0.346 &\ \ 1.172      &\ \ 0.023 &\ \ $\!-$0.048 &\ \ 0.007
        &\ \ $\!-$0.068 &\ \ $\!-$0.031 &\ \ 0.103 &\ \ $\!-$1.423 &\ \ 0.451 \\
325~MeV &\ \ $\!-$0.415 &\ \ 1.216      &\ \ 0.029 &\ \ $\!-$0.057 &\ \ 0.005
        &\ \ $\!-$0.089 &\ \ $\!-$0.030 &\ \ 0.136 &\ \ $\!-$1.479 &\ \ 0.547 \\
435~MeV &\ \ $\!-$0.490 &\ \ 1.265      &\ \ 0.036 &\ \ $\!-$0.068 &\ \ 0.001
        &\ \ $\!-$0.118 &\ \ $\!-$0.029 &\ \ 0.179 &\ \ $\!-$1.548 &\ \ 0.664 \\
\hline
\end{tabular} 
\end{center}
\caption{Wilson coefficient functions \protect\cite{Buch-Buras-Lauten95}.} 
\label{Wilsoncoeff}
\end{table}

\begin{table}[h]
\begin{center}
\begin{tabular}{cllll|llll}
\hline
& \multicolumn{4}{c|}{$16^3\times 32$}& \multicolumn{4}{c}{$24^3\times 32$}\\
&\multicolumn{1}{c}{$a_0$}&\multicolumn{1}{c}{$a_1 [{\rm GeV}^{-2}]$}&\multicolumn{1}{c}{$a_2 [{\rm GeV}^{-4}]$}& 
$\chi^2/$dof
&\multicolumn{1}{c}{$a_0$}&\multicolumn{1}{c}{$a_1 [{\rm GeV}^{-2}]$}&\multicolumn{1}{c}{$a_2 [{\rm GeV}^{-4}]$}& 
$\chi^2/$dof\\
\hline
$Q_1^{(0)}$ &\,$-$0.007(38) &\,$-$0.09(17)   &\ \ \ 0.04(17)  &  0.63
            &\ \ \ 0.017(24)&\,$-$0.10(11)   &\ \ \ 0.01(12)  &  1.88\\
$-\alpha_1Q_{\rm sub}$ 
            &\ \ \ 0.004(25)&\,$-$0.17(12)   &\ \ \ 0.15(13)  &  0.19
            &\,$-$0.041(15) &\ \ \ 0.037(74) &\,$-$0.064(81)  &  0.06\\
$Q_1^{(0)}-\alpha_1Q_{\rm sub}$       
            &\,$-$0.007(51) &\,$-$0.24(24)   &\ \ \ 0.17(24)  &  0.12
            &\,$-$0.024(31) &\,$-$0.06(15)   &\,$-$0.06(16)   &  1.16\\
\hline
$Q_2^{(0)}$ &\ \ \,0.021(34)&\,$-$0.51(15)   &\ \ \ 0.25(15)  &  0.07
            &\,$-$0.002(23) &\,$-$0.50(11)   &\ \ \ 0.26(12)  &  0.35\\
$-\alpha_2Q_{\rm sub}$ 
            &\ \ \,0.000(26)&\ \ \,0.82(13)  &\,$-$0.37(13)   &  2.43
            &\ \ \,0.017(18)&\ \ \ 0.802(92) &\,$-$0.37(10)   &  0.71\\
$Q_2^{(0)}-\alpha_2Q_{\rm sub}$       
            &\ \ \,0.019(40)&\ \ \,0.31(18)  &\,$-$0.13(18)   &  2.09
            &\ \ \,0.024(25)&\ \ \ 0.26(11)  &\,$-$0.06(12)   &  1.40\\
\hline
$Q_3^{(0)}$ &\ \ \,0.02(15) & $-$1.36(64)    &\ \ \ 0.59(65)  &  0.42
            &\ \ \,0.063(90)&\,$-$1.43(44)   &\ \ \,0.58(47)  &  0.92\\
$-\alpha_3Q_{\rm sub}$ 
            &\ \ \,0.0014(95)&\ \ \,1.12(46) &\,$-$0.31(49)   &  0.89
            &\,$-$0.089(59) &\ \ \,1.71(30)  &\,$-$0.92(33)   &  0.34\\
$Q_3^{(0)}-\alpha_3Q_{\rm sub}$       
            &\ \ \,0.02(19) &$-$0.19(86)     &\ \ \ 0.23(89)  &  0.58
            &\,$-$0.02(11)  &\ \ \,0.27(56)  &\,$-$0.32(60)   &  0.87\\
\hline
$Q_5^{(0)}$ &\ \ \,0.37(48) &\ 14.7(2.3)     &\,$-$3.7(2.5)   &  2.56
            &\ \ \,0.27(34) &\ 16.8(1.7)     &\,$-$6.4(1.9)   &  0.48\\
$-\alpha_5Q_{\rm sub}$ 
            &$-$0.15(42)    &$\!\!\!-$16.1(2.0)&\ \ \,4.7(2.2)&  2.56
            &$-$0.11(29)    &$\!\!\!-$17.8(1.5)&\ \ \,7.0(1.7)&  0.40\\
$Q_5^{(0)}-\alpha_5Q_{\rm sub}$       
            &\ \ \,0.020(20)&$-$1.27(90)     &\ \ \,0.95(92)  &  0.17
            &\ \ \,0.13(12) &$\!-$0.85(57)   &\ \ \,0.51(61)  &  0.48\\
\hline
$Q_6^{(0)}$ &\ \ \,0.8(1.4) &\,44.1(6.6)     &$\!\!-$12.4(7.1)&  2.96
            &\ \ \,0.86(97) &\,47.6(5.0)     &$\!\!-$17.2(5.6)&  0.31\\
$-\alpha_6Q_{\rm sub}$ 
            &$-$0.1(1.2)    &$\!\!\!\!-$47.8(5.8)&\ \,14.8(6.3) &  2.61
            &$-$0.19(85)    &$\!\!\!\!-$52.1(4.5)&\ \,21.0(5.0) &  0.42\\
$Q_6^{(0)}-\alpha_6Q_{\rm sub}$       
            &\ \ \,0.053(38)&$\!-$3.0(1.6)   &\ \ \ 1.7(1.6)  &  2.27
            &\ \ \,0.59(22) &$\!-$4.0(1.0)   &\ \ \ 3.3(1.1)  &  0.62\\
\hline
$Q_1^{(2)}$ & $-$0.0023(13) &\ \ 0.0727(64)  &\ \ \ 0.0178(68)&  0.19
            & \,$-$0.00264(65) &\ \ 0.0751(33)&\ \ \ 0.0140(37)& 0.28\\
\hline\hline
&\multicolumn{1}{c}{$b_0$}
&\multicolumn{1}{c}{$b_1\ [{\rm GeV^{-2}}]$}
&\multicolumn{1}{c}{$b_2\ [{\rm GeV^{-4}}]$}
&$\chi^2/$dof
&\multicolumn{1}{c}{$b_0$}
&\multicolumn{1}{c}{$b_1\ [{\rm GeV^{-2}}]$}
&\multicolumn{1}{c}{$b_2\ [{\rm GeV^{-4}}]$}
&$\chi^2/$dof\\
\hline
$\bar{s}d $ & $-$170(11)   &\ \ 116(46)  &$-$64(45) & 2.21
            & $-$186.2(4.0)&\ \ 151(19)  &$-$82(19) & 3.72 \\
\hline
\end{tabular}
\end{center}
\caption{Fit parameters for the chiral extrapolation of the
 $K\to\pi$ matrix elements defined by (\protect\ref{chiral-ratio}) 
which should vanish in the chiral limit. The parameters 
$(a_0,a_1,a_2)$ are determined by the fit function
 $a_0+a_1m_M^2+a_2(m_M^2)^2$.}
\label{quadtab}
\end{table}

\begin{table}[h]
\begin{center}
\begin{tabular}{cllll|llll}
\hline
& \multicolumn{4}{c|}{$16^3\times 32$}& \multicolumn{4}{c}{$24^3\times 32$}\\
& $a_1 [{\rm GeV}^{-2}]$& $a_2 [{\rm GeV}^{-4}]$& $a_4 [{\rm GeV}^{-6}]$& $\chi^2/$dof& $a_1 [{\rm GeV}^{-2}]$& $a_2 [{\rm GeV}^{-4}]$& $a_4 [{\rm GeV}^{-6}]$& $\chi^2/$dof\\
\hline
$Q_1^{(0)}$&\ $-$0.10(11)     &\ \ 0.01(44)     &\ \ 0.06(41)    & 0.64
           &\ \ \ \ 0.004(67) & $-$0.18(28)     &\ \ 0.10(27)    & 2.07\\
$-\alpha_1Q_{\rm sub}$
           &\ $-$0.159(72)    &\ \ 0.14(30)     &$-$0.01(29)     & 0.20         
           &\ \ $-$0.271(40)  &\ \ 0.64(18)     &$-$0.49(18)     & 0.12\\
$Q_1^{(0)}-\alpha_1Q_{\rm sub}$      
           &\ $-$0.28(15)     &\ \ 0.21(60)     &\ \ 0.00(57)    & 0.13
           &\ \ $-$0.257(87)  &\ \ 0.42(37)     &$-$0.36(36)     & 0.98\\
\hline
$Q_2^{(0)}$&\ $-$0.37(10)     & $-$0.06(39)     &\ \ 0.21(37)    & 0.10
           &\ \ $-$0.499(63)  &\ \ \,0.23(27)   &\ \ 0.02(28)    & 0.35\\
$-\alpha_2Q_{\rm sub}$
           &\ \ \ 0.826(75)   &$-$0.39(31)      &\ \ 0.02(30)    & 2.43
           &\ \ \ \ 0.942(47) &$-$0.72(21)      &\ \ 0.26(22)    & 0.43\\
$Q_2^{(0)}-\alpha_2Q_{\rm sub}$      
           &\ \ \ 0.46(12)    &$-$0.46(45)      &\ \ 0.23(42)    & 2.07         
           &\ \ \ \ 0.461(67) &$-$0.56(28)      &\ \ 0.37(27)    & 0.98\\
\hline
$Q_3^{(0)}$&\ $-$1.09(43)     & $-$0.2(1.7)     &\ \ 0.6(1.6)    & 0.35
           &\ \ $-$1.00(25)   &\,$-$0.3(1.0)    &\ \ 0.6(1.0)    & 1.02\\
$-\alpha_3Q_{\rm sub}$
           &\ \ \ \,1.20(27)  &$-$0.4(1.1)      &\ \ 0.1(1.1)    & 0.90        
           &\ \ \ \ 1.07(16)  &\ \ \,0.51(71)   & $-$0.97(71)    & 0.57\\
$Q_3^{(0)}-\alpha_3Q_{\rm sub}$      
           &\ \ \ \,0.05(55)  &$-$0.4(2.2)      &\ \ 0.5(2.1)    & 0.55         
           &\ \ \ \ 0.10(32)  &\ \ \,0.1(1.3)   & $-$0.3(1.3)    & 0.87\\
\hline
$Q_5^{(0)}$&\ \ 17.7(1.4)     &$\!\!\!-$11.1(5.7)&\ \ 5.6(5.6)   & 2.36
           &\ \ 18.82(88)     & $\!\!-$11.0(4.0)&\ \ 3.3(4.1)    & 0.50\\
$-\alpha_5Q_{\rm sub}$
           &$-$17.5(1.2)      &\ \ 8.5(5.0)     &$-$3.0(5.0)     & 2.44         
           &$-$18.68(76)      &\ \ 9.2(3.5)     &$-$1.6(3.7)     & 0.38\\
$Q_5^{(0)}-\alpha_5Q_{\rm sub}$      
           &\ \ \ \,0.08(59)  &$-$1.9(2.3)      &\ \ 1.9(2.2)    & 0.23         
           &\ \ \ \,0.03(34)  &$-$1.3(1.4)      &\ \ 1.2(1.4)    & 0.68\\
\hline
$Q_6^{(0)}$&\ \ 50.9(3.9)     & $\!\!\!-$29(16) &\,13(16)        & 2.80
           &\ \ 54.0(2.5)     & $\!\!\!-$32(12) &\,10(12)        & 0.34 \\
$-\alpha_6Q_{\rm sub}$
           &$-$49.6(3.4)      &\ 21(14)         &$-$6(14)        & 2.54         
           &$-$53.6(2.2)      &\ 25(10)         &$-$2.8(1.1)     & 0.41 \\
$Q_6^{(0)}-\alpha_6Q_{\rm sub}$      
           &\ \ \ \ 0.6(1.1)  &$-$5.9(4.1)      &\ \ 5.0(3.8)    & 2.36         
           &\ \ \ \ 0.04(63)  &$-$5.3(2.6)      &\ \ 5.7(2.5)    & 1.54\\
\hline
$Q_1^{(2)}$&\ \ \ \ 0.0555(38)&\ \ \,0.057(16)  & $-$0.027(15)   & 0.18
           &\ \ \ \ 0.0557(17)&\ \,0.0573(77)   & $\!-$0.0299(79)& 1.50\\
\hline
\end{tabular}
\end{center}
\caption{Same as Table~\ref{quadtab} for the fit function 
$a_1m_M^2+a_2(m_M^2)^2+a_4(m_M^2)^3$.} 
\label{cubic-tab}
\end{table}

\begin{table}
\begin{center}
\begin{tabular}{cllll|llll}
\hline
& \multicolumn{4}{c|}{$16^3\times 32$}& \multicolumn{4}{c}{$24^3\times 32$}\\
&\multicolumn{1}{c}{$a_1 [{\rm GeV}^{-2}]$}&\multicolumn{1}{c}{$a_2 [{\rm GeV}^{-4}]$}&\multicolumn{1}{c}{$a_3 [{\rm GeV}^{-4}]$}& 
$\chi^2/$dof
&\multicolumn{1}{c}{$a_1 [{\rm GeV}^{-2}]$}&\multicolumn{1}{c}{$a_2 [{\rm GeV}^{-4}]$}&\multicolumn{1}{c}{$a_3 [{\rm GeV}^{-4}]$}& 
$\chi^2/$dof\\
\hline
$Q_1^{(0)}$ &\ \ $-$0.11(20)    &\ \ 0.07(14)       &\ \ \ 0.02(39)  & 0.65
            &\ \ \ \ 0.04(12)   & $-$0.110(84)      &\ \ 0.13(25)    & 2.01\\
$-\alpha_1Q_{\rm sub}$ &\ \ $-$0.15(13)    &\ \ 0.128(90)      &\ \ \ 0.01(27)  & 0.20   
            &\ \ $-$0.369(74)   &\ \ 0.268(49)      & $-$0.46(16)    & 0.06\\
$Q_1^{(0)}-\alpha_1Q_{\rm sub}$       &\ \ $-$0.29(27)    &\ \ 0.21(19)       &\,$-$0.03(54)   & 0.13   
            &\ \ $-$0.32(16)    &\ \ 0.15(11)       &$-$0.31(34)     & 1.05\\
\hline
$Q_2^{(0)}$ &\ \ $-$0.32(18)    &\ \ 0.09(13)       &\ \ \,0.20(35)  & 0.09
            &\ \ $-$0.50(12)    &\ \ 0.259(77)      &\ \ 0.01(25)    & 0.35\\
$-\alpha_2Q_{\rm sub}$ &\ \ \ \ 0.83(13)   &$-$0.376(93)       &\ \ \,0.01(28)  & 2.43   
            &\ \ \ \ 0.987(88)  &$-$0.515(56)       &\ \ 0.23(20)    & 0.52\\
$Q_2^{(0)}-\alpha_2Q_{\rm sub}$       &\ \ \ \ 0.50(21)   &$-$0.29(15)        &\ \ \,0.21(41)  & 2.08   
            &\ \ \ \ 0.53(12)   &$-$0.276(85)       &\ \ 0.33(25)    & 1.11\\ 
\hline
$Q_3^{(0)}$ &\ \ $-$1.01(76)    &\ \ 0.32(56)       &\ \ \,0.5(1.5)  & 0.38
            &\ \ $-$0.85(45)    &\ \ 0.10(31)       &\ \ 0.59(97)    & 0.98\\
$-\alpha_3Q_{\rm sub}$ &\ \ \ \ 1.22(49)   &$-$0.40(34)        &\ \ \,0.1(1.0)  & 0.89  
            &\ \ \ \ 0.86(29)   &$-$0.21(19)        & $-$0.93(66)    & 0.49\\
$Q_3^{(0)}-\alpha_3Q_{\rm sub}$       &\ \ \ \ 0.12(98)   &$-$0.01(71)        &\ \ \,0.4(2.0)  & 0.56   
            &\ \ \ \ 0.05(58)   &$-$0.14(39)        & $-$0.2(1.2)    & 0.87\\
\hline
$Q_5^{(0)}$ &\ \ \,18.7(2.4)    & $-$6.8(1.7)       &\ \ \,4.8(5.2)  & 2.44
            &\ \ 19.5(1.7)      & $-$8.6(1.0)       &\ \ 3.1(3.8)    & 0.48\\
$-\alpha_5Q_{\rm sub}$ &$-$17.9(2.1)       &\ \ 6.1(1.4)       &\,$-$2.4(4.6)   & 2.49   
            &$-$19.0(1.4)       &\ \ 7.99(88)       &$-$1.4(3.4)     & 0.38\\
$Q_5^{(0)}-\alpha_5Q_{\rm sub}$       &\ \ \ \,0.5(1.1)   & $-$0.54(77)       &\ \ \,1.9(2.1)  & 0.21   
            &\ \ \ \,0.32(61)   & $-$0.46(42)       &\ \ 1.2(1.3)    & 0.61\\
\hline
$Q_6^{(0)}$ &\ \ \,52.9(7.1)    &$\!\!\!-$19.5(4.9) &\ 11(15)        & 2.87
            &\ \ \,56.1(4.8)    &$\!\!\!-$24.2(3.0) &\,10(11)        & 0.32\\
$-\alpha_6Q_{\rm sub}$ &$-$50.1(6.1)       &\ 16.5(4.1)        & $-$4(13)       & 2.58   
            &$-$54.1(4.2)       &\ 22.6(2.5)        &$\!-$2.5(9.7)   & 0.42\\ 
$Q_6^{(0)}-\alpha_6Q_{\rm sub}$       &\ \ \ \,1.8(1.9)   &$-$2.2(1.5)        &\ \ 5.0(3.7)    & 2.35  
            &\ \ \ \,1.4(1.1)   & $-$1.16(80)       &\ \ 5.7(2.3)    & 1.21\\ 
\hline
$Q_1^{(2)}$ &\ \ \ \,0.0498(68) &\ \ \,0.0364(47)   & $-$0.026(15)   & 0.15
            &\ \ \ \,0.0494(32) &\ \ \,0.0351(20)   & $-$0.0285(73)  & 0.99\\
\hline
\end{tabular}
\end{center}
\caption{Same as Table~\ref{quadtab} for the fit function 
$a_1m_M^2+a_2(m_M^2)^2+a_3(m_M^2)^2\ln m_M^2$ including a chiral logarithm 
term.} 
\label{chlogtab}
\end{table}

\begin{table}
\begin{center}
\begin{tabular}{cccccc}
\hline 
$m_fa$ & 0.02 & 0.03 & 0.04 & 0.05 & 0.06 \\
\hline
$16^3 \times 32$\ \ & $-145.0(3.1)$ & $-135.1(2.5)$ & $-132.6(2.3)$ 
& $-120.9(1.9)$ & $-120.2(1.8)$ \\
$ 24^3\times 32$\ \ & $-154.7(1.6)$ & $-142.1(2.1)$ & $-131.9(1.9)$ 
& $-127.4(1.7)$ & $-119.3(1.6)$ \\
\hline
\end{tabular}
\end{center}
\caption{$\displaystyle\frac{1}{2 f_M^2a} \VEV{\pi^+}{ \bar{s}d }{K^+}$
  as a function of $m_fa$.}
\label{sbard}
\end{table}

\begin{table}
\begin{center}
\begin{tabular}{cclllll}
\hline
&$m_fa$& \multicolumn{1}{c}{0.02}& \multicolumn{1}{c}{0.03}& \multicolumn{1}{c}{0.04}
      & \multicolumn{1}{c}{0.05}& \multicolumn{1}{c}{0.06}\\
\hline
  bare         &$\vev{Q_1}_0$&$\!-$0.0329(69)&$\!-$0.0266(48)&$\!-$0.0242(37)&$\!-$0.0222(29)&$\!-$0.0176(25)\\
$[{\rm GeV}^3]$&$\vev{Q_2}_0$&\ \ 0.0508(57)&\ \ 0.0412(35)&\ \ 0.0442(27)&\ \ 0.0347(20)&\ \ 0.0345(17)\\
               &$\vev{Q_3}_0$&$\!-$0.006(26)& $\!-$0.008(18)&\ \ 0.005(13)& $\!-$0.009(10)&\ \ 0.0039(91)\\
               &$\vev{Q_4}_0$&\ \ 0.078(24)&\ \ 0.059(16)&\ \ 0.074(12)&\ \ 0.0475(93)&\ \ 0.0560(82)\\
               &$\vev{Q_5}_0$& $\!-$0.030(29)& $\!-$0.059(18)& $\!-$0.051(13)& $\!-$0.054(10)&$\!-$0.0439(88)\\
               &$\vev{Q_6}_0$& $\!-$0.031(62)& $\!-$0.157(31)& $\!-$0.109(23)&  $\!-$0.161(15)&$\!-$0.138(12)\\
               &$\vev{Q_7}_0$&\ \ 1.635(30)&\ \ 2.043(33)&\ \ 2.574(42)&\ \ 2.835(43)&\ \ 3.328(49)\\
               &$\vev{Q_8}_0$&\ \ 5.012(91)&\ \ 6.25(10)&\ \ 7.90(13)&\ \ 8.66(13)&\ \ 10.18(15)\\
               &$\vev{Q_9}_0$&$\!-$0.0464(58)&$\!-$0.0357(35)&$\!-$0.0389(28)& $\!-$0.0285(20)& $\!-$0.0284(18)\\
               &$\vev{Q_{10}}_0$&\ \ 0.0372(69)&\ \ 0.0321(48)&\ \ 0.0294(38)&\ \ 0.0284(30)&\ \ 0.0237(25)\\ 
               &$\vev{Q_1}_2$&\ \ 0.01314(15)&\ \ 0.01402(12)&\ \ 0.01487(11)&\ \ 0.015399(98)&\ \ 0.015957(90)\\
               &$\vev{Q_2}_2$&\ \ 0.01314(15)&\ \ 0.01402(12)&\ \ 0.01487(11)&\ \ 0.015399(98)&\ \ 0.015957(90)\\
               &$\vev{Q_7}_2$&\ \ 0.4110(42)&\ \ 0.4292(34)&\ \ 0.4656(28)&\ \ 0.4863(27)&\ \ 0.5264(24)\\
               &$\vev{Q_8}_2$&\ \ 1.238(13)&\ \ 1.261(11)&\ \ 1.3357(87)&\ \ 1.3639(77)&\ \ 1.4451(70)\\
               &$\vev{Q_9}_2$&\ \ 0.01971(23)&\ \ 0.02103(18)&\ \ 0.02231(16)&\ \ 0.02310(15)&\ \ 0.02393(13)\\
               &$\vev{Q_{10}}_2$&\ \ 0.01971(23)&\ \ 0.02103(18)&\ \ 0.02231(16)&\ \ 0.02310(15)&\ \ 0.02393(13)\\
\hline\hline
&$m_fa$& \multicolumn{1}{c}{0.02}& \multicolumn{1}{c}{0.03}& \multicolumn{1}{c}{0.04}
      & \multicolumn{1}{c}{0.05}& \multicolumn{1}{c}{0.06}\\
\hline
renormalized   &$\vev{Q_1}_0$&$\!-$0.0291(68)&$\!-$0.0234(47)&$\!-$0.0206(37)&$\!-$0.0191(29)&$\!-$0.0144(25)\\   
at 1.3 GeV     &$\vev{Q_2}_0$&\ \ 0.0510(69) &\ \  0.0360(43)&\ \ 0.0415(33) &\ \  0.0291(24) &\ \  0.0301(20)\\   
$[{\rm GeV}^3]$&$\vev{Q_3}_0$&\ \ 0.004(28)  &$\!-$0.012(20) &\ \ 0.007(14)  &$\!-$0.015(11)  &\ \  0.0002(99)\\   
               &$\vev{Q_4}_0$&\ \ 0.082(27)  &\ \  0.049(18) &\ \ 0.069(13)  &\ \ 0.035(10)  &\ \  0.0460(92)\\   
               &$\vev{Q_5}_0$&$\!-$0.026(26) &$\!-$0.032(16) &$\!-$0.033(12) &$\!-$0.0253(94)& $\!-$0.0187(82)\\   
               &$\vev{Q_6}_0$&$\!-$0.012(48) &$\!-$0.111(24) &$\!-$0.071(18) &$\!-$0.115(12) & $\!-$0.0960(90)\\   
               &$\vev{Q_7}_0$&\ \  0.797(17) &\ \ 1.021(18)  &\ \ 1.269(21)  &\ \  1.417(21) &\ \  1.640(23)\\    
               &$\vev{Q_8}_0$&\ \  3.428(69) &\ \ 4.374(73)  &\ \ 5.469(86)  &\ \  6.046(87) &\ \  7.024(94)\\    
               &$\vev{Q_9}_0$&$\!-$0.0453(70)&$\!-$0.0287(43)&$\!-$0.0341(34)&$\!-$0.0205(24)& $\!-$0.0212(21)\\   
               &$\vev{Q_{10}}_0$&\ \ 0.0347(68)&\ \ 0.0306(48) &\ \ 0.0278(37) &\ \ 0.0275(29) &\ \ 0.0231(25)\\   
               &$\vev{Q_1}_2$&\ \ 0.01345(16)  &\ \ 0.01436(13)&\ \ 0.01524(11)&\ \ 0.01578(10)&\ \ 0.016361(91)\\    
               &$\vev{Q_2}_2$&\ \ 0.01328(16)  &\ \ 0.01417(12)&\ \ 0.01504(11)&\ \ 0.015571(99)&\ \ 0.016137(91)\\    
               &$\vev{Q_3}_2$&$\!-$0.00002740(31)&$\!-$0.00003058(27)&$\!-$0.00003395(25)&$\!-$0.00003677(24)&$\!-$0.00004007(23)\\ 
               &$\vev{Q_4}_2$&$\!-$0.0002198(36) &$\!-$0.0002349(30) &$\!-$0.0002521(25) &$\!-$0.0002652(21) &$\!-$0.0002830(20)\\  
               &$\vev{Q_5}_2$&\ \  0.0002056(37)&\ \ 0.0002196(31)   &\ \ 0.0002357(25)  &\ \ 0.0002483(22)  &\ \ 0.0002656(20)\\   
               &$\vev{Q_6}_2$&\ \  0.000758(14) &\ \ 0.000789(11)    &\ \ 0.0008274(91)  &\ \ 0.0008517(77)  &\ \ 0.0008913(70)\\   
               &$\vev{Q_7}_2$&\ \  0.2045(36)   &\ \ 0.2243(30)      &\ \ 0.2466(25)     &\ \ 0.2655(22)    &\ \ 0.2897(21)\\    
               &$\vev{Q_8}_2$&\ \  0.846(16)    &\ \ 0.880(13)       &\ \ 0.922(10)      &\ \ 0.9488(86)    &\ \ 0.9922(79)\\    
               &$\vev{Q_9}_2$&\ \  0.02026(24)  &\ \ 0.02161(19)     &\ \ 0.02295(16)    &\ \ 0.02376(15)   &\ \ 0.02464(14)\\     
               &$\vev{Q_{10}}_2$&\ \ 0.02006(24)&\ \ 0.02141(19)     &\ \ 0.02272(16)    &\ \ 0.02353(15)   &\ \ 0.02439(14)\\     
\hline
\end{tabular}
\end{center}
\caption{Hadronic matrix elements $\vev{Q_i}_0$ and $\vev{Q_i}_2$ 
($i=1,\cdots 10$) in units of ${\rm GeV}^3$ at each $m_fa$ 
on a $16^3\times 32$ lattice.  The upper half of the table 
lists the bare values.  The lower half are those 
renormalized in the $\ovl{\rm MS}$ scheme at $\mu=1/a$ and run to 
$\mu=1.3$ GeV for $N_f=3$ using $\Lambda_{\overline{\rm MS}}^{(3)}=372$ MeV, 
which corresponds to $\Lambda_{\overline{\rm MS}}^{(4)}=325$ MeV. }
\label{HMEmf-16}
\end{table}

\begin{table}
\begin{center}
\begin{tabular}{cclllll}
\hline
&$m_fa$& \multicolumn{1}{c}{0.02}& \multicolumn{1}{c}{0.03}& \multicolumn{1}{c}{0.04}
      & \multicolumn{1}{c}{0.05}& \multicolumn{1}{c}{0.06}\\
\hline
bare           &$\vev{Q_1}_0$&$\!-$0.0235(34) &$\!-$0.0205(37)&$\!-$0.0217(25)&$\!-$0.0161(22)& $\!-$0.0196(19)\ \\
$[{\rm GeV}^3]$&$\vev{Q_2}_0$&\ \ 0.0460(28)&\ \ 0.0457(27)&\ \ 0.0378(19)&\ \ 0.0351(16)&\ \ 0.0345(11)\\
               &$\vev{Q_3}_0$&\ \ 0.013(12)&\ \ 0.021(13)&\ \ 0.0010(98)&\ \ 0.0116(81)&\ \ 0.0003(68)\\
               &$\vev{Q_4}_0$&\ \ 0.082(11)&\ \ 0.087(12)&\ \ 0.0600(91)&\ \ 0.0627(74)&\ \ 0.0544(58)\\
               &$\vev{Q_5}_0$& $\!-$0.027(14)& $\!-$0.044(13)& $\!-$0.0515(97)&$\!-$0.0393(82)&$\!-$0.0442(65) \\
               &$\vev{Q_6}_0$& $\!-$0.105(26) & $\!-$0.183(24)& $\!-$0.167(17) &$\!-$0.136(14)& $\!-$0.117(11)\\
               &$\vev{Q_7}_0$&\ \ 1.697(15)&\ \ 2.157(29)&\ \ 2.563(35)&\ \ 2.990(40)&\ \ 3.295(44)\\
               &$\vev{Q_8}_0$&\ \ 5.211(44)&\ \ 6.584(85)&\ \ 7.84(11) &\ \ 9.13(12) &\,10.08(13)\\
               &$\vev{Q_9}_0$&$\!-$0.0417(28)& $\!-$0.0412(28)&$\!-$0.0324(20)&  $\!-$0.0299(17)& $\!-$0.0295(12)\\
               &$\vev{Q_{10}}_0$&\ \ 0.0278(34)&\ \ 0.0250(38)&\ \ 0.0267(26)&\ \  0.0212(22)&\ \ 0.0246(19)\\
               &$\vev{Q_1}_2$&\ \ 0.013154(43)&\ \ 0.014163(52)&\ \ 0.014781(48)&\ \ 0.015335(45)&\ \ 0.015853(43)\\
               &$\vev{Q_2}_2$&\ \ 0.013154(43)&\ \ 0.014163(52)&\ \ 0.014781(48)&\ \ 0.015335(45)&\ \ 0.015853(43)\\
               &$\vev{Q_7}_2$&\ \ 0.3996(15)&\ \ 0.4222(18)&\ \ 0.4559(15)&\ \ 0.4900(14)&\ \ 0.5184(13)\\
               &$\vev{Q_8}_2$&\ \ 1.2119(48)&\ \ 1.2444(55)&\ \ 1.3128(45)&\ \ 1.3783(41)&\ \ 1.4271(40)\\
               &$\vev{Q_9}_2$&\ \ 0.019730(65)&\ \ 0.021244(78)&\ \ 0.022172(72)&\ \ 0.023003(67)&\ \ 0.023779(65)\\
               &$\vev{Q_{10}}_2$&\ \ 0.019730(65)&\ \ 0.021244(78)&\ \ 0.022172(72)&\ \ 0.023003(67)&\ \ 0.023779(65)\\
\hline\hline
&$m_fa$& \multicolumn{1}{c}{0.02}& \multicolumn{1}{c}{0.03}& \multicolumn{1}{c}{0.04}
      & \multicolumn{1}{c}{0.05}& \multicolumn{1}{c}{0.06}\\
\hline
renormalized   &$\vev{Q_1}_0$&$\!-$0.0203(34)&$\!-$0.0173(36)&$\!-$0.0182(25)&$\!-$0.0130(21)&$\!-$0.0163(19)\\  
at 1.3 GeV     &$\vev{Q_2}_0$&\ \ 0.0433(33) &\ \ 0.0401(34) &\ \ 0.0322(23)&\ \  0.0309(19) &\ \  0.0310(14)\\  
$[{\rm GeV}^3]$&$\vev{Q_3}_0$&\ \ 0.015(14)  &\ \ 0.017(14)  &$\!-$0.004(11)&\ \ 0.0084(88)  &$\!-$0.0014(73)\\  
               &$\vev{Q_4}_0$&\ \ 0.078(13)  &\ \ 0.076(14)  &\ \ 0.048(10) &\ \ 0.0535(84)  &\ \  0.0466(65)\\  
               &$\vev{Q_5}_0$&$\!-$0.008(12) &$\!-$0.011(11) &$\!-$0.0214(90)&$\!-$0.0144(74)&$\!-$0.0231(57)\\  
               &$\vev{Q_6}_0$&$\!-$0.072(20) &$\!-$0.132(19) &$\!-$0.120(14) &$\!-$0.095(11) &$\!-$0.0790(84)\\  
               &$\vev{Q_7}_0$&\ \ 0.8415(80) &\ \ 1.072(15)  &\ \ 1.271(17)  &\ \  1.488(19) &\ \  1.637(21)\\   
               &$\vev{Q_8}_0$&\ \ 3.631(33)  &\ \ 4.566(60)  &\ \ 5.434(70)  &\ \  6.352(79) &\ \  7.002(86)\\   
               &$\vev{Q_9}_0$&$\!-$0.0376(33)&$\!-$0.0338(34)&$\!-$0.0247(24)&$\!-$0.0233(20)&$\!-$0.0231(14)\\  
               &$\vev{Q_{10}}_0$&\ \ 0.0259(34)&\ \ 0.0234(37)&\ \ 0.0256(26)&\ \ 0.0205(21) &\ \ 0.0239(19)\\  
               &$\vev{Q_1}_2$&\ \ 0.013469(44) &\ \ 0.014499(53) &\ \ 0.015140(49)&\ \ 0.015717(46)&\ \ 0.016253(44)\\  
               &$\vev{Q_2}_2$&\ \ 0.013295(44) &\ \ 0.014317(53) &\ \ 0.014944(49)&\ \ 0.015507(45)&\ \ 0.016031(43)\\  
               &$\vev{Q_3}_2$&$\!-$0.00002694(11)&$\!-$0.00003018(14)&$\!-$0.00003331(13)&$\!-$0.00003662(13)&$\!-$0.00003960(13)\\
               &$\vev{Q_4}_2$&$\!-$0.0002180(15) &$\!-$0.0002301(17) &$\!-$0.0002476(15) &$\!-$0.0002660(13) &$\!-$0.0002807(13)\\ 
               &$\vev{Q_5}_2$&\ \ 0.0002037(15) &\ \ 0.0002142(17)&\ \ 0.0002312(15)&\ \ 0.0002492(13)&\ \ 0.0002634(13)\\ 
               &$\vev{Q_6}_2$&\ \ 0.0007562(57) &\ \ 0.0007728(63)&\ \ 0.0008144(54) &\ \  0.0008575(48)&\ \ 0.0008866(46)\\ 
               &$\vev{Q_7}_2$&\ \ 0.2010(15)    &\ \ 0.2180(17)   &\ \ 0.2409(15)  &\ \  0.2658(14)   &\ \ 0.2866(14)\\  
               &$\vev{Q_8}_2$&\ \ 0.8440(64)    &\ \ 0.8618(70)   &\ \ 0.9078(60)  &\ \  0.9552(53)   &\ \ 0.9872(51)\\  
               &$\vev{Q_9}_2$&\ \ 0.020281(66)  &\ \ 0.021830(80) &\ \ 0.022796(74)&\ \ 0.023666(69)  &\ \ 0.024474(66)\\   
               &$\vev{Q_{10}}_2$&\ \ 0.020083(66)&\ \ 0.021623(80)&\ \ 0.022575(73)&\ \ 0.023431(68)  &\ \ 0.024228(66)\\   
\hline
\end{tabular}
\end{center} 
\caption{Same as Table~\ref{HMEmf-16} for the $24^3\times 32$ lattice.}
\label{HMEmf-24}
\end{table}

\begin{table}
\begin{center}
\begin{tabular}{ccll}
\hline
&\ \ \ &\hspace*{0.3cm} $q^* =1/a$\hspace*{1cm} &\hspace*{0.5cm} $q^* =\pi/a$\hspace*{1cm} \\
\hline
$\vev{Q_1}_0$ & & $\!-$0.0203(34)& $\!-$0.0152(33)\\
$\vev{Q_2}_0$ & &\ \ 0.0433(33)&\ \ 0.0424(33)\\
$\vev{Q_3}_0$ & &\ \ 0.015(14)&\ \  0.019(14)\\
$\vev{Q_4}_0$ & &\ \ 0.078(13)& \ \ 0.076(13)\\
$\vev{Q_5}_0$ & &  $\!-$0.008(12)&  $\!-$0.005(12)\\
$\vev{Q_6}_0$ & &  $\!-$0.072(20)&  $\!-$0.050(16)\\
$\vev{Q_7}_0$ & &\ \ 0.8415(80)&\ \  0.7986(77)\\
$\vev{Q_8}_0$ & &\ \ 3.631(33)&\ \  2.873(26)\\
$\vev{Q_9}_0$ & & $\!-$0.0376(33)& $\!-$0.0317(33)\\
$\vev{Q_{10}}_0$& &\ \ 0.0259(34)& \ \ 0.0257(34)\\
$\vev{Q_1}_2$ & &\ \  0.013469(44)&\ \ 0.014314(46)\\
$\vev{Q_2}_2$ & &\ \  0.013295(44)&\ \  0.013760(45)\\
$\vev{Q_7}_2$ & &\ \  0.2010(15)&\ \    0.1912(14)\\
$\vev{Q_8}_2$ & &\ \   0.8440(64)&\ \   0.6678(51)\\
$\vev{Q_9}_2$ & &\ \   0.020281(66)&\ \ 0.021754(70)\\
$\vev{Q_{10}}_2$& &\ \ 0.020083(66)&\ \ 0.021149(69)\\
\hline
\end{tabular}
\end{center}
\caption{Renormalized hadronic matrix elements at $\mu = 1.3$ GeV
in units of ${\rm GeV}^3$ from different
matching point $q^*=1/a$ (left column) and
$\pi/a$ (right column). Values are taken 
at $m_fa=0.02$ on a $24^3\times 32$ lattice. }
\label{matching-diff}
\end{table}

\begin{table}
\begin{center}
\begin{tabular}{llc|lc}
\hline 
&\ \ quadratic\ &$\chi^2/$dof\ &\ \ chiral log.\ &$\chi^2/$dof\ \ \\
\hline 
$\vev{Q_1}_0$ &$\!-$0.034(20) & 0.14 &$\!-$0.035(36) &0.14  \\ 
$\vev{Q_2}_0$ &\ \ 0.070(19)  & 3.03 &\ \ 0.083(34)  &3.05  \\ 
$\vev{Q_3}_0$ &\ \ 0.033(82)  & 0.91 &\ \ 0.06(15)   &0.94  \\ 
$\vev{Q_4}_0$ &\ \ 0.131(78)  & 1.68 &\ \ 0.17(14)   &1.71  \\ 
$\vev{Q_5}_0$ &$\!-$0.008(72) & 0.03 &\ \ 0.03(13)   &0.02  \\ 
$\vev{Q_6}_0$ &\ \ 0.08(12)   & 2.64 &\ \ 0.20(21)   &2.63  \\ 
$\vev{Q_6}_0$ (4pts.)&$\!-$0.04(17)&4.32  &$\!-$0.02(31)&4.38\\ 
$\vev{Q_7}_0$ &\ \ 0.247(78)  & 1.78 &\ \ 0.11(15)   &1.69 \\ 
$\vev{Q_8}_0$ &\ \ 1.07(32)   & 2.87 &\ \ 0.48(60)   &2.77 \\ 
$\vev{Q_9}_0$ &$\!-$0.067(19) & 3.32 &$\!-$0.082(35) &3.35 \\ 
$\vev{Q_{10}}_0$&\ \ 0.037(21)& 0.17 &\ \ 0.037(37)  &0.17 \\ 
$\vev{Q_1}_2$ &\ \ 0.01102(54)& 0.34 &\ \ 0.00990(98)&0.29 \\ 
$\vev{Q_2}_2$ &\ \ 0.01087(54)& 0.33 &\ \ 0.00975(97)&0.28 \\ 
$\vev{Q_3}_2$ &$\!-$0.0000203(12)&0.49&$\!-$0.0000195(21)&0.46\\ 
$\vev{Q_4}_2$ &$\!-$0.000188(12) &0.33&$\!-$0.000187(22) &0.37\\ 
$\vev{Q_5}_2$ &\ \ 0.000177(12)  &0.33&\ \ 0.000179(23)  &0.34\\ 
$\vev{Q_6}_2$ &\ \ 0.000694(46)  &0.31&\ \ 0.000694(83)  &0.31\\ 
$\vev{Q_7}_2$ &\ \ 0.164(12)     &0.36&\ \ 0.167(23)     &0.38\\ 
$\vev{Q_8}_2$ &\ \ 0.776(51)     &0.30&\ \ 0.775(92)     &0.31\\ 
$\vev{Q_9}_2$ &\ \ 0.01660(81)   &0.34&\ \ 0.0149(15)    &0.29\\ 
$\vev{Q_{10}}_2$&\ \ 0.01642(81) &0.33&\ \ 0.0147(15)    &0.29\\ 
\hline
\end{tabular}
\end{center}
\caption{Hadronic matirx elements in units of ${\rm GeV}^3$
 in the chiral limit $m_M^2\to 0$ on a $16^3\times 32$ lattice. 
 The column named ``quadratic'',``chilal log.''or ``(quadratic) + breaking''
 corresponds to three types of fit forms described in the text.
 Chiral extrapolations are made using data
 at all $m_fa=0.02$--$0.06$ (5 points) except for an alternative 
 extraporation of $\vev{Q_6}_0$ exclluding the point at $m_fa=0.02$(4 points).}
\label{HME0-16}
\end{table}

\begin{table}
\begin{center}
\begin{tabular}{llc|lc}
\hline 
& quadratic & $\chi^2/$dof &\ chiral log. & $\chi^2/$dof\\ 
\hline 
$\vev{Q_1}_0$&$\!-$0.031(12)&   0.98 &$\!-$0.039(21)  &  1.04 \\ 
$\vev{Q_2}_0$&\ \  0.066(11)&   0.36 &\ \ 0.082(20)   &  0.43 \\
$\vev{Q_3}_0$&\ \  0.032(48)&   0.83 &\ \ 0.044(86)   &  0.82 \\
$\vev{Q_4}_0$&\ \  0.124(45)&   0.69 &\ \ 0.155(82)   &  0.69 \\ 
$\vev{Q_5}_0$&\ \  0.001(40)&   0.44 &\ \ 0.003(73)   &  0.43 \\ 
$\vev{Q_6}_0$&\ \  0.014(66)&   1.53 &\ \ 0.16(12)    &  1.20 \\ 
$\vev{Q_6}_0$(4pts.)&$\!-$0.19(13)& 0.10 &$\!-$0.18(25)&0.10\\ 
$\vev{Q_7}_0$&\ \  0.252(53)&   0.43 &\ \ 0.07(11)    &  0.45\\ 
$\vev{Q_8}_0$&\ \  1.23(22) &   0.27 &\ \ 0.58(43)    &  0.35\\ 
$\vev{Q_9}_0$&$\!-$0.063(11)&   0.45 &$\!-$0.082(20)  &  0.53\\ 
$\vev{Q_{10}}_0$&\ \ 0.034(12)& 1.09 &\ \ 0.039(22)   & 1.14\\ 
$\vev{Q_1}_2$&\ \  0.01104(19)& 4.55 &\ \ 0.00979(36) & 2.99\\ 
$\vev{Q_2}_2$&\ \  0.01089(19)& 4.79 &\ \ 0.00964(35) & 3.16\\ 
$\vev{Q_3}_2$&$\!-$0.00001942(50)&0.25&$\!-$0.00001832(96)&0.18\\ 
$\vev{Q_4}_2$&$\!-$0.0001848(59)& 1.28 &$\!-$0.000187(11)&1.25\\ 
$\vev{Q_5}_2$&\ \  0.0001737(61)& 1.67 &\ \ 0.000179(11) &1.57\\ 
$\vev{Q_6}_2$&\ \  0.000691(22) & 2.08  &\ \ 0.000708(42) &1.98\\ 
$\vev{Q_7}_2$&\ \  0.1580(60)   & 1.27 &\ \ 0.163(11)    &1.18\\ 
$\vev{Q_8}_2$&\ \  0.772(25)    & 2.09 &\ \ 0.792(47)    &1.99\\ 
$\vev{Q_9}_2$&\ \  0.01663(28)  & 4.48 &\ \ 0.01476(54)  &2.94\\ 
$\vev{Q_{10}}_2$&\ \ 0.01646(28)& 4.66 &\ \ 0.01458(53)  & 3.07\\ 
\hline
\end{tabular}
\end{center}
\caption{Same as Table~\ref{HME0-16} for the $24^3\times 32$ 
lattice.}
\label{HME0-24}
\end{table}


\begin{table}
\begin{center}
\begin{tabular}{cll|ll}
\hline
& {$16^3\times 32$}& & {$24^3\times 32$}& \\
&{quadratic} &{chiral log.} &{\ quadratic} &{chiral log.} \\
\hline
$B^{(1/2)}_1$&\ \ 8.3(5.0) &\ \ 8.6(8.9)   &\ \ 7.7(2.9) &\ \ 9.6(5.2) \\
$B^{(1/2)}_2$&\ \ 3.43(95) &\ \ 4.1(1.7)   &\ \ 3.23(55) &\ \ 4.04(98) \\ 
$B^{(1/2)}_3$&\ \ 2.7(6.7) &\ \ 5(12)      &\ \ 2.6(3.9) &\ \ 3.6(7.1) \\
$B^{(1/2)}_4$&\ \ 3.6(2.1) &\ \ 4.5(3.8)   &\ \ 3.4(1.2) &\ \ 4.3(2.3) \\
$B^{(1/2)}_5$&\ \ 0.04(40) &$\!-$0.15(71)  &\ \ 0.01(22) &$\!-$0.02(41)\\
$B^{(1/2)}_6$&$\!-$0.14(22)  &$\!-$0.38(38)&$\!-$0.03(12)&$\!-$0.29(22)\\ 
$B^{(1/2)}_6$(4pts.)&\ \ 0.07(31)&\ \ 0.03(58)&\ \ 0.35(25)&\ \ 0.34(47)\\ 
$B^{(1/2)}_7$&\ \ 0.49(15) &\ \ 0.22(29)   &\ \ 0.50(10) &\ \ 0.14(21) \\ 
$B^{(1/2)}_8$&\ \ 0.73(22) &\ \ 0.32(41)   &\ \ 0.83(15) &\ \ 0.39(29) \\
$B^{(1/2)}_9$&\ \ 5.5(1.6) &\ \ 6.8(2.8)   &\ \ 5.19(92) &\ \ 6.7(1.7) \\ 
$B^{(1/2)}_{10}$&\ \ 3.0(1.7) &\ \ 3.0(3.0)&\ \ 2.78(98) &\ \ 3.2(1.8) \\ 
$B^{(3/2)}_1$&\ \ 0.480(24)&\ \ 0.431(43)  &\ \ 0.4809(82)&\ \ 0.426(16)\\ 
$B^{(3/2)}_2$&\ \ 0.473(23)&\ \ 0.425(42)  &\ \ 0.4745(81)&\ \ 0.420(15)\\ 
$B^{(3/2)}_7$&\ \ 0.640(49)&\ \ 0.651(88)  &\ \ 0.616(23) &\ \ 0.634(44)\\
$B^{(3/2)}_8$ &\ \ 0.924(61)&\ \ 0.92(11)  &\ \ 0.920(30) &\ \ 0.944(55)\\
$B^{(3/2)}_9$&\ \ 0.482(24)&\ \ 0.433(43)  &\ \ 0.4830(82)&\ \ 0.429(16)\\ 
$B^{(3/2)}_{10}$&\ \ 0.477(24)&\ \ 0.428(43)&\ \ 0.4779(82)&\ \ 0.423(15)\\
\hline
\end{tabular}
\end{center}
\caption{B parameters in the chiral limit with the chiral logarithm fit.}
\label{Bparams}
\end{table}

\begin{table}
\begin{center}
\begin{tabular}{clll|lll}
\hline 
&\multicolumn{3}{c|}{$16^3\times 32$}&\multicolumn{3}{c}{$24^3\times 32$}\\
\hline
& $\re A_0 [10^{-8}{\rm GeV}]$  & $\re A_2 [10^{-8}{\rm GeV}] $ & 
$\ \ \ \omega^{-1}\ \ \ $
& $\re A_0 [10^{-8}{\rm GeV}]$  & $\re A_2 [10^{-8}{\rm GeV}] $ & 
$\ \ \ \omega^{-1}\ \ \ $\\
\hline\hline
$\Lambda_{\overline{\rm MS}}^{(4)}=325$~MeV\\
\hline
0.02 &\ \    13.1(1.4) & 1.867(22) & 7.01(78) &\ \ \  10.80(69) & 1.8689(62) & 5.78(37)\\
0.03 &\ \ \ \ 9.45(84) & 1.992(17) & 4.75(43) &\ \ \ \ 9.90(69) & 2.0129(74) & 4.92(35)\\
0.04 &\ \    10.42(68) & 2.114(15) & 4.93(33) &\ \ \ \ 8.26(45) & 2.1006(68) & 3.93(22)\\
0.05 &\ \ \ \ 7.66(49) & 2.188(14) & 3.50(22) &\ \ \ \ 7.61(38) & 2.1792(64) & 3.49(17)\\
0.06 &\ \ \ \ 7.52(40) & 2.267(13) & 3.32(17) &\ \ \ \ 7.86(28) & 2.2527(61) & 3.49(12)\\
\hline\hline
$\Lambda_{\overline{\rm MS}}^{(4)}=215$~MeV\\
\hline
0.02 &\ \    12.5(1.4) & 1.911(23) & 6.55(72)  &\ \   10.40(66) & 1.9130(63) & 5.43(34)\\
0.03 &\ \ \ \ 9.12(81) & 2.039(18) & 4.47(40) &\ \ \ \ 9.59(66) & 2.0602(76) & 4.66(32)\\
0.04 &\ \    10.04(64) & 2.164(16) & 4.64(31) &\ \ \ \ 8.01(43) & 2.1500(70) & 3.72(20)\\
0.05 &\ \ \ \ 7.41(47) & 2.240(14) & 3.31(21) &\ \ \ \ 7.38(36) & 2.2306(65) & 3.31(16)\\
0.06 &\ \ \ \ 7.30(38) & 2.321(13) & 3.15(16) &\ \ \ \ 7.60(26) & 2.3058(63) & 3.29(12)\\
\hline\hline
$\Lambda_{\overline{\rm MS}}^{(4)}=435$~MeV\\
\hline
0.02 &\ \    13.7(1.5) & 1.821(22) & 7.52(84) &\ \    11.20(72) & 1.8228(60) & 6.14(40)\\
0.03 &\ \ \ \ 9.78(89) & 1.943(17) & 5.03(46) &\ \    10.18(73) & 1.9635(72) & 5.19(38)\\
0.04 &\ \    10.80(71) & 2.062(15) & 5.24(35) &\ \ \ \ 8.50(48) & 2.0489(67) & 4.15(23)\\
0.05 &\ \ \ \ 7.87(51) & 2.134(14) & 3.69(24) &\ \ \ \ 7.82(40) & 2.1254(62) & 3.68(19)\\
0.06 &\ \ \ \ 7.74(42) & 2.211(12) & 3.50(19) &\ \ \ \ 8.11(29) & 2.1970(60) & 3.69(13)\\
\hline
\end{tabular}
\end{center}
\caption{Values of $\re A_0$, $\re A_2$ and $\omega^{-1}$ obtained at
 each $m_fa$ for both lattice sizes,
with $\Lambda_{\overline{\rm MS}}^{(4)}$=325 MeV, 215 MeV and 435 MeV.}
\label{delI-tab}
\end{table}

\begin{table}
\begin{center}
\begin{tabular}{cc|llllc}
\hline 
\multicolumn{2}{c}{$24^3\times 32$}&\ \ $\xi_0$&\ \ $\xi_1$&\ \ $\xi_2$&
\ \ $\xi_3$& \ \ $\chi^2/$dof\\
\hline\hline
quadratic  & $\re A_0\ [10^{-8}{\rm GeV}]$&\,16.5(2.2) &$\!-$27.7(9.2)&
\ \ \ \ - &\,21.8(8.8) &0.34\\
           & $\re A_2\ [10^{-8}{\rm GeV}]$&\ \ 1.531(26) &\ \ \ 1.62(12)&
\ \ \ \ -&$\!-$0.86(13)&  4.91\\
           & $\omega^{-1}$\ \      &\ \ 9.5(1.1)& $\!-$18.2(4.6)&\ \ \ \ -&
 13.7(4.3)& 0.13\\
\hline
chiral log.& $\re A_0\ [10^{-8}{\rm GeV}]$&    20.7(4.0)  & $\!-$11.4(2.8) &  
 \,20.1(8.3) & \ \ \ \ - & 0.50 \\
           & $\re A_2\ [10^{-8}{\rm GeV}]$& \ \,1.353(50) & \ \ \ 0.977(31) 
  & $\!-$0.82(11) & \ \ \ \ - & 3.25 \\
           & $\omega^{-1}$\ \             &    12.3(2.0)  & \ \,$\!-$8.0(1.5)
  &   \,12.9(4.1) & \ \ \ \ - & 0.26\\
\hline
\end{tabular}
\end{center}
\caption{Fit parameters for $\re A_0$, $\re A_2$ and $\omega^{-1}$ with
$\xi_0 + \xi_1 m_M^2 + \xi_3 (m_M^2)^2 $ (quadratic fit) and
$\xi_0 + \xi_1 m_M^2 + \xi_2 m_M^2\ln m_M^2 $ (chiral logarithm fit).
Results on a $24^3\times 32$ lattice with 
$\Lambda_{\overline{\rm MS}}^{(4)}=325$~MeV are shown.}
\label{fitp-a0a2o24}
\end{table}

\begin{table}
\begin{center}
\begin{tabular}{clll|lll}
\hline 
&\multicolumn{3}{c}{$16^3\times 32$} &\multicolumn{3}{c}{$24^3\times 32$}\\
\hline
& $P^{(1/2)}$ & $P^{(3/2)} $ &$\varepsilon'/\varepsilon\ [10^{-4}]$
&\ \ $P^{(1/2)}$ & $P^{(3/2)} $ &$\varepsilon'/\varepsilon\ [10^{-4}]$\\
\hline\hline
$\Lambda_{\overline{\rm MS}}^{(4)}=325$~MeV\\
\hline
0.02 & 0.1(1.2)   & 4.93(11) & $\!-$6.3(1.5) &\ \ 1.69(50) &\ \ 4.923(45)&\ \ $\!-$4.21(64)\\ 
0.03 & 2.97(61)   & 5.084(88)& $\!-$2.74(78) &\ \ 3.36(47) &\ \ 4.944(49)&\ \ $\!-$2.06(60)\\ 
0.04 & 2.19(44)   & 5.291(70)& $\!-$4.03(56) &\ \ 3.50(33) &\ \ 5.200(41)&\ \ $\!-$2.21(41)\\ 
0.05 & 3.65(29)   & 5.416(59)& $\!-$2.30(37) &\ \ 3.19(26) &\ \ 5.470(37)&\ \ $\!-$2.96(33)\\ 
0.06 & 3.42(22)   & 5.657(53)& $\!-$2.90(28) &\ \ 3.01(19) &\ \ 5.632(35)&\ \ $\!-$3.41(24)\\ 
\hline\hline
$\Lambda_{\overline{\rm MS}}^{(4)}=215$~MeV\\
\hline
0.02 & 0.06(94)   & 3.713(87) & $\!-$4.7(1.2) &\ \ 1.34(41) &\ \ 3.707(36)&\ \ $\!-$3.07(52)\\
0.03 & 2.38(50)   & 3.815(70) & $\!-$1.86(63) &\ \ 2.70(38) &\ \ 3.701(39)&\ \ $\!-$1.31(49)\\
0.04 & 1.74(36)   & 3.962(56) & $\!-$2.89(45) &\ \ 2.81(26) &\ \ 3.892(33)&\ \ $\!-$1.40(33)\\
0.05 & 2.93(24)   & 4.049(47) & $\!-$1.45(30) &\ \ 2.56(21) &\ \ 4.094(27)&\ \ $\!-$1.99(27)\\
0.06 & 2.75(18)   & 4.228(42) & $\!-$1.92(22) &\ \ 2.41(16) &\ \ 4.211(28)&\ \ $\!-$2.34(19)\\
\hline\hline
$\Lambda_{\overline{\rm MS}}^{(4)}=435$~MeV\\
\hline
0.02 & 0.1(1.4)  & 6.16(13) & $\!-$7.8(1.8)  &\ \ 2.05(61) &\ \ 6.150(54)&\ \ $\!-$5.33(78)\\
0.03 & 3.63(75)  & 6.36(11) & $\!-$3.56(95)  &\ \ 4.09(58) &\ \ 6.197(58)&\ \ $\!-$2.74(73)\\
0.04 & 2.67(54)  & 6.629(84)& $\!-$5.15(68)  &\ \ 4.26(40) &\ \ 6.518(50)&\ \ $\!-$2.93(50)\\
0.05 & 4.43(36)  & 6.790(71)& $\!-$3.06(46)  &\ \ 3.88(32) &\ \ 6.853(44)&\ \ $\!-$3.87(40)\\
0.06 & 4.16(27)  & 7.091(64)& $\!-$3.82(34)  &\ \ 3.65(24) &\ \ 7.059(42)&\ \ $\!-$4.43(29)\\
\hline
\end{tabular}
\end{center}
\caption{Values of $P^{(1/2)}$, $P^{(3/2)}$ and $\varepsilon'/\varepsilon$ 
at each $m_fa$ for both lattice volumes,
with $\Lambda_{\overline{\rm MS}}^{(4)}$=325 MeV, 215 MeV and 435 MeV.}
\label{epep-tab}
\end{table}


\clearpage
%

\begin{figure}
 \begin{center}
  \leavevmode
\hspace*{-0.7cm}
  \includegraphics[width=8cm, clip]{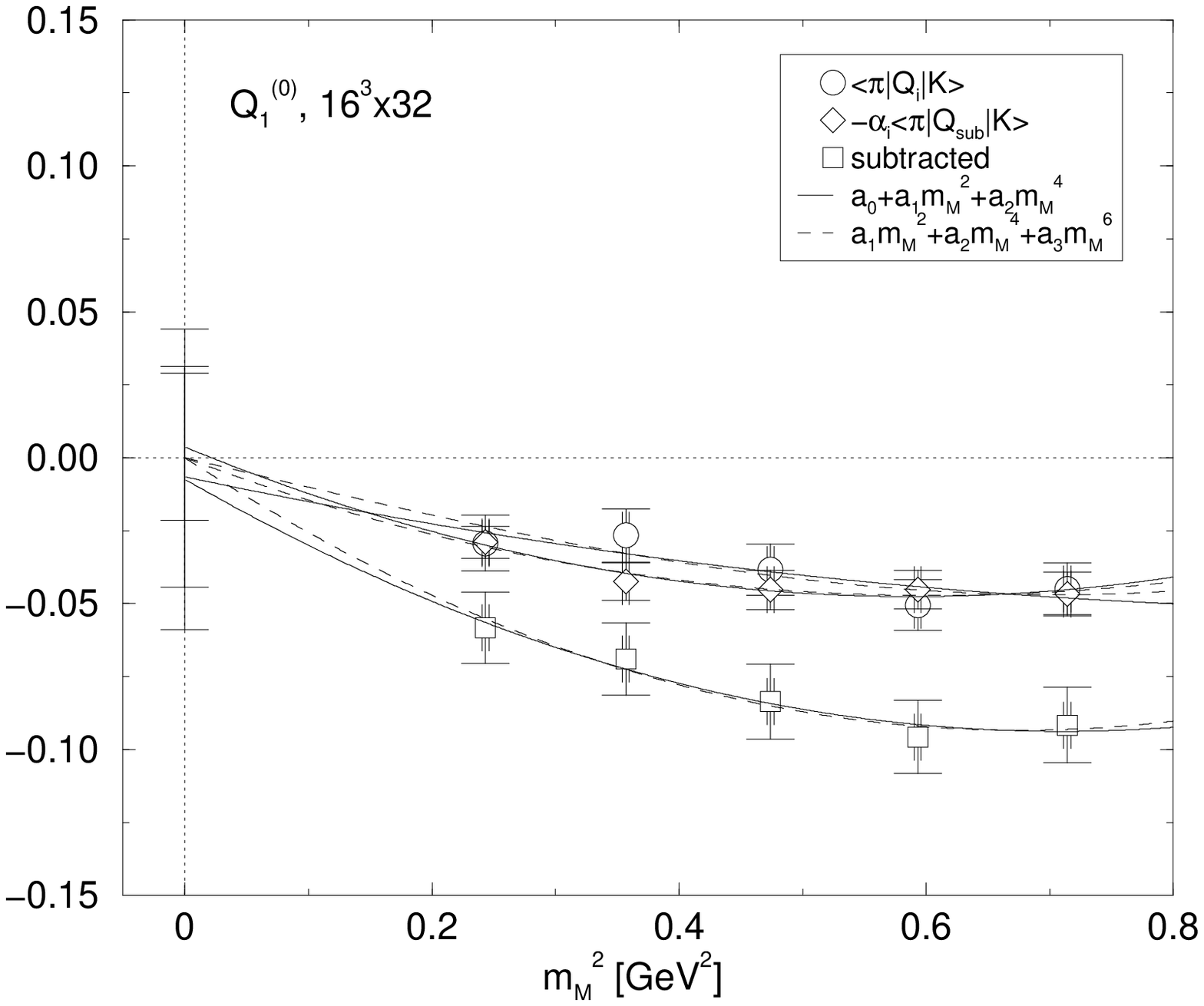}
\hspace{0.4cm}
  \includegraphics[width=8cm, clip]{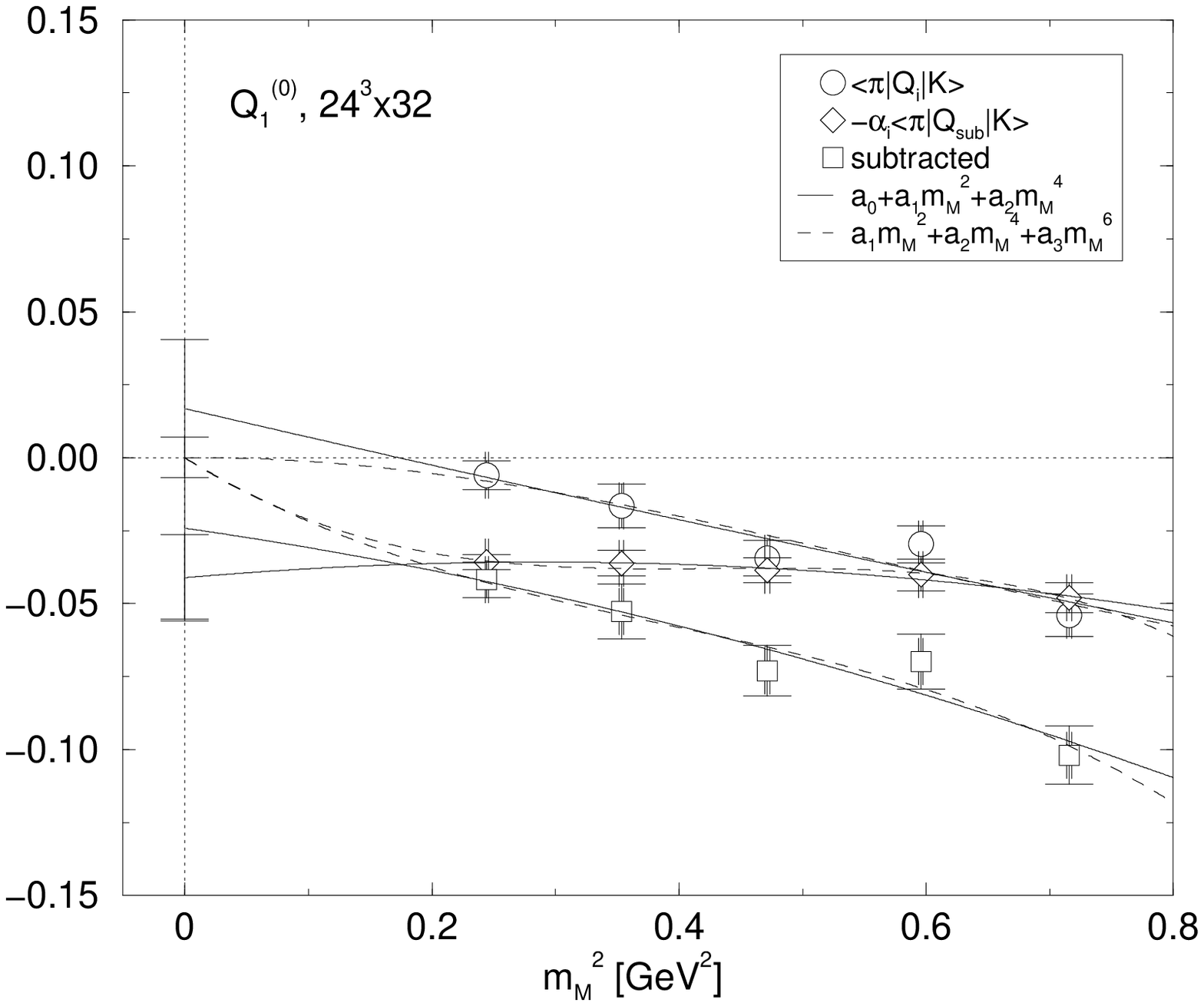}
 \end{center}
 \begin{center}
  \leavevmode
  \includegraphics[width=8.3cm, clip]{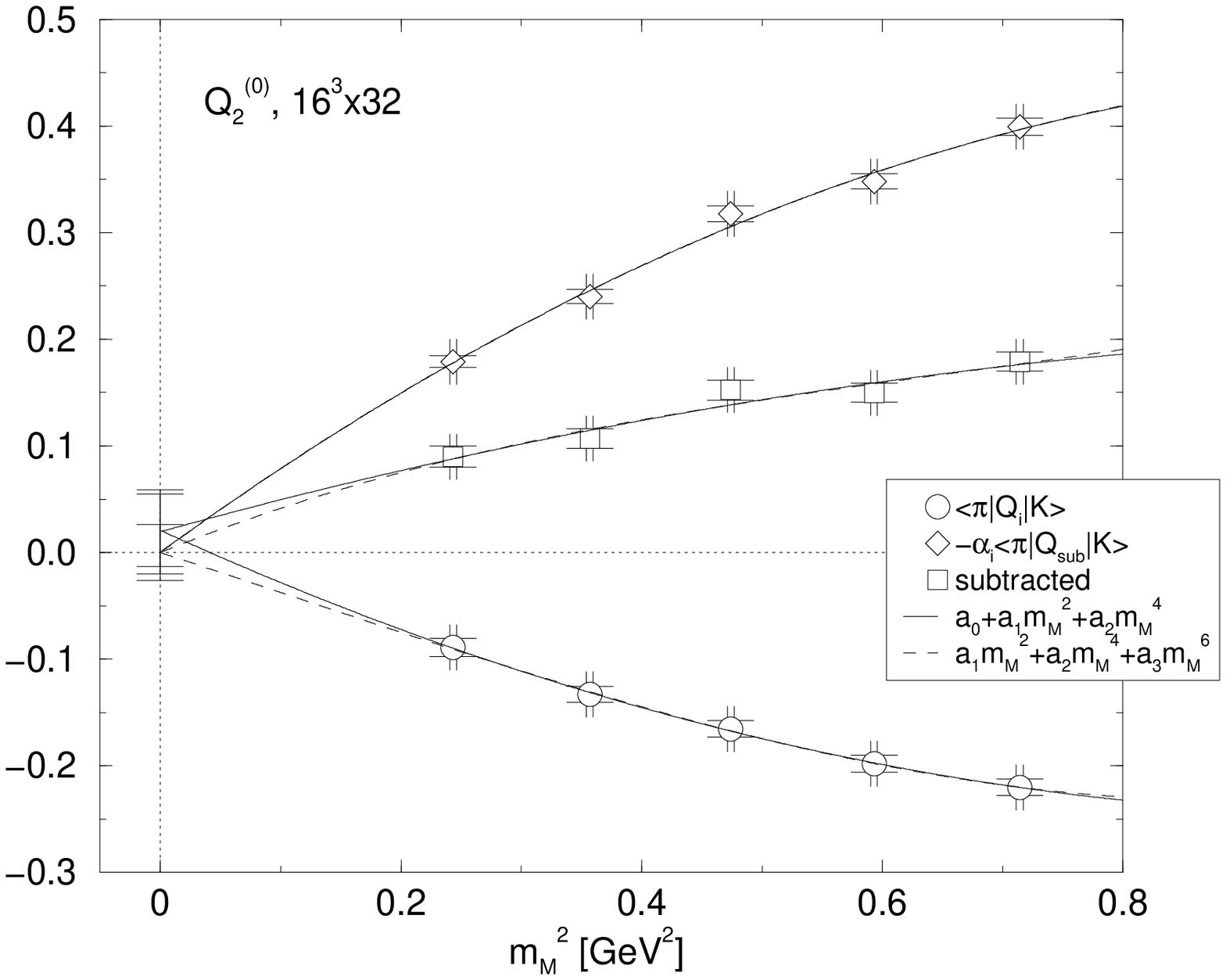}
\hspace{0.1cm}
  \includegraphics[width=8.3cm, clip]{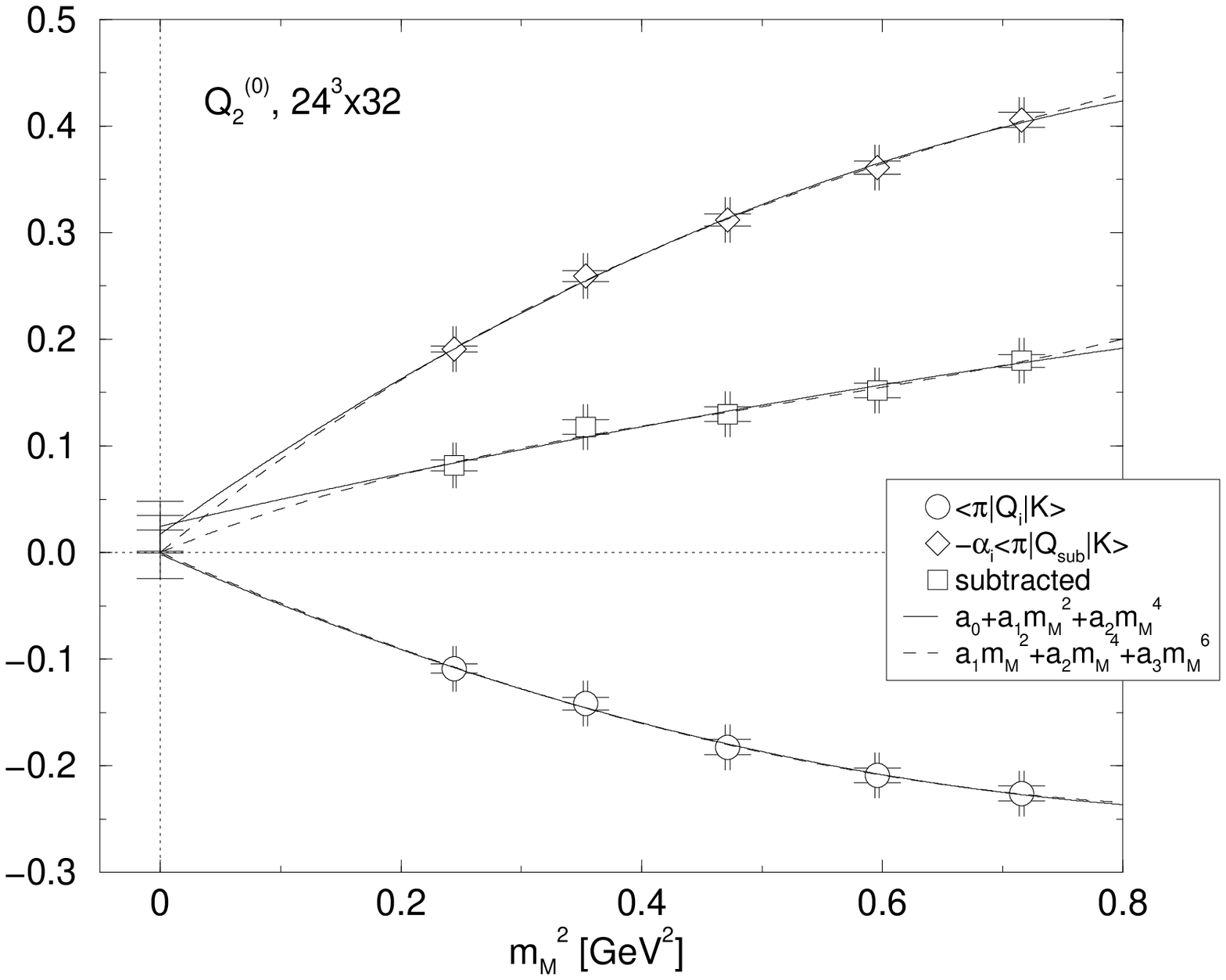}
 \end{center}
 \begin{center}
  \leavevmode
  \includegraphics[width=8.3cm, clip]{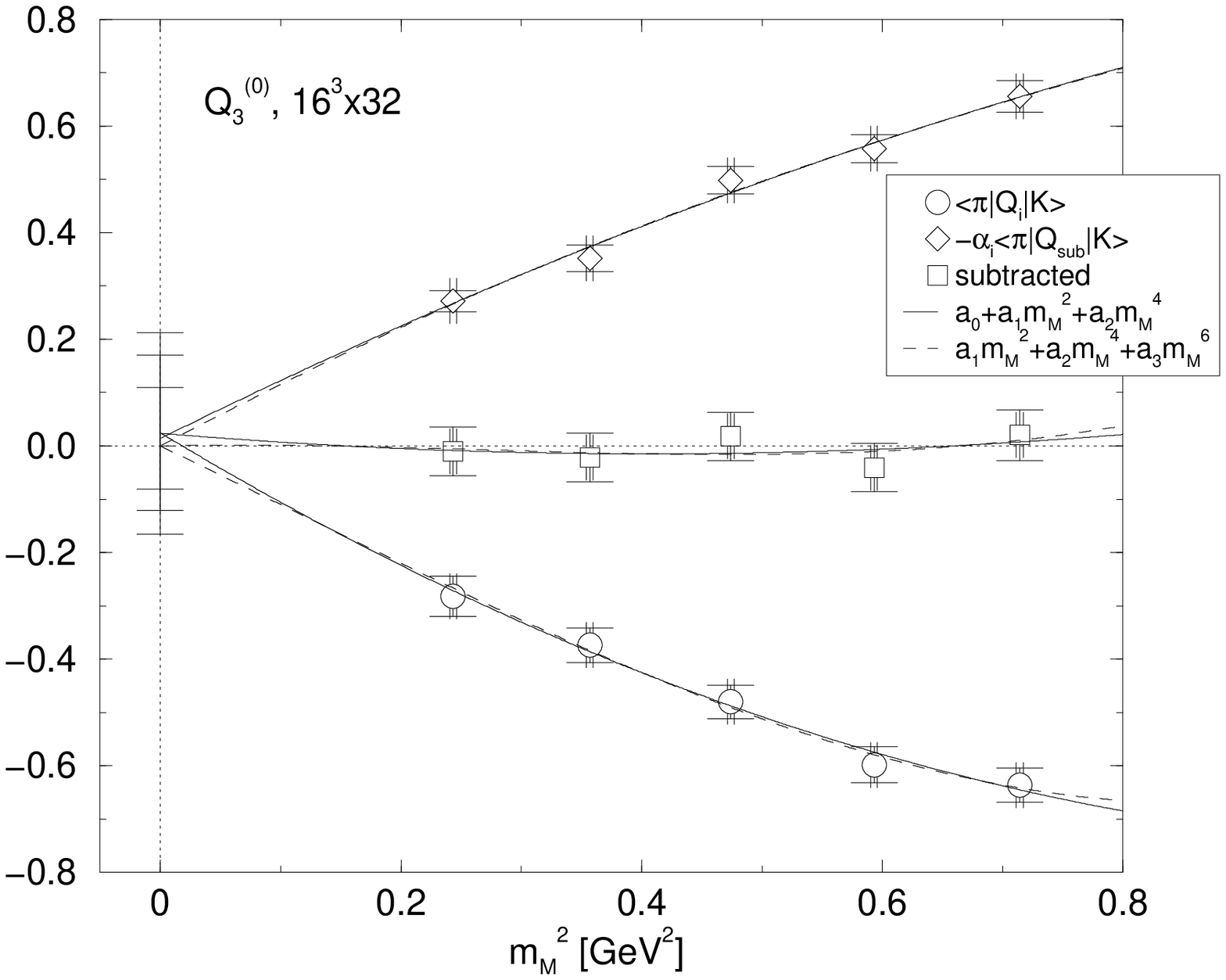}
\hspace{0.1cm}
  \includegraphics[width=8.3cm, clip]{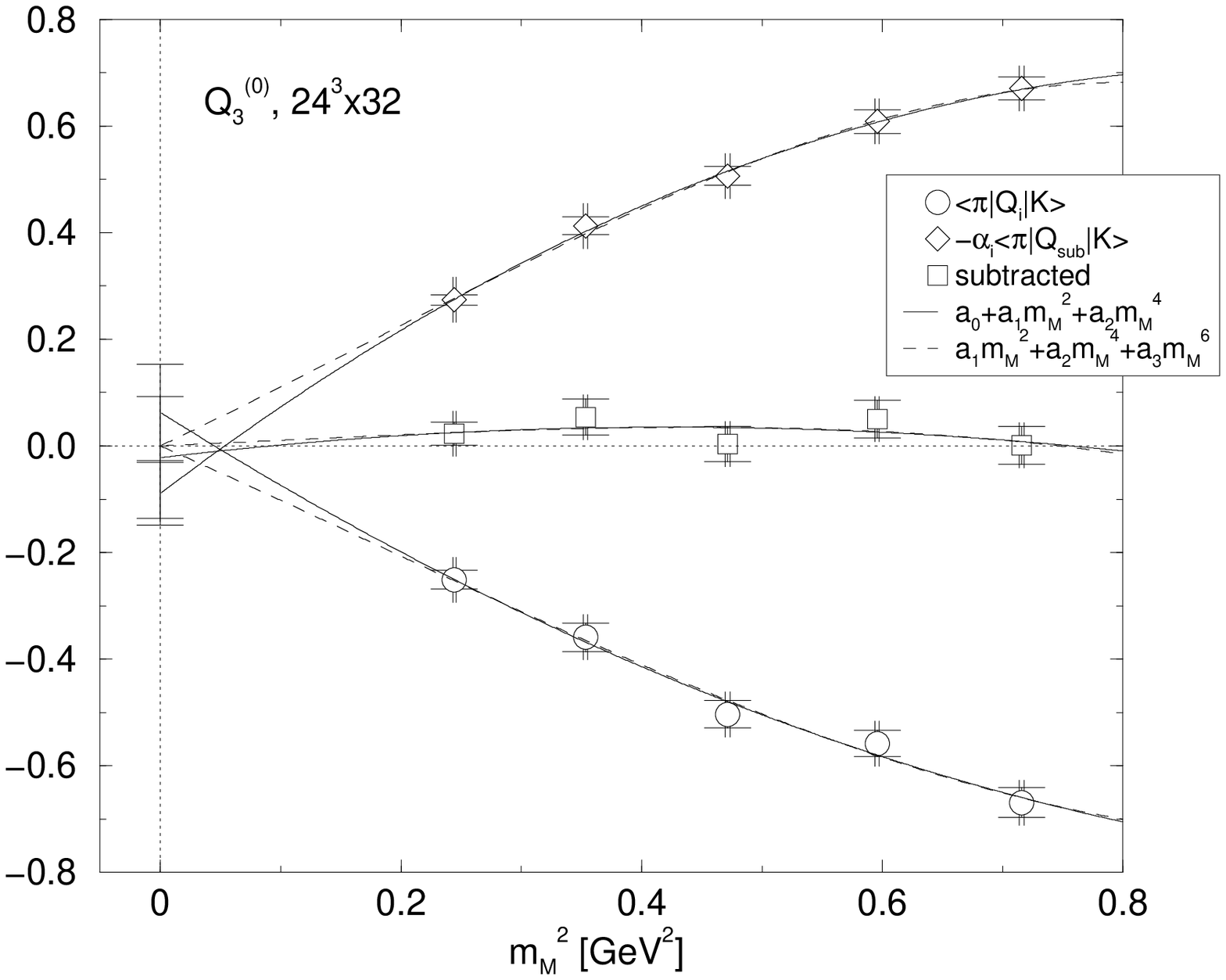}
 \end{center}
\end{figure}

\begin{figure}
 \begin{center}
  \leavevmode
  \includegraphics[width=8.3cm, clip]{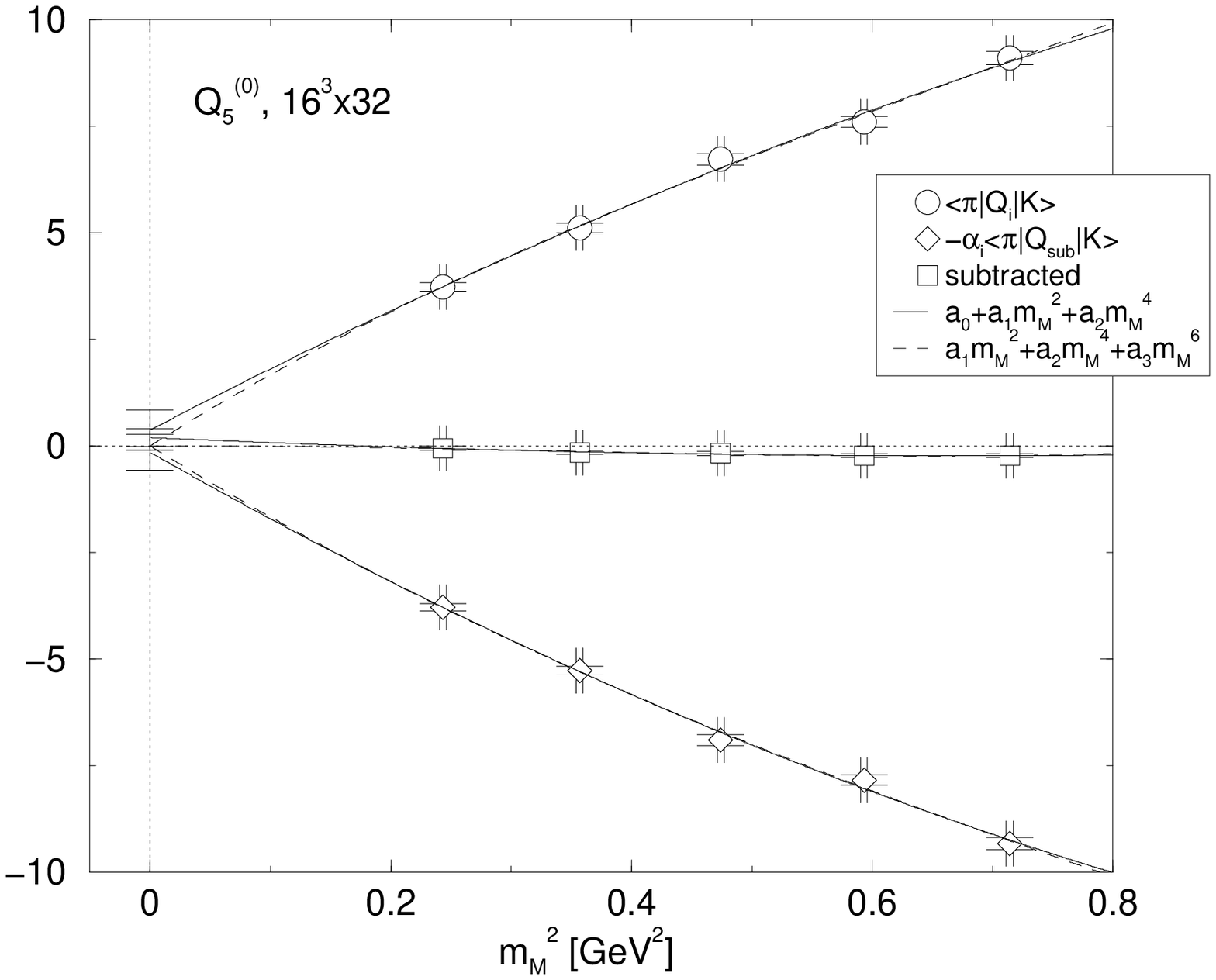}
\hspace{0.1cm}
  \includegraphics[width=8.3cm, clip]{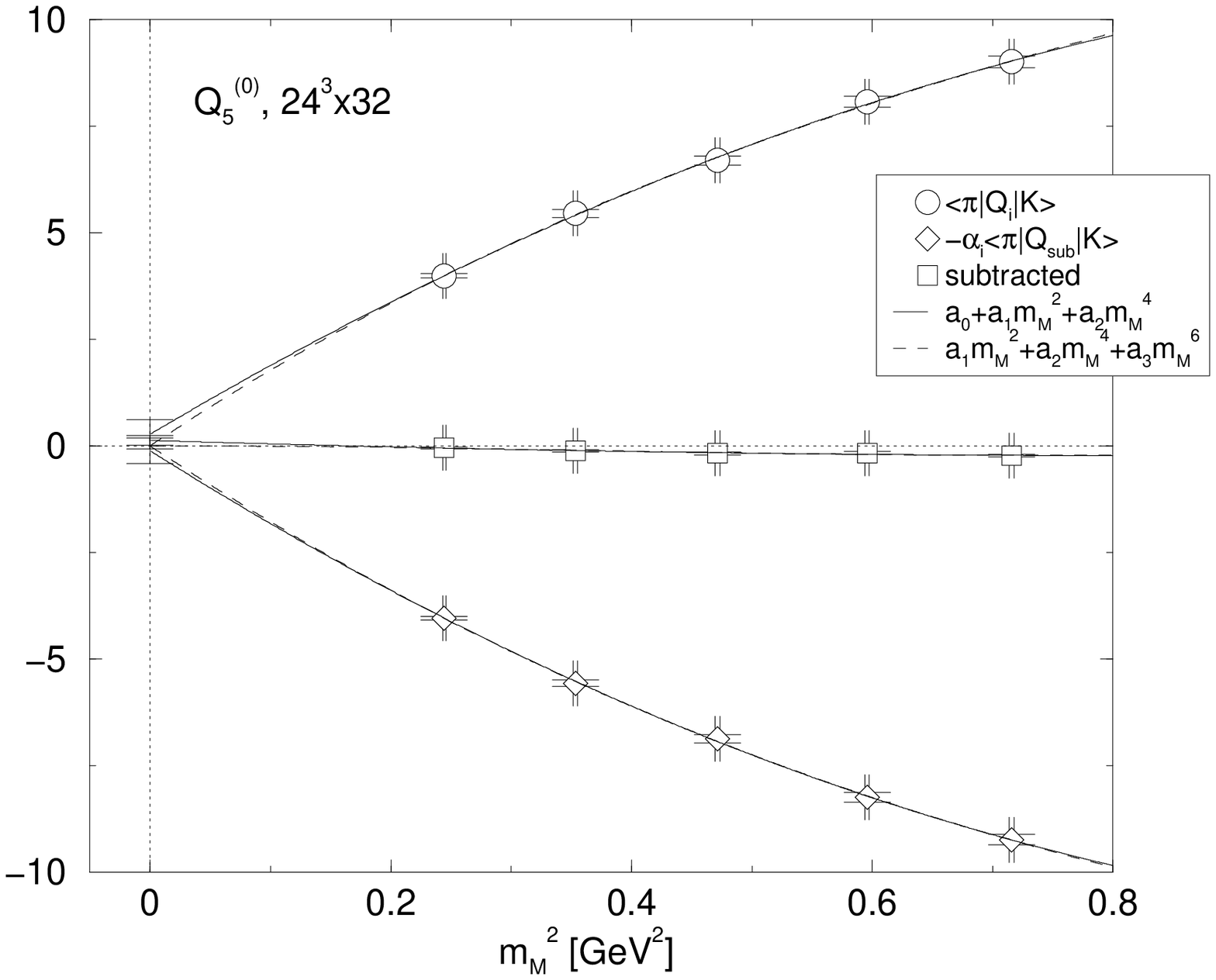}
 \end{center}
\end{figure}

\clearpage

\begin{figure}
 \begin{center}
  \includegraphics[width=8.1cm, clip]{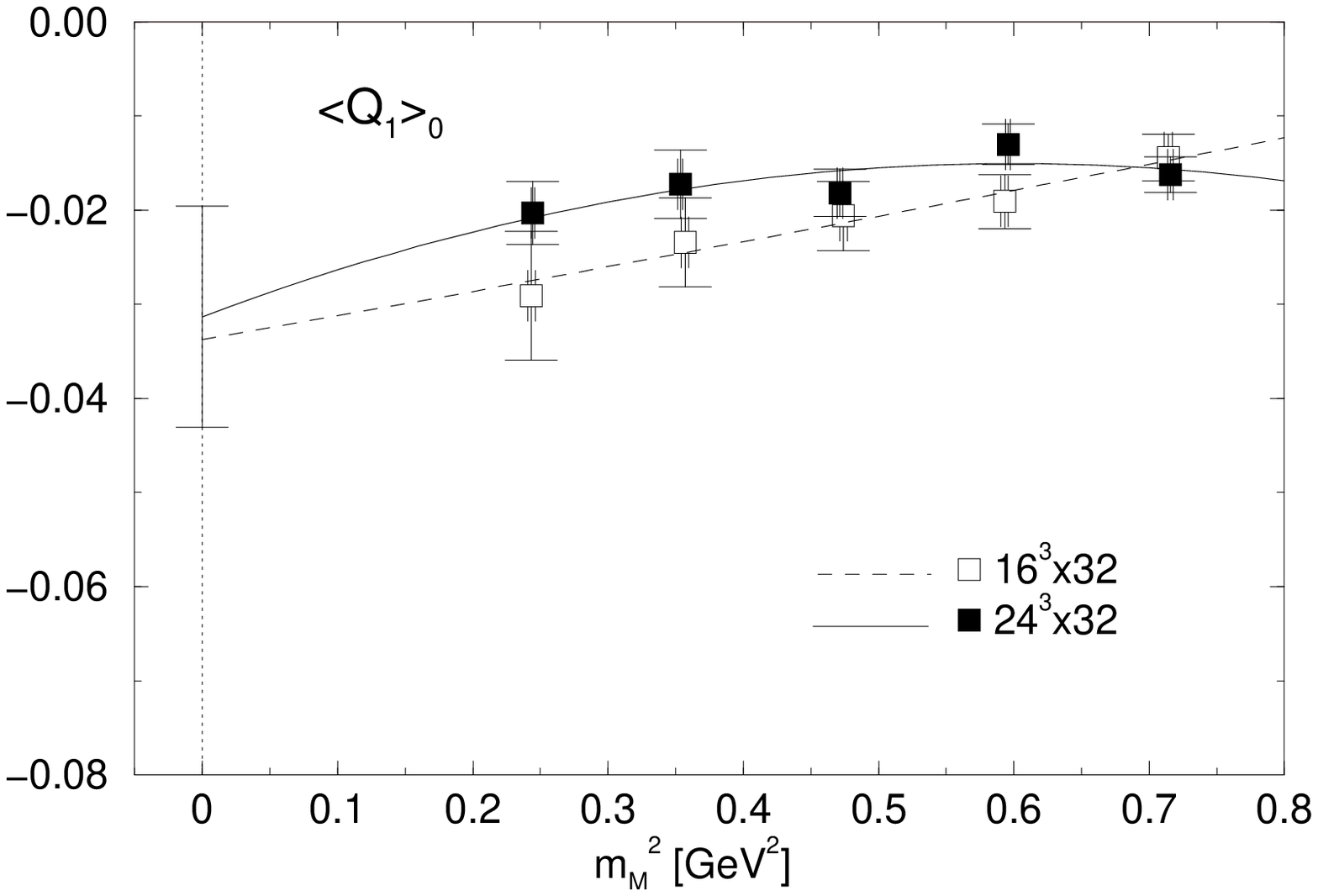}
  \includegraphics[width=8cm, clip]{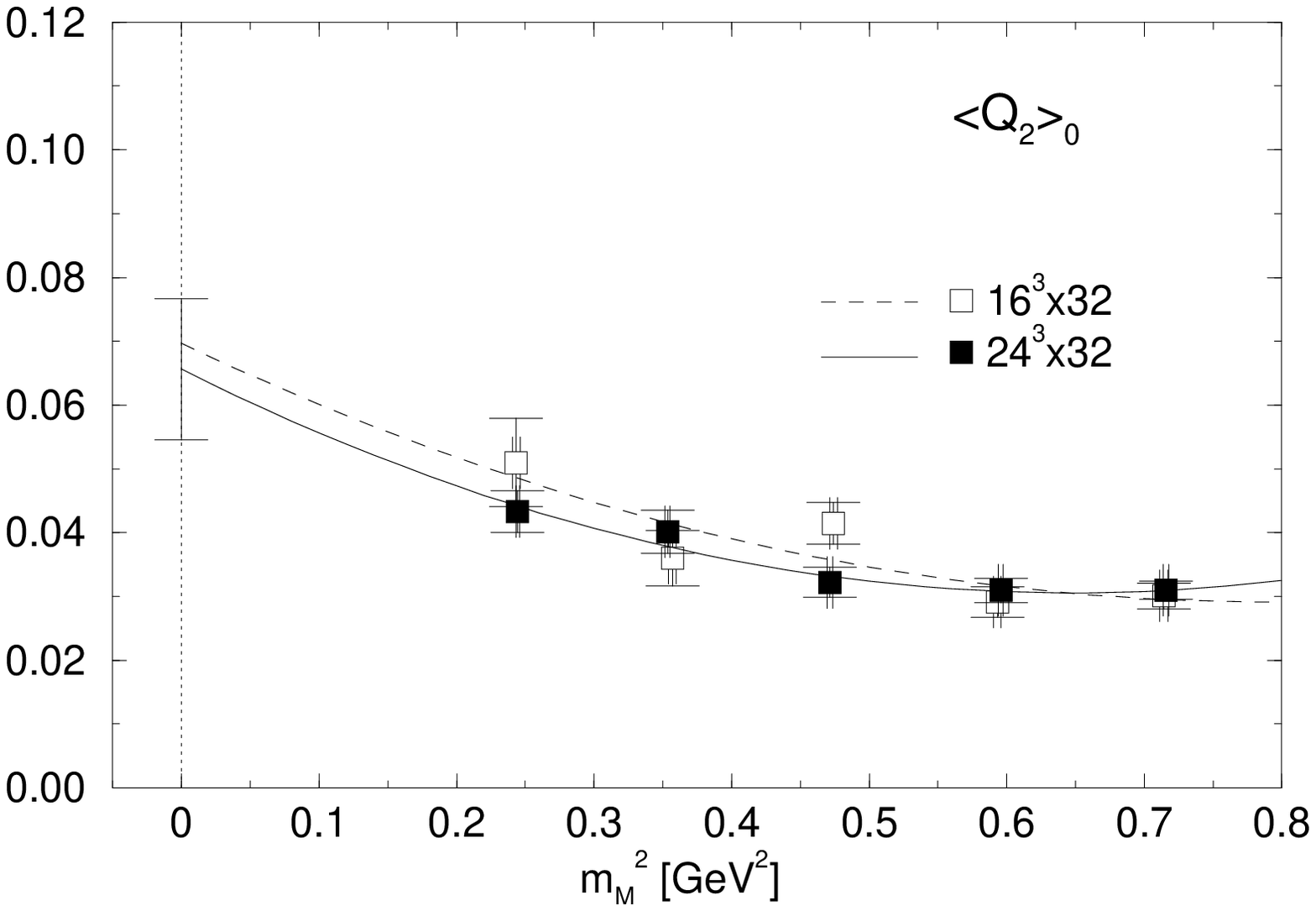}
 \end{center}
\hspace*{0.8cm}
 \begin{center}
  \includegraphics[width=8.3cm, clip]{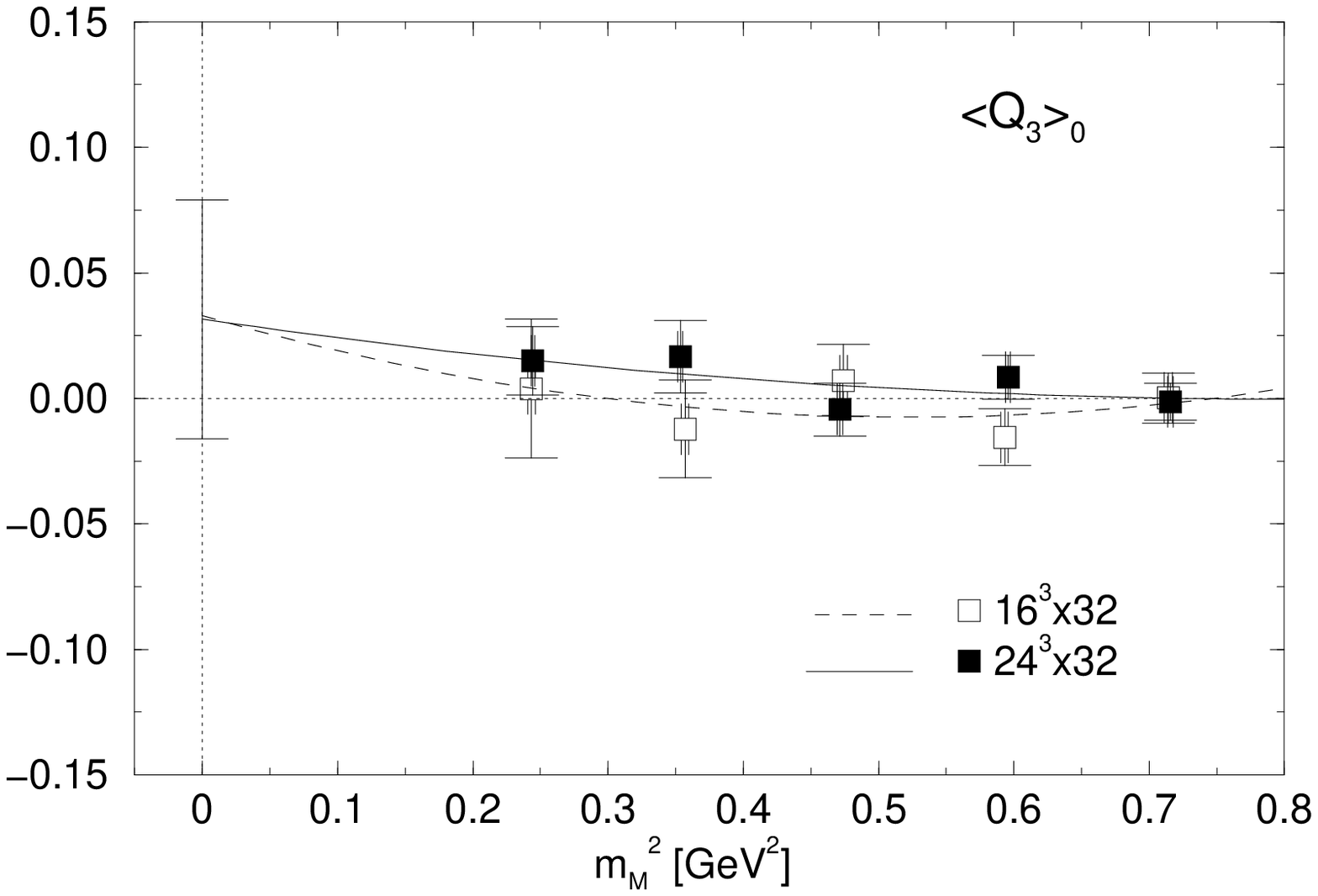}
  \includegraphics[width=8cm, clip]{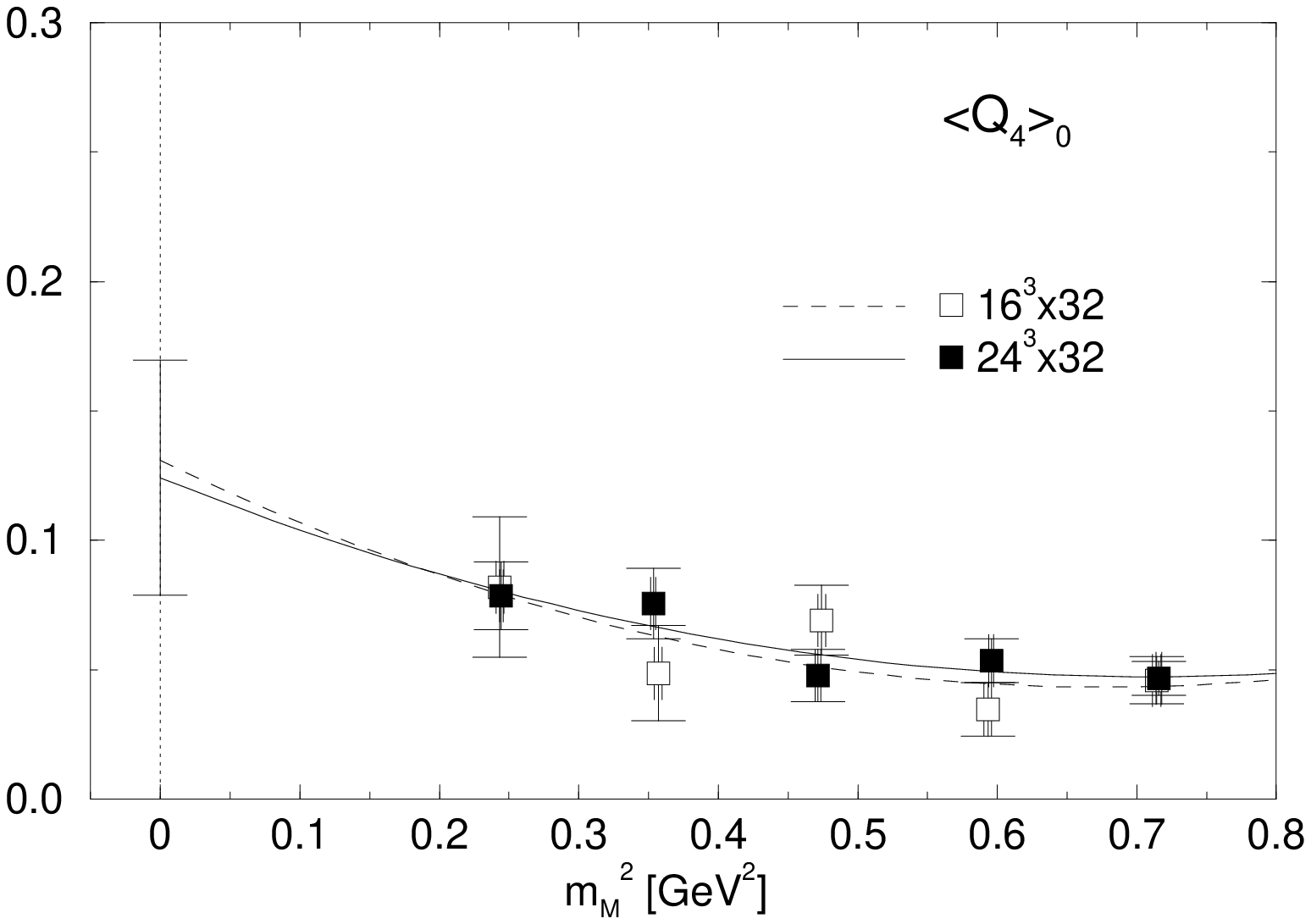}
 \end{center}
 \begin{center}
  \includegraphics[width=8.2cm, clip]{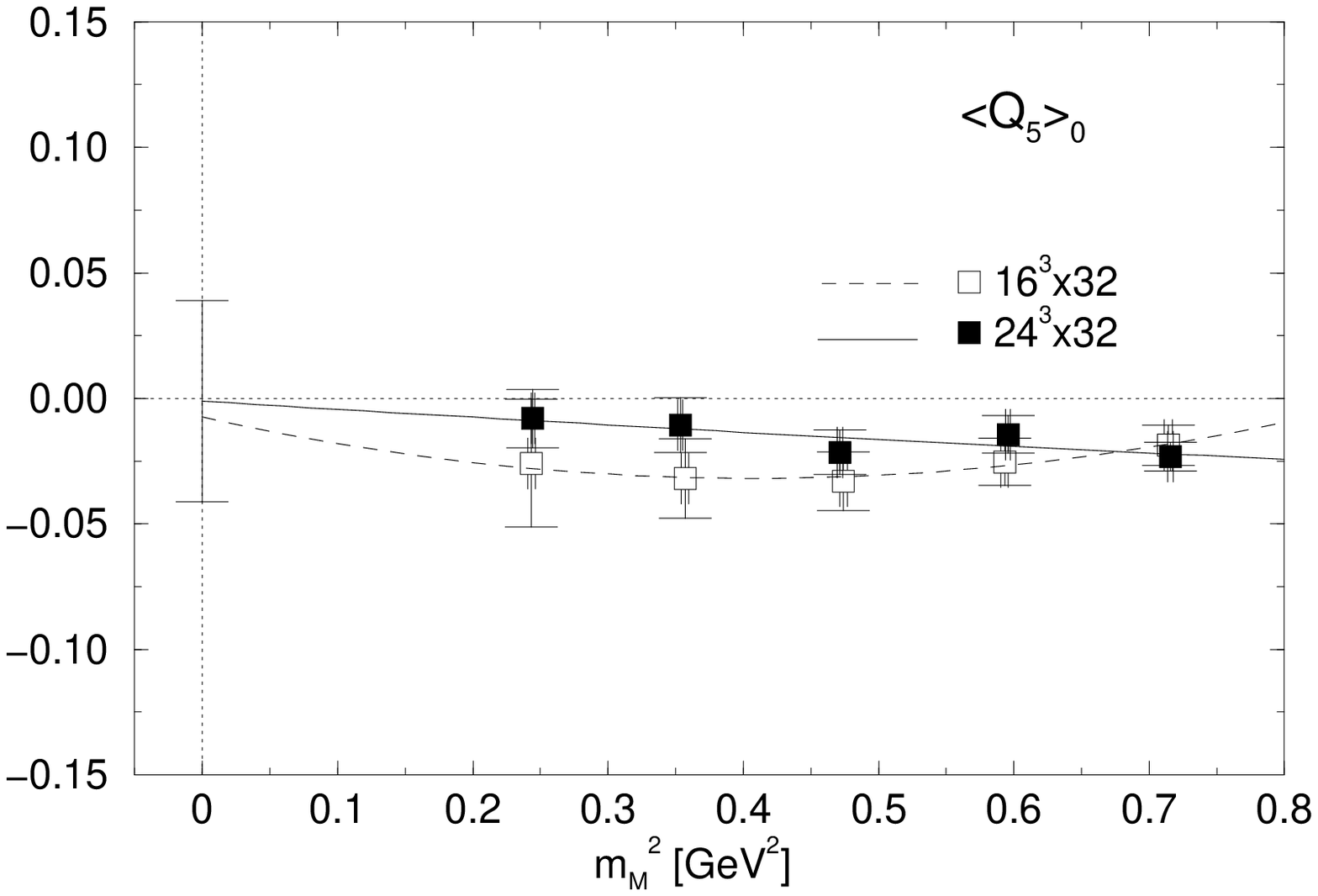}
  \includegraphics[width=8.1cm, clip]{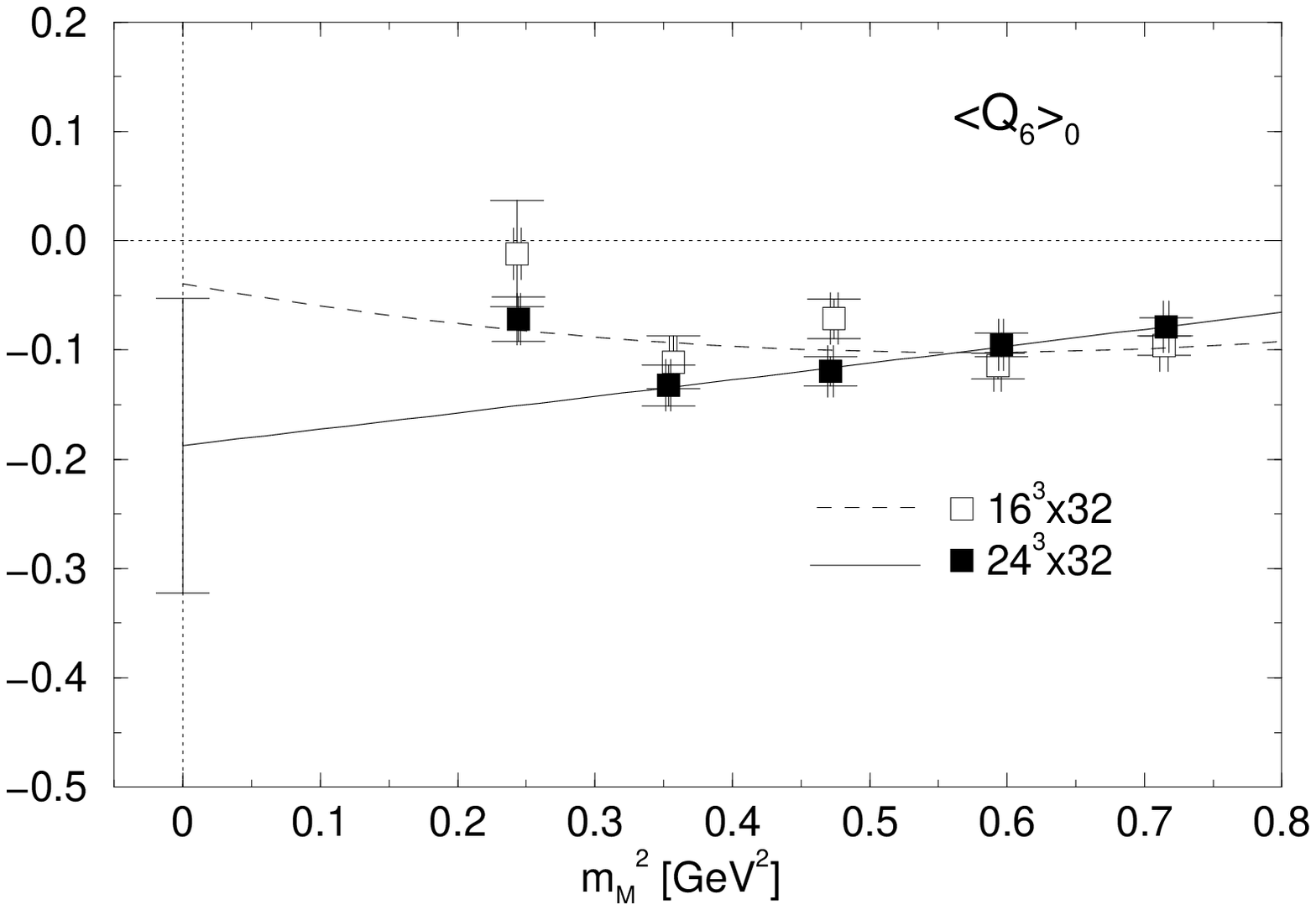}
 \end{center}
\end{figure}

\begin{thebibliography}{200}

\bibitem{ALA-HARA99}A.~Alavi-Harati {\it et al.}, \J{\PRL}{83}{1999}{22};
J. Graham, ``{\it A new measurement of $\varepsilon'/\varepsilon$}'', 
Fermilab Seminar (June 8, 2001), 
http://kpasa.fnal.gov:8080/public/ktev.html

\bibitem{FANTI99}V.~Fanti {\it et al.}, \J{\PL}{B465}{1999}{335};
G. Unal, ``{\it A new measurement of direct CP violation by NA48}'', 
CERN Particle Physics Seminar (May 10, 2001),
http://www.cern.ch/NA48/

\bibitem{KobayashiMaskawa} M.~Kobayashi and T.~Maskawa,
\J{\PTP}{49}{1973}, 652.

\bibitem{Buras-lec}For reviews, see, 
A.~J.~Buras, Lectures given at the Les Housche Summer 
School '98 (hep-ph/9806471); Lectures given at the 38th Erice 
International School of Subnuclear Physics '00 (hep-ph/0101336).

\bibitem{cabibbo84} 
N. Cabibbo, G. Martinelli, R. Petronzio, 
\J{\NP}{B244}{1984}{381}.

\bibitem{brower84}
R. Brower, M. Gavela, R. Gupta, G. Maturana, \J{\PRL}{53}{1984}{1318}.

\bibitem{bernard84}
C. Bernard, in {\it Gauge Theory on a Lattice: 1984}, ed. C. Zachos 
{\it et al.} (National Technical Information Service, Virginia, 1984). 

\bibitem{lellouch} For a recent review, see, 
L.~Lellouch, \J{\NPSup}{94}{2001}{142}. 

\bibitem{Kaplan92}D.~Kaplan, \J{\PL}{B288}{1992}{342}.

\bibitem{Shamir93}Y.~Shamir, \J{\NP}{B406}{1993}{90}.

\bibitem{Shamir95}V.~Furman and Y.~Shamir, \J{\NP}{B439}{1995}{54}.

\bibitem{Kuramashi99} For a review, see Y. Kuramashi, 
\J{\NPSup}{83}{2000}{24}.

\bibitem{MaianiTesta90}L.~Maiani and M.~Testa, \J{\PL}{B245}{1990}{585}.

\bibitem{Historic-reviews} Early attempts were reviewed in,  
C. Bernard and A. Soni, \J{\NPSup}{9}{1989}{155}.
For recent studies, see, L. Lellouch, \cite{lellouch}. 

\bibitem{Bernard85}C.~Bernard, T.~Draper, A.~Soni, H.~D.~Politzer 
and M.~B.~Wise, \J{\PR}{D32}{1985}{2343}.

\bibitem{JLQCD-ReA2}JLQCD Collaboration, S.~Aoki {\it et al.}, 
\J{\PR}{58}{1998}{054503}.

\bibitem{Ciuchini-etal96}M.~Ciuchini, E.~Franco, G.~Martinelli
and L.~Silvestrini, \J{\PL}{B380}{1996}{353}.

\bibitem{Lellouch-Luscher}L.~Lellouch and M.~L\"uscher, 
Commun. Math. Phys. {\bf 219} (2001) 31.

\bibitem{Marctinelli-etal2001}C.-J.~D.~Lin, G.~Martinelli, 
C.~T.~Sacharajda and M.~Testa, \J{\NP}{B619}{2001}{467}.

\bibitem{Gupta-etal}R.~Gupta, T.~Bhattacharya and S.~R.~Sharpe, 
\J{\PR}{D55}{1997}{4036}.

\bibitem{Donini-etal99}C.~Donini, V.~Gimenez, L.~Giusti and 
G.~Martinelli, \J{\PL}{B470}{1999} 233.

\bibitem{Lell-David} Laurent~Lellouch and C.-J.~David~Lin,
        \J{\NPSup}{73}{1999} 312.

\bibitem{PK}D.~Pekurovsky and G.~Kilcup, \J{\PR}{D64}{2001}{074502}; 
            D.~Pekurovsky, hep-lat/9909141.

\bibitem{Blum-Soni} T.~Blum and A.~Soni,
\J{\PR}{D56}{1997}{174}; \J{\PRL}{79}{1997}{3595}. 

\bibitem{AIKT00}S.~Aoki, T.~Izubuchi, Y.~Kuramashi and Y.~Taniguchi,
 \J{\PR}{D62}{2000}{094502}.

\bibitem{CPPACSdwqcd}CP-PACS Collabolation, A. Ali Khan {\it et al.}, 
\J{\PR}{D63}{2001}{114504}.

\bibitem{RBCcollab}T.~Blum {\it et al.}, hep-lat/0007038.

\bibitem{BK-CPPACS}CP-PACS Collabolation, A. Ali Khan {\it et al.}, 
\J{\PR}{D64}{2001}{114506}.

\bibitem{BK-JLQCD}JLQCD Collaboration, S.~Aoki {\it et al.},
        \J{\PRL}{81}{1778}{1998}; \J{\PR}{D60}{034511}{1999}.

\bibitem{CPPACSepep}CP-PACS Collaboration, 
A. Ali Khan {\it et al.}, \J{\NPSup}{94}{2001}{283}. 

\bibitem{Blum00}RBC Collaboration (T.~Blum {\it et al.}), 
\J{\NPSup}{94}{2001}{291}.

\bibitem{Mawhiney00}RBC Collaboration (R.~D.~Mawhinney for the
collaboration), \J{\NPSup}{94}{2001}{315}.

\bibitem{Vranas-ple} P.~Vranas, \J{\NPSup}{94}{2001}{177}.

\bibitem{Bij-Wise84}J. Bijnens and M. B. Wise, 
Phys. Lett. {\bf B137} (1984) 245. 

\bibitem{Ciri-Golo00} V. Cirigliano and E. Golowich, Phys. Lett. 
{\bf B475} (2000) 351.

\bibitem{ishizuka}N.~Ishizuka, unpublished.

\bibitem{Iwasaki83}Y.~Iwasaki, preprint, UTHEP-118 (Dec. 1983), unpublished;
\J{\NP}{B258}{1985}{141}.

\bibitem{IKKY}Y.~Iwasaki, K.~Kanaya, T.~Kaneko and T.~Yoshie,
\J{\PR}{D56}{1997}{151}.

\bibitem{lat99Kaneko}CP-PACS Collaboration, A.~Ali Khan {\it et al.},
\J{\NPSup}{83-84}{2000}{176}.

\bibitem{okamoto}CP-PACS Collaboration, M.~Okamoto {\it et al.},
\J{\PR}{D60}{1999}{094510}.

\bibitem{Buch-Buras-Lauten95}
For a review, see, G.~Buchalla, A.~J.~Buras and
M.~E.~Lautenbacher, \J{\RMP}{68}{1996}{1125}.

\bibitem{Bethke00}
See, {\it e.g.}, S. Bethke, J. Phys. G: Nucl. Part. Phys. {\bf 26} (2000) R27.

\bibitem{AIKT98}S.~Aoki, T.~Izubuchi, Y.~Kuramashi and Y.~Taniguchi,
 \J{\PR}{D59}{1999}{094505}.

\bibitem{AIKT99}S.~Aoki, T.~Izubuchi, Y.~Kuramashi and Y.~Taniguchi, 
Phys.\ Rev.\ {\bf D60} (1999) 114504.  

\bibitem{AK00}S.~Aoki and Y.~Kuramashi, Phys.\ Rev.\ {\bf D63} (2001) 054504.

\bibitem{Taniguchi01} S.~Aoki, T.~Izubuchi, Y.~Kuramashi and Y.~Taniguchi,
in preparation.

\bibitem{Golterman:2000fw}
M.~Golterman and E.~Pallante, JHEP {\bf 0008} (2000) 023.

\bibitem{Donoghue-Golowich}
J.~Donogue and E.~Golowich, \J{\PL}{B478}{2000}{172}.
See V.~Cirigliano, J.~Donogue, E.~Golowich and K.~Maltman,
\J{\PL}{B522}{2001}{245} for the latest result.

\bibitem{Golterman:2001}
M.~Golterman and E.~Pallante, JHEP {\bf 0110} (2001) 037. 

\bibitem{KS2}
T.~Bhattacharya {\it et al.}, \J{\NPSup}{106}{2002}{311}.

\bibitem{PDG} D.~E.~Groom {\it et al.}, The European Physical Journal 
{\bf C15}, 1 (2000). 

\bibitem{Bosch-etal99} S.~Bosch {\it et al.}, \J{\NP}{B565}{2000}{3}.

\bibitem{Martinelli-etal}
G.~Altarelli, G.~Curci, G.~Martinelli and S.~Petrarca, 
\J{\NP}{B187}{1981}{461};
M.~Ciuchini, E.~Franco, G.~Martinelli, L.~Reina, \J{\PL}{B301}{1993}{263};
\J{\NP}{B415}{1994}{403}.

\bibitem{Buras-etal}A.~J.~Buras and P.~H.~Weisz, \J{\NP}{B333}{1990}{66};
A.~J.~Buras, M.~Jamin, M.~E.~Lautenbacher and P.~H~.Weisz, 
\J{\NP}{B370}{1992} 69; addendum \J{\NP}{B375}{1992} 501;
\J{\NP}{B400}{1993}{37};
A.~J.~Buras, M.~Jamin and M.~E.~Lautenbacher, \J{\NP}{B400}{1993} 75;
\J{\NP}{B408}{1993}{209}.

\end{thebibliography}
\end{document}